\RequirePackage{fix-cm}

\documentclass[smallextended]{svjour3}
\smartqed  
 \usepackage{mathptmx}      
\usepackage[utf8]{inputenc}
\usepackage[english]{babel}
\usepackage{amsmath, array}
\usepackage{amssymb}
 
\usepackage{amsthm}
\usepackage{graphicx}
\usepackage[caption=false]{subfig}
\usepackage{enumitem}
\usepackage{bm}
\usepackage{float}
\usepackage{xcolor}
\usepackage{ wasysym }
\usepackage{verbatim}
\usepackage{polynom}
\usepackage{multicol}
\newcommand{\ds}{\displaystyle}
\renewcommand{\l}{\left(}
\renewcommand{\r}{\right)}

\newcommand{\p}{\partial}

\DeclareMathOperator*{\argmin}{argmin}
\usepackage{natbib}
\usepackage{pdflscape}
\usepackage{afterpage}
\usepackage{makecell}

 \journalname{Bulletin of Mathematical Biology Preprint}
\begin{document}

\title{Parameter Estimation for Mixed-Mechanism Tear Film Thinning}

\author{Rayanne A. Luke         \and
        Richard J. Braun \and Tobin A. Driscoll \and Deborah Awisi-Gyau \and Carolyn G. Begley 
}

\authorrunning{R. Luke \textit{et al.}} 

\institute{R. A. Luke \at
              Department of Mathematical Sciences, University of Delaware, Newark, DE, 19716, USA \\
              \email{rayanne@udel.edu}
}

\date{\today}

\maketitle

\begin{abstract}

Etiologies of tear breakup include evaporation-driven, divergent flow-driven, and a combination of these two. A mathematical model incorporating evaporation and lipid-driven tangential flow is fit to fluorescence imaging data. The lipid-driven motion is hypothesized to be caused by localized excess lipid, or ``globs.'' Tear breakup quantities such as evaporation rates and tangential flow rates cannot currently be directly measured during breakup. We determine such variables by fitting mathematical models for tear breakup and the computed fluorescent intensity to experimental intensity data gathered \textit{in vivo}. Parameter estimation is conducted via least squares minimization of the difference between experimental data and computed answers using either the trust-region-reflective or Levenberg-Marquardt algorithm. Best-fit determination of tear breakup parameters supports the notion that evaporation and divergent tangential flow can cooperate to drive breakup. The resulting tear breakup is typically faster than purely evaporative cases.  Many instances of tear breakup may have similar causes, which suggests that interpretation of experimental results may benefit from considering multiple mechanisms.

\end{abstract}

\section{Introduction}

The tear film (TF) provides necessary moisture and nutrients to the ocular surface and, when its thickness is uniform, a smooth optical surface for clear vision. The TF is classically considered a three-layered film, composed of a thin, 20 to 100 nm or more thick oily lipid layer (\citealt{king2011,braun2015}), an aqueous layer a few microns thick (\citealt{king2004,lu2014,huang2016}), and the glycocalyx, a half-micron thick bound mucin layer that sits atop the ocular surface  (\citealt{king2004}). Evaporation of water from the TF is thought to be retarded by the lipid layer (\citealt{mishima61,king2010, dursch2018}), although there is some disagreement on this issue (\citealt{georgiev2017}). A healthy glycocalyx promotes wetting of the ocular surface (\citealt{gipson2004,argueso2001,tiffany1990,tiffany1990b}). The lipid layer comprises mostly nonpolar lipids, but surface-active polar lipids exist at the aqueous/lipid interface that can act as a surfactant and drive aqueous flow (\citealt{johnson2004,mcculley1997,butovich2013}). Structure and function of the lipid layer is an active area of research (\citealt{borchman2019,paananen2020}).

The majority of the aqueous layer of the TF is supplied by the lacrimal gland near the temporal canthus (\citealt{dartt2009}), with the puncta draining the excess near the nasal canthus during the opening interblink phase (\citealt{doane81}). The meibomian glands of the eyelids secrete the lipid layer and cells in the conjunctival epithelium supply the soluble mucins (\citealt{aydemir2010}). Tangential flow along the corneal surface can be directed inward if driven by pressure-induced capillary flow (\citealt{oron1997}). Alternatively, the Marangoni effect may drive outward flow, whereby surface concentration gradients induce shear stress at the aqueous/lipid interface (\citealt{craster2009}). Evaporation of water from the tear film into the air decreases the fluid volume (\citealt{mishima61}; \citealt{tomlinson09}; \citealt{kimball2010}).  Osmosis supplies water from the ocular epithelia (\citealt{braun2012}; \citealt{cerretani2014}; \citealt{braun2015}).

Tear film breakup (TBU) is considered to occur when a dark spot appears in the fluorescent tear film,also called a dry spot (\citealt{norn1969}). Tear breakup time (TBUT) is the time required to produce the first dark spot in the tear film, as judged clinically. A related term is full-thickness tear breakup (FT-TBU) (\citealt{begley13}), when there is effectively no aqueous layer between the lipid layer and glycocalyx. This term is referred to as ``touchdown'' in \cite{king2018}. We refer to the first occurrence of this in the trial as full-thickness breakup time (FT-TBUT). \cite{king2018} theorize that at FT-TBUT, the inner polar lipids of the tear film lipid layer touch the outer tips of the glycocalyx. Tear film breakup time (TBUT) and FT-TBUT can differ by as much as minutes if the dark spot appears and then thins very slowly. 

Studying TBU and FT-TBU is important to understanding dry eye syndrome (DES), as TF instability has been suggested to play an important etiological role in the disease (\citealt{craig2017defn,willcox2017tf}). DES affects between 5 and 50\% of the population depending on the diagnostic criteria used, and diminishes quality of life, vision, and ocular comfort (\citealt{nelson2017}). 
TBU is considered an etiological factor that may induce DES via inadequate lubrication of the ocular surface, hyperosmolarity of the TF (\citealt{gilbard1978,lemp2007,willcox2017tf}) and imflammation (\citealt{mertzanis2005,miljanovic2007}).

Osmolarity is defined as a combined osmotically-active solute concentration that comprises mostly salt ions in the aqueous layer (\citealt{stahl2012}). Osmotic flow from the cornea to the tear film is generated by a concentration difference between the aqueous layer and the corneal epithelium (\citealt{peng2014,braun2015}).
Lab-on-a-chip technology allows osmolarity to be measured in the inferior meniscus in a clinical setting (\citealt{lemp2011}). The osmolarity of a normal (non-dry eye) tear film is in the range 296–302 mOsM (\citealt{lemp2011}; \citealt{tomlinson2006}; \citealt{versura2010}); healthy blood ranges from 285–295 mOsM (\citealt{tietz1995}). Some observations show that meniscus osmolarity levels reach 316–360 mOsM in DES (\citealt{gilbard1978};
\citealt{tomlinson2006}; \citealt{sullivan2010}; \citealt{dartt2013}). Since clinical measurements of osmolarity cannot target the cornea, estimates from experiment or mathematical models are useful. \cite{liu09} experimentally estimated values as high as 900 mOsM. \citet{braun2015} and \cite{peng2014} computed similar or higher values in mathematical models of TBU. \cite{li2016} did so for models of the whole ocular surface. \cite{luke2020} estimated maximum values ranging between 645-864 mOsM by fitting experimental data with the evaporation-driven thinning model given in \cite{braun2017}. Of interest in this article is the osmolarity of the tear film during thinning up to FT-TBU and how the dominant mechanism causing thinning or the trial length may affect the maximum salt concentration attained in a breakup region. \cite{braun2015} found osmolarity values up to ten times the isotonic concentration when modeling the TF as a spatially uniform film that thinned due to evaporation. 

Imaging is an important tool for analyzing TF dynamics. Fluorescence imaging (\citealt{king2013}), spectral interferometry (\citealt{king2004, king2009, nichols2005}) and optical coherence tomography (\citealt{wang2003}) are all common imaging techniques.  Insertion of dyes such as fluorescein have been used to stain epithelial cells (\citealt{norn1970,bron2015}, e.g.), estimate tear drainage rates or turnover times (\citealt{webber86}), visualize general TF dynamics (\citealt{benedetto86}; \citealt{begley13}; \citealt{king2013q}; \citealt{li2014}), estimate TF first breakup times (\citealt{norn1969}), and capture the progression of breakup regions (\citealt{liu06}).  We will refer to the fluorescent quantities (e.g., concentration and intensity) using FL. An FL concentration below the critical concentration is in the dilute regime; above the critical concentration, it is in the self-quenching regime (\citealt{webber86}). FL intensity is proportional to TF thickness in the dilute regime; TF thickness is approximately proportional to the square root of FL intensity in the self-quenching regime (\citealt{nichols2012,braun2014}).  Simultaneous imaging via interferometry for the lipid layer thickness and FL intensity for the aqueous layer found TBU is caused by different mechanisms (\citealt{king2013}). 
\citet{braun2015} found that flow inside the
tear film during TBU can advect fluorescein and thereby change the expected appearance of the TBU; this can complicate interpretation of FL imaging. Simultaneous imaging can help interpret TF dynamics (\citealt{himebaugh2012,king2013,arnold2010}).

Models with a single independent space dimension incorporating surface tension, viscosity, gravity, evaporation, and wetting forces have been developed to study breakup. \cite{sharma85, sharma86} extended previous work to include a fluid mucus layer with van der Waals-driven breakup. Two-layer film theory has been extended by \cite{zhang03,zhang04} to include van der Waals forces in both mucus and aqueous layers, as well as surfactant transport. The authors found that van der Waals forces drove the fluid mucus layer to instability and the TF to rupture; both were interpreted as breakup.

To better to understand hyperosmolarity in breakup regions, recent TF thinning models have included osmolarity. \cite{braun2012,braun2015} studied an ordinary differential equation model with constant evaporation at the tear/air interface and osmotic flow at the tear/cornea interface proportional to the osmolarity increase above the isotonic value. 
 \cite{braun2012} stopped thinning at the glycocalyx by including van der Waals forces, which allowed a zero permeability condition to be used at the tear/cornea interface.
\citet{peng2014} extended these models to include space-dependent evaporation with two parts: a stationary, variable-thickness lipid layer with fixed resistance to diffusion of water, and air resistance, which included convective and diffusive transport outside the tear film. The authors found evaporation-driven elevated osmolarity levels in breakup regions, and that diffusion of solutes out of the breakup region prevented osmosis from stopping thinning. \citet{braun2015,braun2017} used a Gaussian or hyperbolic tangent evaporation profile with a central peak rate larger than the surrounding constant rate. All of the models found elevated osmolarity levels caused by evaporation-driven thinning in the breakup region that could reach several times the isotonic value. \citet{braun2017} were able to determine when capillary flow balances or dominates viscous effects by their scaling choices. Diffusion is shown to have a larger magnitude near breakup than that of advection for both solute distributions. In the center of breakup, diffusion of salt ions is four times faster than that of fluorescein, which causes the maximum osmolarity to fall short of the limiting value set by the flat film result (\citealt{peng2014,braun2014}).

Various mechanisms causing breakup have been proposed and studied, and can roughly be categorized by the time until FT-TBU is reached. Evaporation causes relatively slow TF thinning (\citealt{king2010}) and cannot explain rapid TBU, in which a dry spot may form in a few tenths of a second (\citealt{king2018,yokoi2013}). \cite{zhong2019} hypothesized that Marangoni-driven tangential flow can drive rapid thinning, while \cite{yokoi2013,yokoi2019} suggest that dewetting causes some instances of rapid circular thinning.
\cite{zhong2019} derived a model that includes rapid breakup induced by strong tangential flow caused by ``globs,'' or relatively thick areas of the lipid layer. 
The Marangoni effect induces flow due to a reduction in the aqueous/air surface tension brought about by the increase in surfactant concentration in the glob. 
The authors noted that if FT-TBUT (TBUT in their terminology) occurs in over 4 s, the cause of thinning is cooperative: tangential flow dominates early on but evaporation becomes the main mechanism later.  This cutoff is similar to that for short breakup time used by \cite{yokoi2019} and others.

Many experiments to measure TF thinning rates have been conducted and we discuss a sample here. \citet{hamano1981} used an invasive method in an open chamber to determine thinning rates. King-Smith and coworkers used spectral interferometry (\citealt{nichols2005,kimball2010,king2010}). 
 \cite{dursch2018} computed a weighted average of TF thinning rates over the cornea combined with a heat transfer analysis and thermal imaging. Their approach averaged pure water rates and slow rates from a functioning lipid layer. An evaporimeter measured evaporation over the eye palpebral fissure using controlled conditions (\citealt{peng2014b}). \cite{wong2018} reviewed literature values of measured evaporation rates over the palpebral fissure. None of these studies targeted areas of TBU specifically.

\cite{luke2020} developed a parameter estimation scheme for fitting experimental data from FL images with axisymmetric and linear mathematical models given in \cite{braun2017} for evaporation-driven thinning. They found realistic optimal values for peak and background evaporation rates and dry spot sizes, and their thinning rate estimates fell within experimental ranges (\citealt{nichols2005}). Minimum theoretical TF thickness values from the fits leveled off around 1.5 $\mu$m on average and maximum osmolarity estimates clustered a little over twice the isotonic value. Clinician-determined TBUT and FT-TBUT were compared with the first and last fit times from the optimizations. Last fit times were found to be within 20\% of FT-TBUT, whereas TBUT was 40-80\% shorter than FT-TBUT. Theoretical FT-TBUT was estimated using the time scale given by the model and was found to be smaller in comparison to both the fitting data and experimental data. This is likely because the spots and streaks analyzed were characterized as small by the theory (\citealt{braun2017}). Normalized theoretical TF thickness $h$ and FL intensity $I$ were compared for various initial FL concentrations. FL intensity computed using an initial FL concentration between 0.1\% and 0.15\% was shown to most closely match the TF thickness profile.

In this article we present the results of fitting a TF thinning model incorporating both lipid-driven tangential flow and evaporation developed by \cite{zhong2019} to experimental FL intensity data from healthy subjects' TFs. These findings include flow rates and lipid glob sizes that have not been measured \textit{in vivo} in FT-TBU. We believe these results advance the understanding of TF thinning and dry spot formation as they provide evidence that lipid-driven flow can cooperate with evaporation to cause breakup, and can serve as a reference point when comparing to dry eye patient data. 

This article is organized as follows. We describe the data used and present the axisymmetric model for spots with various evaporation distribution options; the corresponding streak version is found in Appendix \ref{s:streak_eqns}, along with a derivation of the circular model. Our fitting procedure is outlined and results are given. Discussion and conclusions follow.

\section{FL Images}
\label{sec:data}

We use data from twenty-five normal, healthy subjects taken in a study conducted at Indiana University (\citealt{awisigyau2020}) as discussed in \cite{braun2017} and \cite{luke2020}; we reiterate a brief description below. The study received approval from the Biomedical Institutional Review Board of Indiana University. Declaration of Helsinki principles were followed during data collection and informed consent was obtained from subjects. We refer to a trial as the sequence of images of the subject's eye. 2\% sodium fluorescein solution is instilled in the patient's eye and a light with a cobalt blue excitation filter is shined on the eye so that the aqueous layer of the tear film (TF) fluoresces green (\citealt{carlson2004}). (The critical FL concentration can also be expressed in molar as 0.0053 M; see Appendix \ref{sec:app_conc}.) 

\begin{figure}
\centering
\subfloat[][S9v1t4]{\includegraphics[scale=.06]{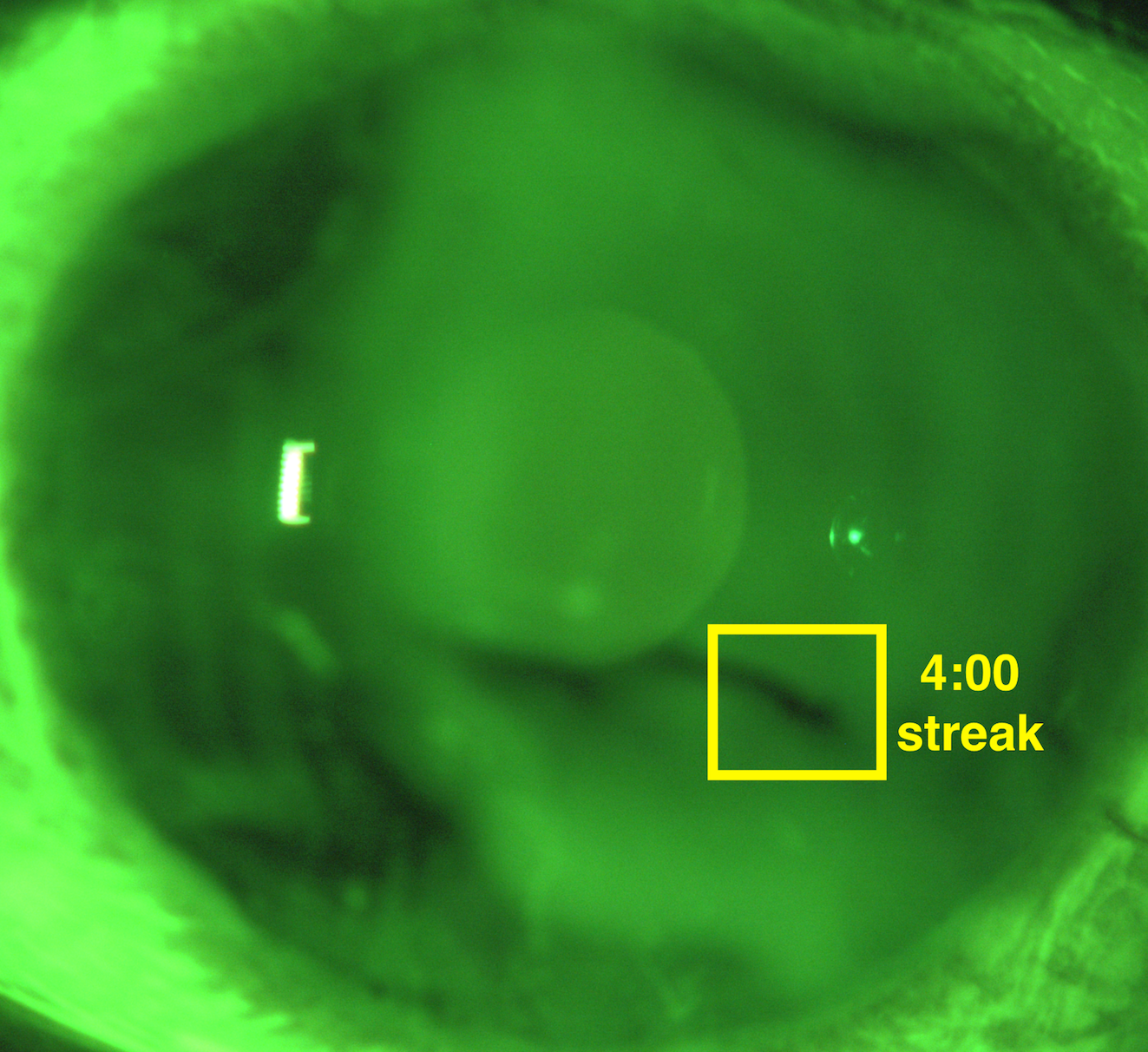}}
\subfloat[][S9v2t1]{\includegraphics[scale=.0592]{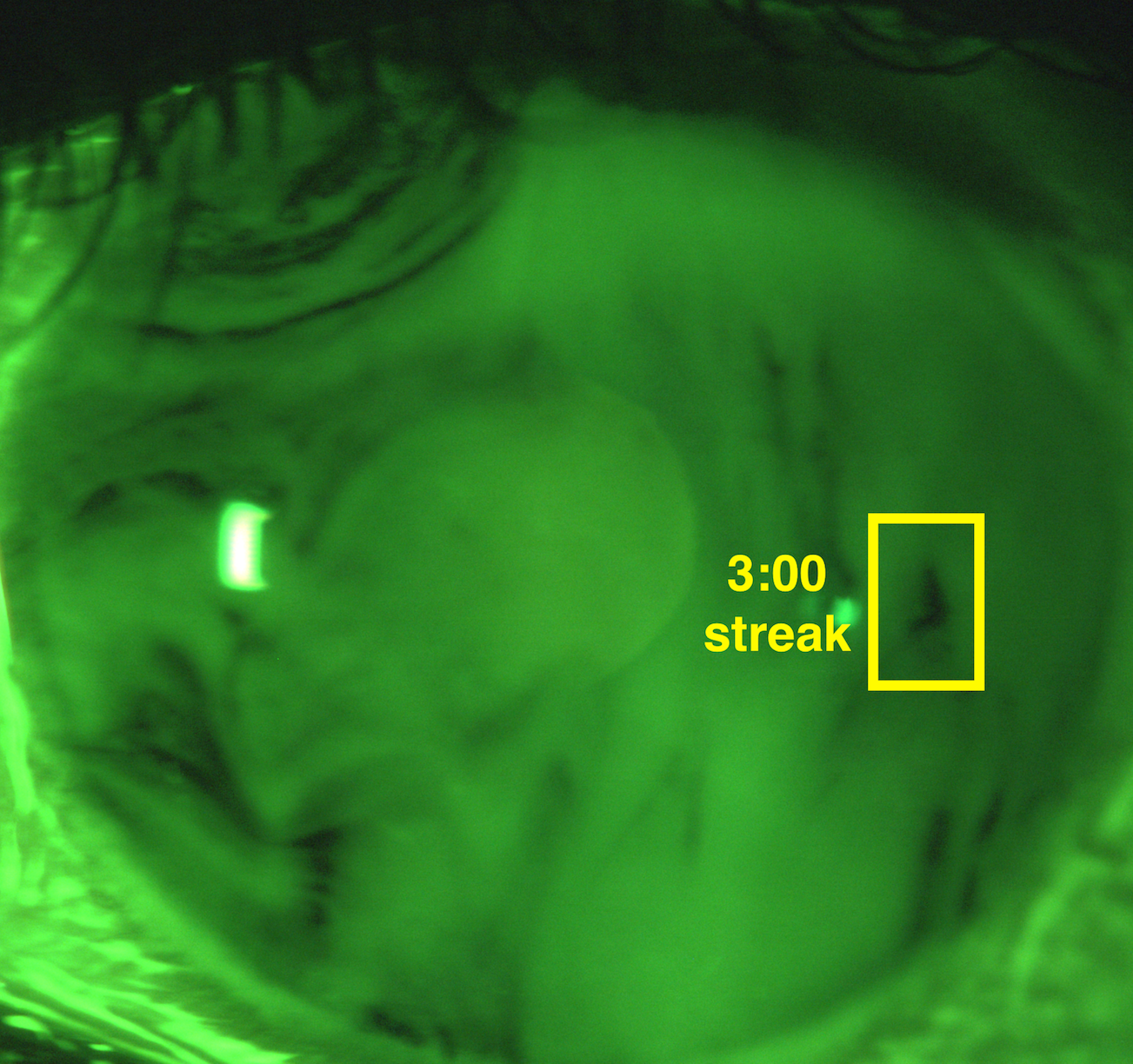}}
\subfloat[][S9v2t5]{\includegraphics[scale=.0595]{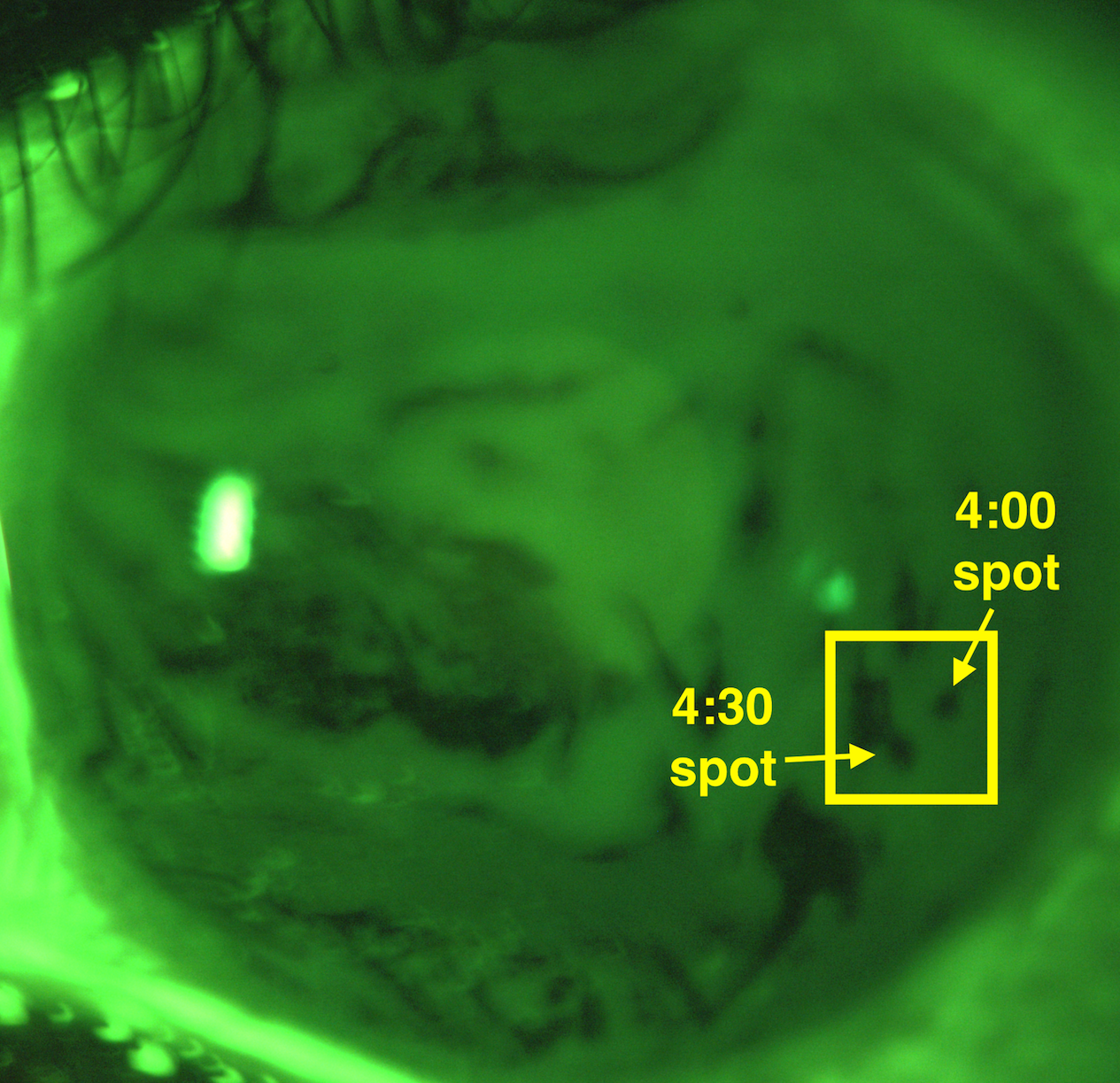}} \\
\subfloat[][S10v1t6]{\includegraphics[scale=.055]{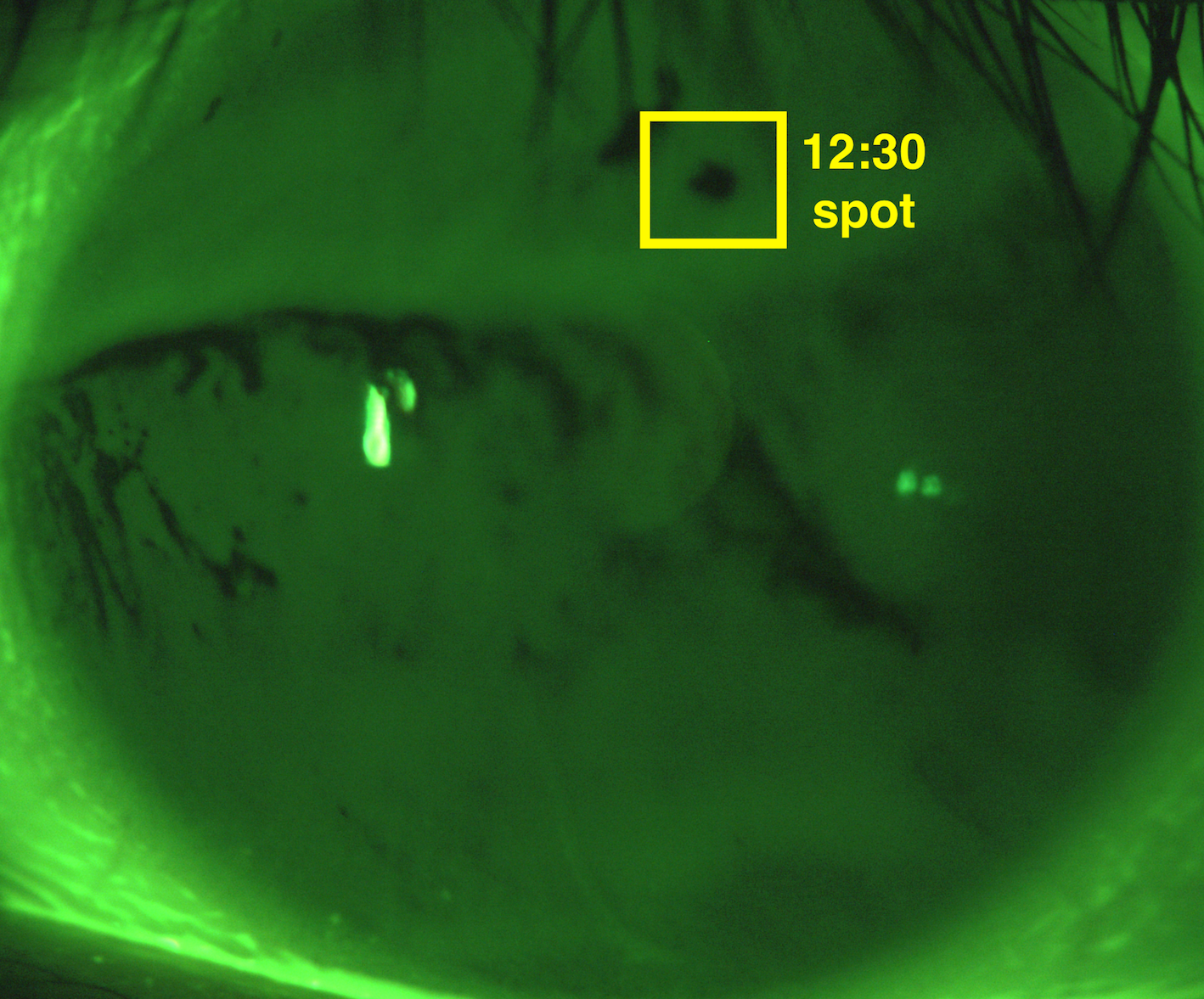}} 
\subfloat[][S13v2t10]{\includegraphics[scale=.062]{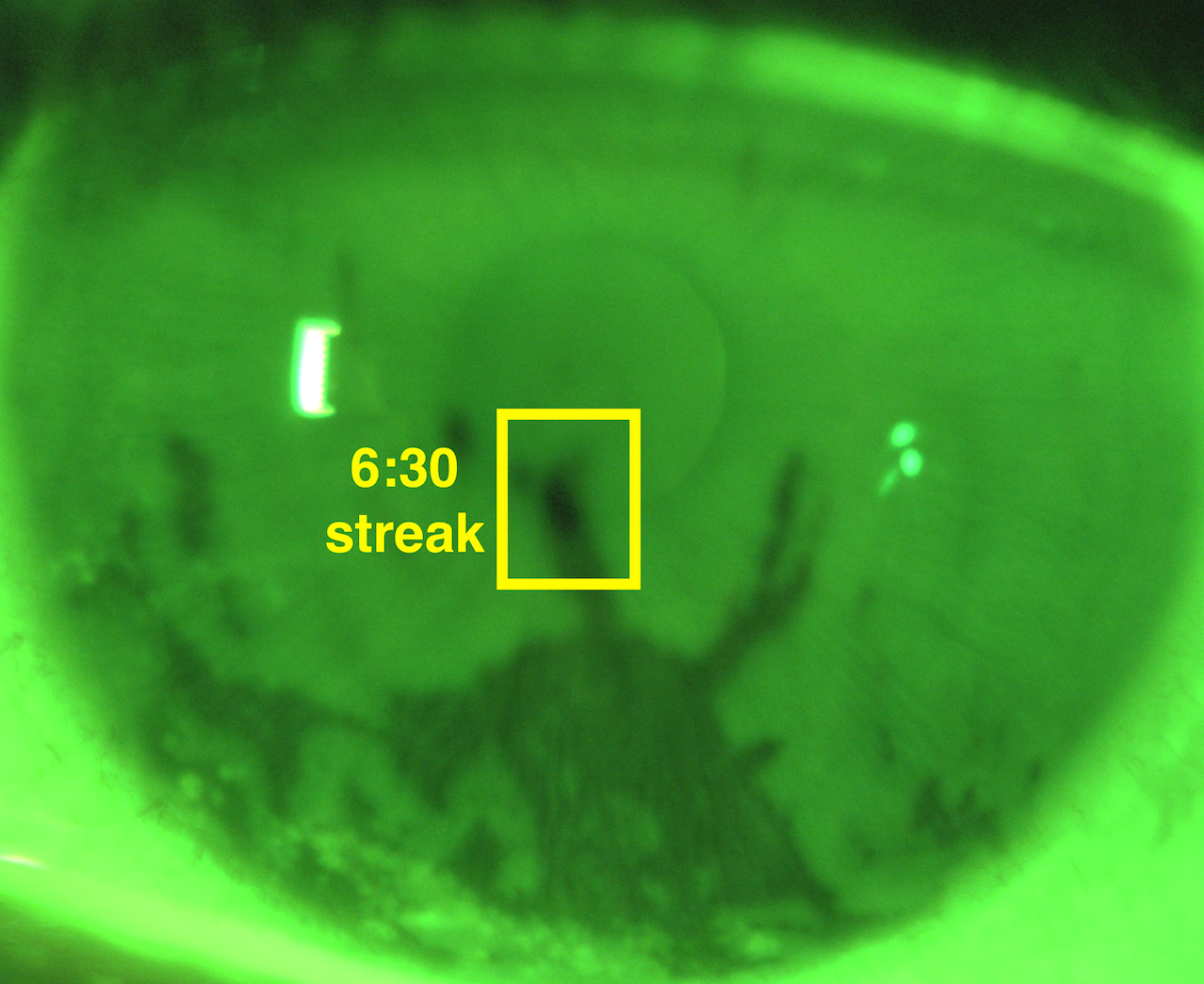}}
\subfloat[][S18v2t4]{\includegraphics[scale=.0505]{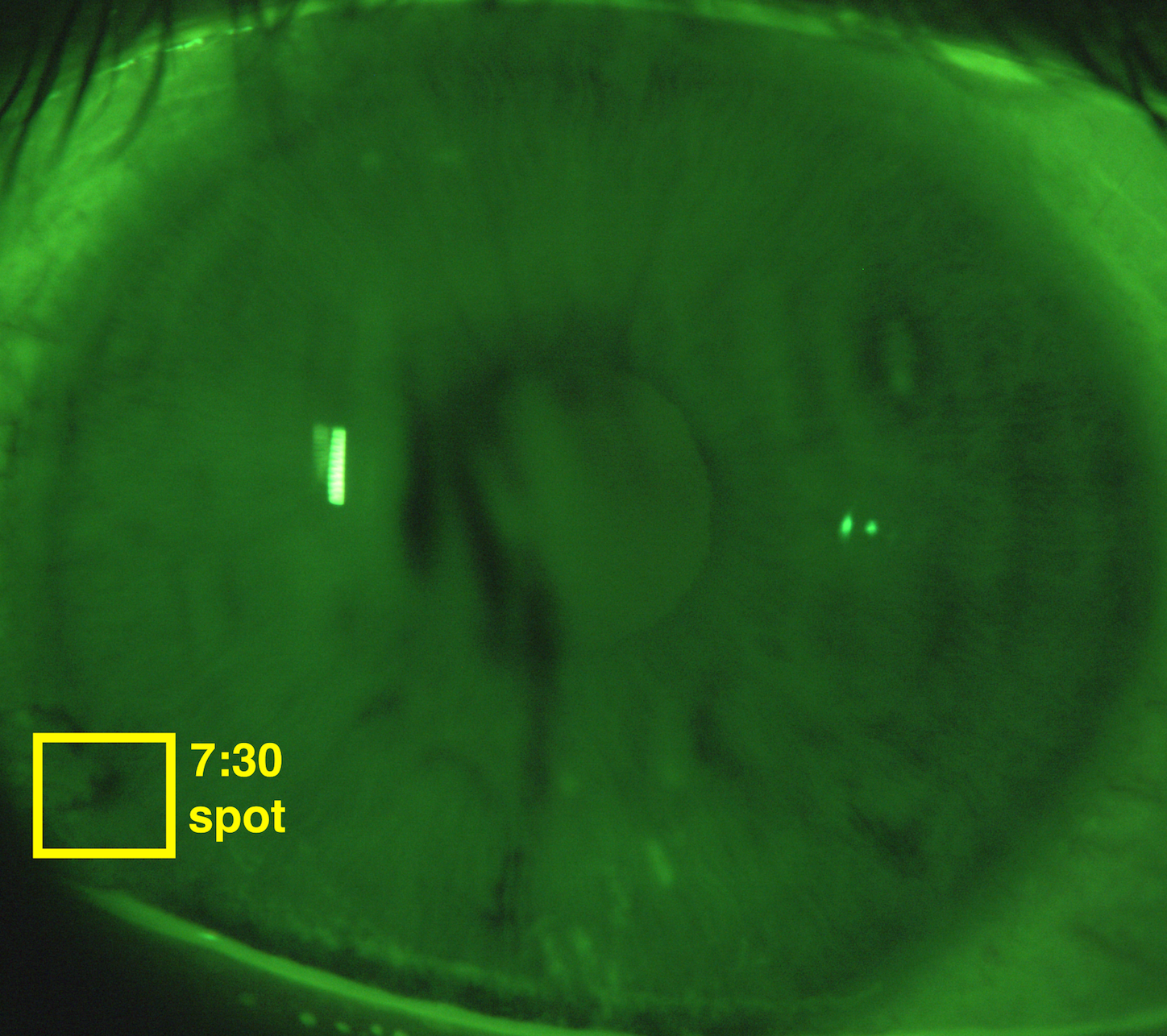}} \\
\subfloat[][S27v2t2]{\includegraphics[scale=.055]{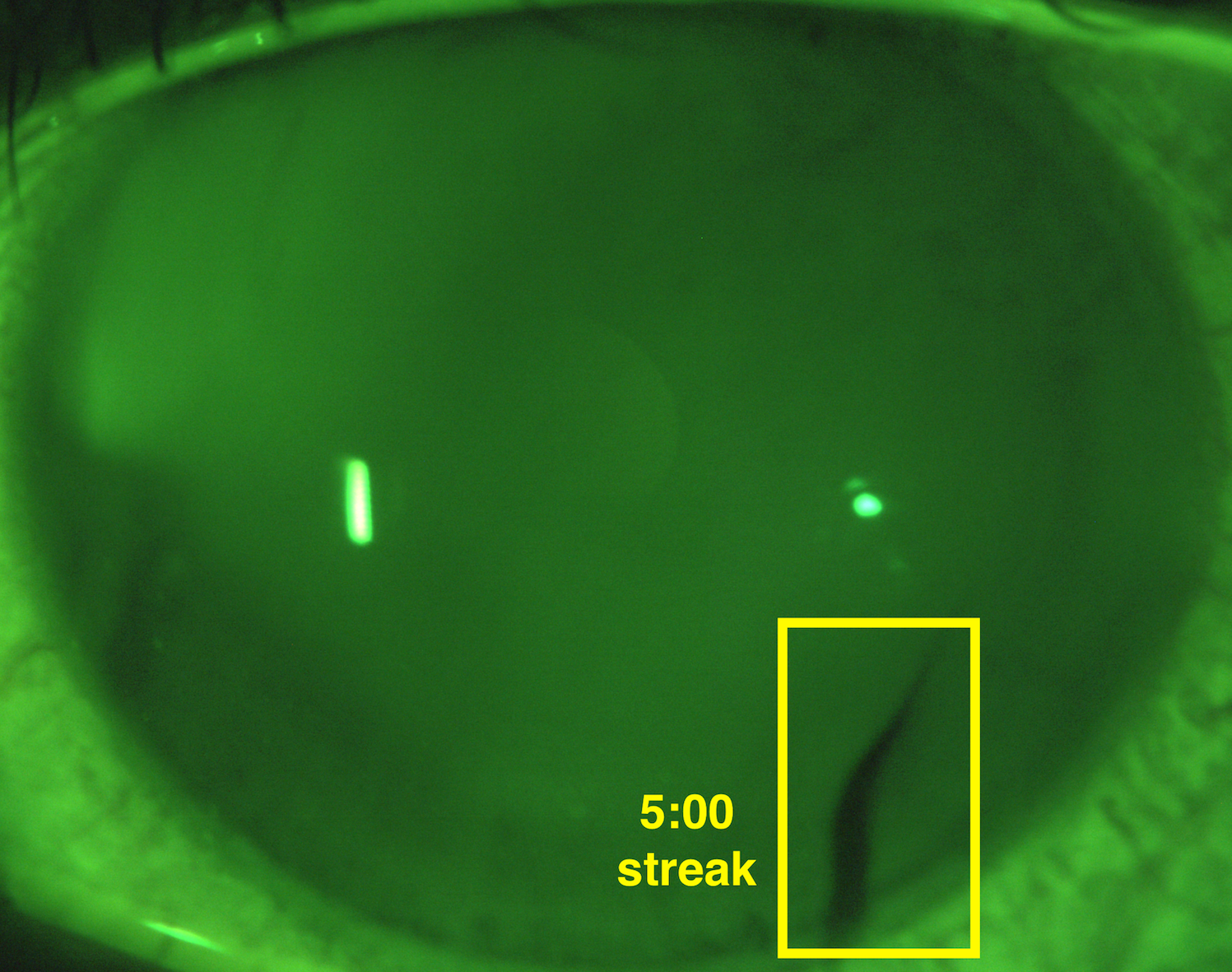}}
\subfloat[][S27v2t2 5:00 streak surface plot]{\includegraphics[scale=.128]{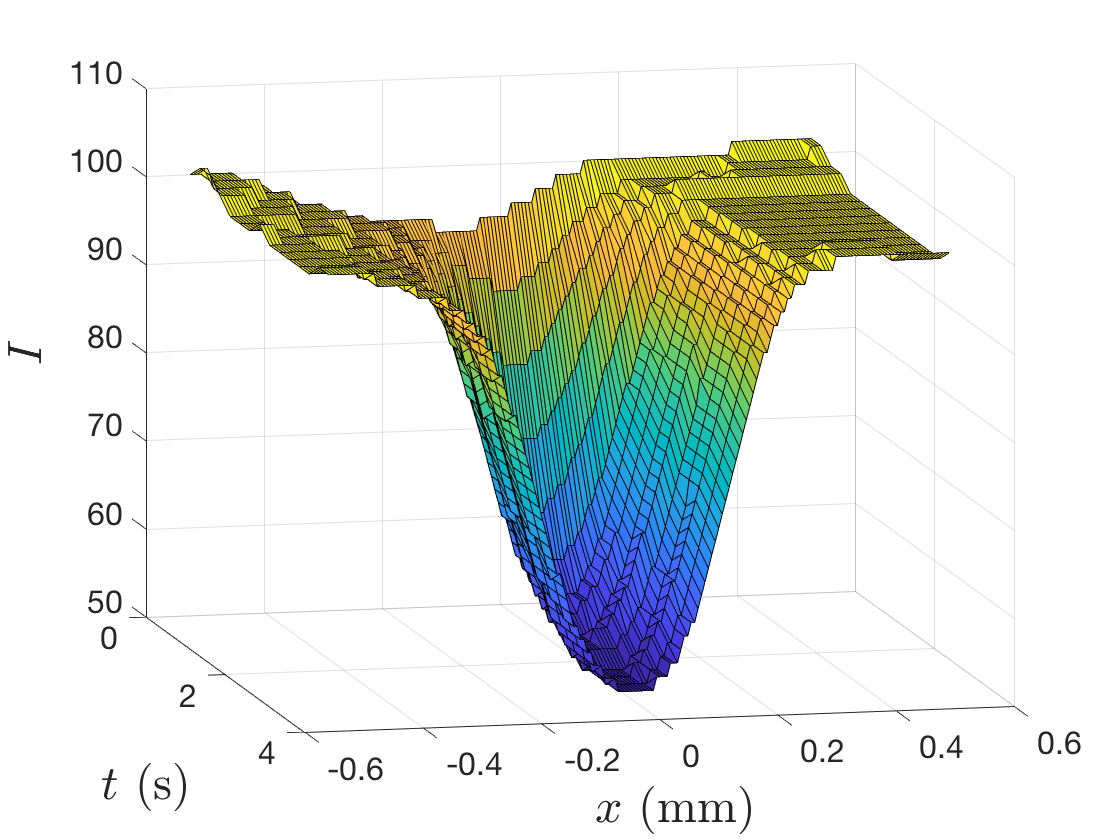}}
\caption{\footnotesize{(a)-(g): the last image in each trial. The bright rectangle, called the Purkinje reflex, is due to the reflection from the light source. (h): Surface plot of the S27v2t2 5:00 shown in (g).}}
\label{fig:images}
\end{figure}
 
FT-TBU is computer-aided determination of thinning to what is evidently a very small aqueous thickness. As in \cite{luke2020}, we select single spot- or streak-shaped FT-TBU instances to extract data from and fit with our circular or linear models. All instances reported in this paper are shown in Figure \ref{fig:images}.
\cite{king2013} recorded several instances of breakup that formed in a few seconds or less using simultaneous imaging of the lipid and aqueous TF layers. \cite{zhong2019} note that evaporation alone cannot produce rapid breakup, citing that it takes at least 8 seconds to observe a dark spot for a 3.5 $\mu$m thick TF with an evaporation rate of 25 $\mu$m/min. This influenced our choice of 8 seconds as an upper bound for the FT-TBUTs to study in this work. Our data varies between a frame rate of 4 or 5 per second; this restricts the time resolution depending on the trial as the data that is extracted from these movies is very dynamic. A comparison of the rate of decrease of FL intensity is explored in Section \ref{sec:results}; see Figures \ref{fig:min_max} and \ref{fig:FL_dec}.

\section{Model}
\label{sec:model}

We hypothesize that intermediate TF breakup, occurring between one and eight seconds, is driven by a combination of Marangoni effect-driven tangential flow and evaporation. We believe the data we work with lacks the time resolution needed to capture rapid TF breakup driven solely by the Marangoni effect, which may occur in under one second (\citealt{zhong2018}).

 We discuss the model in axisymmetric coordinates; the Cartesian case is described in Appendix \ref{s:streak_eqns}. The TF is modeled as a single-layer Newtonian fluid with constant viscosity $\mu$ and density $\rho$.  The mean surface tension at the tear/air interface, $\sigma$, is assumed constant, although the change in surface tension, $(\Delta \sigma)_0$, will be allowed to vary. The surface tension $\sigma$ is given by the linearized equation of state 
 \begin{equation}
 \sigma = \sigma_0 + (\Delta \sigma)_0 (\Gamma' - \Gamma_0), 
 \end{equation}
 where $\Gamma'$ is the surfactant concentration with $\Gamma_0$ its initial condition.
 The solute diffusivities are assumed constant as well.

\subsection{Scalings}
\label{sec:scalings}

The following scalings non-dimensionalize the system of equations governing TF thickness, pressure inside the film, surfactant concentration atop the film, and the transport of solutes in the film. Dimensional quantities are denoted by primes. See Appendix \ref{sec:app_gov} for more details.
\begin{equation}
r' = \ell r, \quad z' = dz, \quad \epsilon = d/\ell, \quad t' = \frac{\ell}{U} t, \quad h' = dh, \quad u' = U u, \quad v' = \epsilon U v,
\label{scale1}
\end{equation}
\begin{equation}
p' = \frac{\mu U}{\ell \epsilon^2} p, \quad J' = \epsilon \rho U  J, \quad \Gamma' = \Gamma_0 \Gamma, \quad c' = c_0 c, \quad f' = f_{cr} f.
\label{scale2}
\end{equation}
Dimensional parameters used in the model are summarized in Table \ref{table:dim}.
\begin{table}
\centering
\begin{tabular}{llll}
\hline
\multicolumn{4}{c}{Dimensional parameters}                                                                                                                           \\ \hline
Parameter       & Description                      & Value                                                & Reference                                                \\ \hline
$\mu$           & Viscosity                        & $1.3 \times 10^{-3}$ Pa $\cdot$ s                    & \cite{tiffany1991} \\
$\rho$          & Density                          & $10^3$ kg $\cdot$ m$^{-3}$                           & Water                                                    \\
$A^*$           & Hamaker constant                 & $6 \pi \times 3.5 \times 10^{-19}$ s $\cdot$ m$^{-1}$  & \cite{ajaev2001} \\
$d$             & Initial TF thickness             & $2-5 \times 10^{-6}$ m                               & Calculated  (\citealt{luke2020})\\
$v'$      & Thinning rate               & $0.5-25 \ \mu$m/min                                      & \cite{nichols2005} \\
$V_w$           & Molar volume of water            & 1.8 $\times 10^{-5}$ m$^3 \cdot $ mol$^{-1}$         & Water                                                    \\
$D_s$ & Surface diffusion coefficient & 3 $\times 10^{-8} \text{m}^2$/s & \cite{casalini2011} \\
$D_f$           & Diffusivity of fluorescein       & 0.39 $\times 10^{-9}$ m$^2$/s                        & \cite{casalini2011} \\
$D_o$           & Diffusivity of salt              & 1.6 $\times 10^{-9}$ m$^2$/s                         & \cite{riquelme2007} \\
$c_0$ & Isotonic osmolarity & 300 mOsM &  \cite{lemp2011} \\
$f_{cr}$ & Critical FL concentration & 0.2 \% & \cite{webber86} \\
$f_0'$ & Initial FL concentration & 0.259-0.4 \% & Calculated \\
$P_o$           & Permeability of cornea           & $12.1 \times 10^{-6}$ m/s                            & \cite{braun2015} \\
$\epsilon_f$    & Napierian extinction coefficient  & $1.75 \times 10^7$ m$^{-1}$ M$^{-1}$                  & \cite{mota1991} \\ 
$\sigma_0$      & Surface tension                  & 0.045 N $\cdot$ m$^{-1}$                             & \cite{nagyova1999}           \\ 
$(\partial_{\Gamma} \sigma)_0$ & Composition dependence & 0.01 N/m & \cite{aydemir2010} \\
$(\Delta \sigma)_0$ & Change in surface tension & 1.74 $-$ 60.3 N/m & Calculated \\
$\ell$          & Characteristic length & 0.138 $-$ 0.412 mm                                       & Calculated                                               \\ 
$U$ & Characteristic velocity & 0.0560 $-$ 0.0990 mm/s & Calculated \\
$t_s$ & Time scale & 1.75 $-$ 6.6 s & Fit interval \\
\hline
\end{tabular}
\caption{\footnotesize{The dimensional parameters used. The range of estimates for thinning rates are from point measurements from published studies. Some ranges are extended in our results below.}}
\label{table:dim}
\end{table}

\cite{zhong2019} choose the scalings for $U$ and $\ell$  based on two dimensionless quantities: $M$, the (reduced) Marangoni number, and $S$, the contribution of surface tension:
$$M = \frac{\epsilon (\Delta \sigma)_0}{\mu U}, \quad S = \frac{\sigma_0 \epsilon^3}{\mu U}.$$
In \cite{zhong2019}, the expressions for $M$ and $S$ are set equal to unity. We seek to determine the Marangoni number via the optimization. \cite{zhong2019} used a time scale of 0.0205 s; this is too fast in the context of our data. Therefore, we adapt the model to our problem by choosing new scalings. We will use the fit interval as the time scale on which we assume breakup occurs. We set $S = 1$ and let the change in surface tension, $(\Delta \sigma)_0$, vary in the optimization instead of $M$ to optimize over dimensional parameters. We determine $U$ through the time scale of the trial, and $S$ relates $U$ and $\ell$. 

The time scale of the model is $\ell/U$; we take the length of the trial as $t_s$, and we find $U = \ell/t_s.$ This gives 
\begin{equation} U = \frac{\sigma_0 \epsilon^3 }{\mu} = \frac{\sigma_0 d^3}{\mu \ell^3}.
\end{equation}
Equating the two expressions for $U$ and solving for the length scale gives
\begin{equation}
\ell = \l \frac{t_s \sigma_0 d^3}{\mu }\r^{1/4}.
\end{equation}
Now knowing $\ell$, this gives $U$ and $M$ as
\begin{equation}
U = \l \frac{\sigma_0 d^3}{\mu t_s^3} \r^{1/4}, \quad M = \epsilon (\Delta \sigma)_0 \l \frac{t_s^3}{\sigma_0 \mu^3 d^3}\r^{1/4}.
\label{U_M_Lan}
\end{equation} 
The non-dimensional parameters that arise as a result of the scalings are given in Table \ref{table:nondim}.

\begin{table} 
\centering
\begin{tabular}{llll}
\hline
\multicolumn{4}{c}{Non-dimensional parameters with typical values} \\ \hline
Parameter & Description   & Expression                                  & Value  \\ \hline
\vspace*{2mm}
$\epsilon$ & Aspect ratio & $d / \ell$                            & 0.0130 \\
\vspace*{2mm}
$S$     & Contribution of surface tension    & $ (\sigma_0 \epsilon^3)/(\mu U)$  & 1      \\
\vspace*{2mm}
$M$ & Contribution of Marangoni effect & $\epsilon (\Delta \sigma)_0 \left[ t_s^3/(\sigma_0 \mu^3 d^3) \right]^{1/4}$ & 0.275 $-$ 5.50 \\ \vspace*{2mm}
$A$   & Non-dimensional Hamaker constant      & $ A^*/\left[ \epsilon (\Delta \sigma)_0 d \ell \right]$           & $6.59 \times 10^{-10}$ \\
\vspace*{2mm}
$P_c$  & Permeability of cornea     & $ (P_o V_w c_0)/(\epsilon U)$              & 0.0653  \\
\vspace*{2mm}
Pe$_f$  & P\'{e}clet number for FL diffusion    & $ U \ell/D_f$        & 45.3   \\
\vspace*{2mm}
Pe$_c$  & P\'{e}clet number for salt iron diffusion    & $U \ell / D_o$        & 11.0  \\
\vspace*{2mm}
Pe$_s$  & P\'{e}clet number for surface diffusion    & $\epsilon (\Delta \sigma)_0 \ell/(\mu D_s)$        & 1.54   \\
\vspace*{2mm}
$\phi$  & \begin{tabular}[c]{@{}c@{}}Non-dimensional Napierian \\ extinction coefficient\end{tabular}    & $ \epsilon_f f_{\text{cr}} d$                & 0.279  \\ \hline
\end{tabular}
\caption{\footnotesize{Dimensionless parameters that arise from scaling the dimensional fluid mechanics problem. The values given are based upon the values of Table \ref{table:dim}, $d = 3 \ \mu$m, $t_s = 3$ s, and $(\Delta \sigma)_0 = 20 \ \mu$N/m.}}
\label{table:nondim}
\end{table}

\subsection{Lubrication Theory}
\label{lub}

Due to the small aspect ratio of the thickness of the film to the length along the film, we use lubrication theory to simplify the governing equations. The derivation is given in Appendix \ref{sec:app_gov}-\ref{s:spot_eqns}. We assume diffusion, advection, osmosis, the Marangoni effect, and evaporation affect the height of the TF in a combination that ultimately leads to FT-TBU. 

In axisymmetric coordinates, the fluid velocity in the film is denoted by $\bm{u} = (u,w)$, where $u$ and $w$ are the radial and vertical velocities, respectively. Conservation of mass and momentum for water and solutes in the TF and surfactant along the surface lead to the following system of equations (\ref{eq:dhdt}--\ref{eq:dfdt}). 
\begin{equation}
\p_t h + J - P_c(c-1) + \frac{1}{r} \p_r ( r h \bar{u}) = 0,
\label{eq:dhdt}
\end{equation}
\begin{equation}
p = - \frac{1}{r} \p_r (r \p_r h) - A h^{-3},
\label{eq:p}
\end{equation}
\begin{equation}
\partial_t \Gamma= \left[ \text{Pe}_s^{-1} \left( \frac{1}{r} \p_r (r \p_r \Gamma) \right) - \frac{1}{r} \p_r (r u_r \Gamma) \right] B,
\label{eq:G}
\end{equation}
\begin{equation}
h(\p_t c + \bar{u}\p_r c) = \text{Pe}_c^{-1} \frac{1}{r} \p_r (rh \p_r c) + Jc -  P_c(c-1)c,
\end{equation}
\begin{equation}
h(\p_t f + \bar{u}\p_r f) = \text{Pe}_f^{-1} \frac{1}{r} \p_r (rh \p_r f) + Jf -  P_c(c-1)f.
\label{eq:dfdt}
\end{equation}
In equation \ref{eq:dhdt}, $J$ and $\bar{u}$ represent the evaporative term and the depth-averaged horizontal fluid velocity, respectively. We discuss options for $J$ in Section \ref{sec:evap_dist}. In equation \ref{eq:G}, $u_r$ is the horizontal surface velocity of fluid, and $B$ is a tanh function used as a smooth approximation to a transition step function between the domains on which different boundary conditions exist for the surfactant concentration, $\Gamma$: 
\begin{equation}
B(r; R_I, R_W) = \frac{1}{2} + \frac{1}{2} \tanh \left( \frac{r - R_I}{R_W} \right).
\end{equation} 
Here, $R_I$ is the glob radius, and the transition width $R_W$ is set to 0.1. Initially, the lipid has a high, constant concentration on $[0, R_I]$ and low outside, but solute is transported due to tangential flow from the Marangoni effect as time progresses.

The horizontal surface TF fluid velocity $u_r(r,h,t)$ and average horizontal TF fluid velocity $\bar{u}$ throughout the film are given by

\begin{equation}
u_r(r,h,t) = - \frac{\frac{1}{2} h^2 (\p_r p)  B + M \p_r \Gamma h B}{B + (1-B)h,}
\end{equation}

\begin{equation}
\bar{u} = -  \frac{\frac{1}{3} h^2 (\p_r p) \left[ B + \frac{1}{4} h (1-B) \right]+ \frac{1}{2} M \p_r \Gamma h B}{B + (1-B)h}.
\end{equation}
Note that the Marangoni number, $M$, multiplies the radial derivative of $\Gamma$. The quantity $\bar{u}$ comprises the combination of pressure gradient-induced Poiseuille flow (due to capillarity), and shear stress-driven Couette flow (due to the Marangoni effect). As the TF is deformed, capillarity increases in relative importance, and the decrease in lipid concentration difference reduces the Marangoni effect.

The FL intensity $I$ is computed from the TF thickness $h$ and the FL concentration $f$:
\begin{equation}
I = I_0 \frac{1 - \exp(-\phi f h)}{ 1 + f^2}.
\label{eq:I_nd}
\end{equation}
Here, $\phi$ is the nondimensional Napierian extinction coefficient, and $I_0$ is a normalization coefficient found using a least squares fit to model eye measurements (\citealt{wu2015}).  

\subsection{Evaporation Distributions}
\label{sec:evap_dist}

Following \cite{zhong2019}, we explore four evaporation distribution choices as listed below. 

\textit{Case (a)}: TF thinning is assumed to be driven only by tangential flow due to the Marangoni effect; we assume evaporation is irrelevant and exclude it from the model. This case assumes that the thinning occurs on too short a time scale for evaporation to play a role.

\textit{Case (b):} The surfactant distribution is assumed to have no effect on evaporation, and we assume a uniform profile: 

\begin{equation}
J' = \rho v',
\end{equation}
where $v'$ is a constant thinning rate. The glob may be poorly organized and thus allow evaporation in an amount equal to the lower concentration lipid surrounding it.

\textit{Case (c):} We hypothesize that lipid with a higher concentration is disorganized, and as such does not protect the TF underneath from evaporation.  Outside the glob, we assume the lower-concentration lipid provides a sufficient barrier against evaporation for the duration of the trial. We let there be nonzero constant evaporation under the glob and zero evaporation outside the glob, given by:

\begin{equation}
J' = \rho (1-B)v'.
\end{equation} 
It may seem counterintuitive that thicker lipid could allow a higher rate of evaporation than in a thinner region, but it may be seen experimentally (\citealt{king2013, king2010}).

\textit{Case (d):} We assume the glob provides an excellent barrier to evaporation whereas the lipid with lower concentration outside allows TF fluid to evaporate. We choose zero evaporation under the glob and nonzero evaporation outside the glob, given by: 

\begin{equation}
J' =\rho Bv'.
\end{equation}
We found that options (b) and (c) were the most successful at fitting breakup instances in the 1-8 second range. 

\subsection{Boundary and Initial Conditions}

We enforce no flux of fluid or solutes at the outer boundary of the domain, $r=R_0$, resulting in homogeneous Neumann conditions for all dependent variables there:
\begin{equation}
\p_r h(R_0,t) = \p_r p(R_0,t) = \p_r c(R_0,t) = \p_r f(R_0,t)  = \p_r \Gamma(R_0, t) = 0.
\end{equation}
Similarly, we enforce symmetry at the origin. We assume that a blink restores the TF thickness and solute concentrations to uniform values across the cornea. Thus, the initial conditions are spatially uniform:
\begin{equation}
h(r,0) = c(r,0) = 1, \quad f(r,0) = f_0.
\end{equation}
The initial pressure is computed from (\ref{eq:p}) using symmetry. We estimate the initial FL concentration via a separate procedure using model eye calculations and a custom MATLAB code following \cite{wu2015}. The initial TF thickness is estimated by a calculation described in our previous paper: 
\begin{equation}
h_0' = - \frac{1}{\epsilon_f f_0'} \log \l 1 - \frac{I_b-I_s}{I_0^*} \left[1 + (f_0'/f_{\text{cr}})^2 \right] \r,
\end{equation}
where $I_b$ and $I_s$ are averages of intensity values in the region of breakup from the high and low light setting image values, respectively, and $I_0^*$ is a scaling of $I_0$ by the ratio of an average of intensity values in the region of breakup from the first high light setting image to the last low light setting image. We use $d=h_0'$ to give the nondimensional initial thickness value $h_0=1$. From this measurement $h_0'$ we subtract one micron for the thickness of the glycocalyx (\citealt{luke2020, king2004}).

The nondimensionalization results in initial values of $\Gamma = 1$ under the glob and $\Gamma = 0.1$ outside the glob. Written using the transition function $B$, the initial condition for $\Gamma$ is 
\begin{equation}
\Gamma(r,0) = 1 \cdot [1 - B(r) ] + 0.1 \cdot B(r).
\end{equation}

\section{Optimization} 

We follow the process described in \cite{luke2020}; a summary is given below.

\subsection{Data Preparation}
\label{data_prep}

The high light setting images in each trial are converted from RGB color images to grayscale, smoothed with a Gaussian filter, and stabilized using the Purkinje reflex (\citealt{awisigyau2020}), a bright artefact reflecting the light source, via custom MATLAB codes. We select a region of interest in the last image where FT-TBU forms. FT-TBU instances are chosen to be a simple shape (roughly linear or circular), dark enough (from monitoring the local minimum FL intensity), and developing on an intermediate time scale (between 1 and 8 seconds). We sample pixel intensities from every bright image in the trial on a line segment across a spot or streak FT-TBU at an orientation that we choose via custom MATLAB codes. These codes have been updated from those used in \cite{luke2020} to allow for drift of the TF to be captured: the line segment for data extraction can be drawn manually to follow the movement of the TF fluid. The data is further stabilized by aligning the minimum of each time level with the origin; the data is shifted by less than 0.1 mm on average.  

We fit the theoretical FL intensity function to a subset of experimental FL intensity data from the video; most optimizations use 6-10 time levels from the trial. The starting frame is the last frame before the FL intensity data starts dropping or the first high light setting image in the trial if thinning begins instantaneously. In some trials, there is evidence that thinning has begun before the light source is turned up, and as a result there is significant decrease in FL intensity in the center of breakup in the first bright image. To remedy this, we introduce one or two ``ghost'' time levels, which allows the model solution to start with a uniform time level that is not compared to the experimental FL data. This is a product of the time scale of our data; sometimes a few seconds elapse between the last blink and full illumination of the cornea. These are added to match the rate of thinning seen between experimental time levels. With this choice we capture additional information about the thinning, such as the initial magnitude of the fluid flow. The last frame is the final frame before the FL intensity data stops decreasing. 

For the purpose of fitting, we define FT-TBUT as the time at which the FL intensity stops decreasing. The pixel intensity values typically stop decreasing between 30 and 50 using a 0-255 (8 bit) scale at the illumination settings used.

 \subsection{Optimization Problem}
\label{optim_prob}

We discuss the optimization problem for spots; the streak version is similar.
Expressed in continuous variables, we seek to minimize $\ds ||I_{\text{th}}(r,t) - I_{\text{ex}}(r,t)||_2^2$ over the parameters $v'$, the evaporation rate, $R_I'$, the radius of the glob of lipid, and $(\Delta \sigma)_0$, the change in surface tension created by lipid concentration gradients. Here, $r$ corresponds to the distance from the center of the spot or streak TBU, and $t$ corresponds to the time after the light source brightness has been increased to the high setting. Both parameters have been nondimensionalized with the scalings given in Section \ref{sec:scalings}. The norm is over all $r \in [0, R]$ and $t \in [0, T]$ excluding any ``ghost'' time levels from the theoretical FL intensity, where $R$ corresponds to the radius of the FT-TBU and $T$ corresponds to the total length of the trial. As in \cite{luke2020}, we widen the computational domain by a factor of three in most cases (in $[0,R_0]$) and compare $I_{ex}$ with the subset of $I_{th}$ corresponding to $[0,R]$ to reduce the sensitivity of our optimization to our choices of initial guesses and boundary conditions.

 The optimization problem for spots may be written
\begin{equation}
\argmin_{v', R_I', (\Delta \sigma)_0} ||I_{\text{th}}(r,t; v', R_I',(\Delta \sigma)_0) - I_{\text{ex}}(r,t)||_2^2,
\end{equation}
where $R_I'$ and $r$ are replaced with $X_I'$ and $x$ in the Cartesian case for fitting linear FT-TBU. The theoretical intensity, $I_{th}$, is computed after solving the coupled partial differential equations system for film thickness, $h$, and fluorescein concentration, $f$.

 \subsection{Numerical Method and Stopping Criterion}

Following our previous work, we solve the TF dynamics model (\ref{eq:dhdt}-\ref{eq:dfdt}) using an application of the method of lines.  The spatial derivatives are discretized using collocation at second-kind Chebyshev points (\citealt{trefethen2000}, \citealt{canuto2012}). We enforce symmetry at the origin to avoid singularities in the axisymmetric case; this is achieved by expanding all operators in $r$ and dropping odd derivatives. The resulting system of differential algebraic equations for the dependent variables at the grid points
is solved using \verb|ode15s| in MATLAB (MathWorks, Natick, MA, USA). For the optimization, we use a nonlinear least squares minimization implemented by \verb|lsqnonlin| in MATLAB with the trust-region reflective algorithm (\citealt{nocedal2006}) and we added a second order finite difference approximation of the Jacobian (\citealt{leveque2007}), which improved performance. In this work, we found that the Levenberg-Marquardt and trust-region reflective algorithms produce similar optimal values, but the latter is often preferable for its reduced average computation time. For the mixed-mechanism fits we report, the computation times for the optimizations varied from 1.5 minutes to 111 minutes.

To generate initial guesses for optimization, forward computations were conducted until the theoretical dynamics were close to the experimental. For each instance, the solver stopped because the change in residual was less than the specified tolerance. Optimization tolerances of roughly the square root of the ODE solver tolerances were used. 

\section{Results}
\label{sec:results}

We begin by presenting characteristic nondimensional solutions with evaporation Cases (b) and (c) presented in Section \ref{sec:evap}. We then show the results of fitting breakup instances with our mixed-mechanism model that incorporates Marangoni effect-induced tangential flow and evaporation. Examples of data extractions and fits of the various dynamics are shown. For comparision, we also fit with evaporation only (\citealt{luke2020}) and zero evaporation, which is Case (a). The fitting results for these models are summarized in Sections \ref{sec:evap} and \ref{sec:mar}, respectively. A study of fluid profiles and the effect of varying the initial FL concentration on the fitting procedure follow. Example fits for the evaporation-only and zero evaporation models are shown in Online Resource 1.

\subsection{Nondimensional Solutions (Without Fitting)}

\begin{figure} 
\centering
\subfloat[][Case (b) evaporation]{\includegraphics[scale=.15]{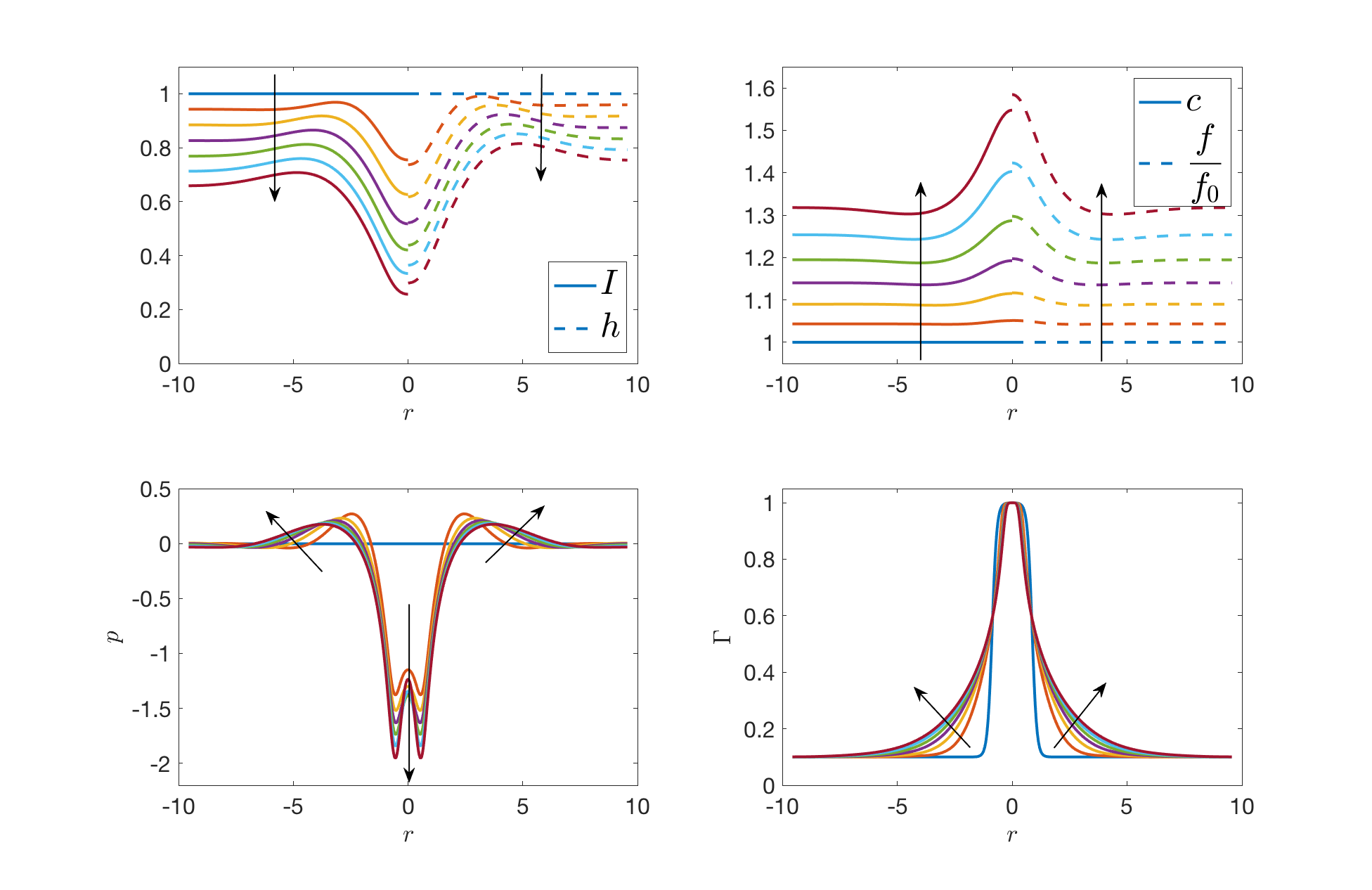} \label{fig:nondim_a}} \\
\subfloat[][Case (c) evaporation]{\includegraphics[scale=.15]{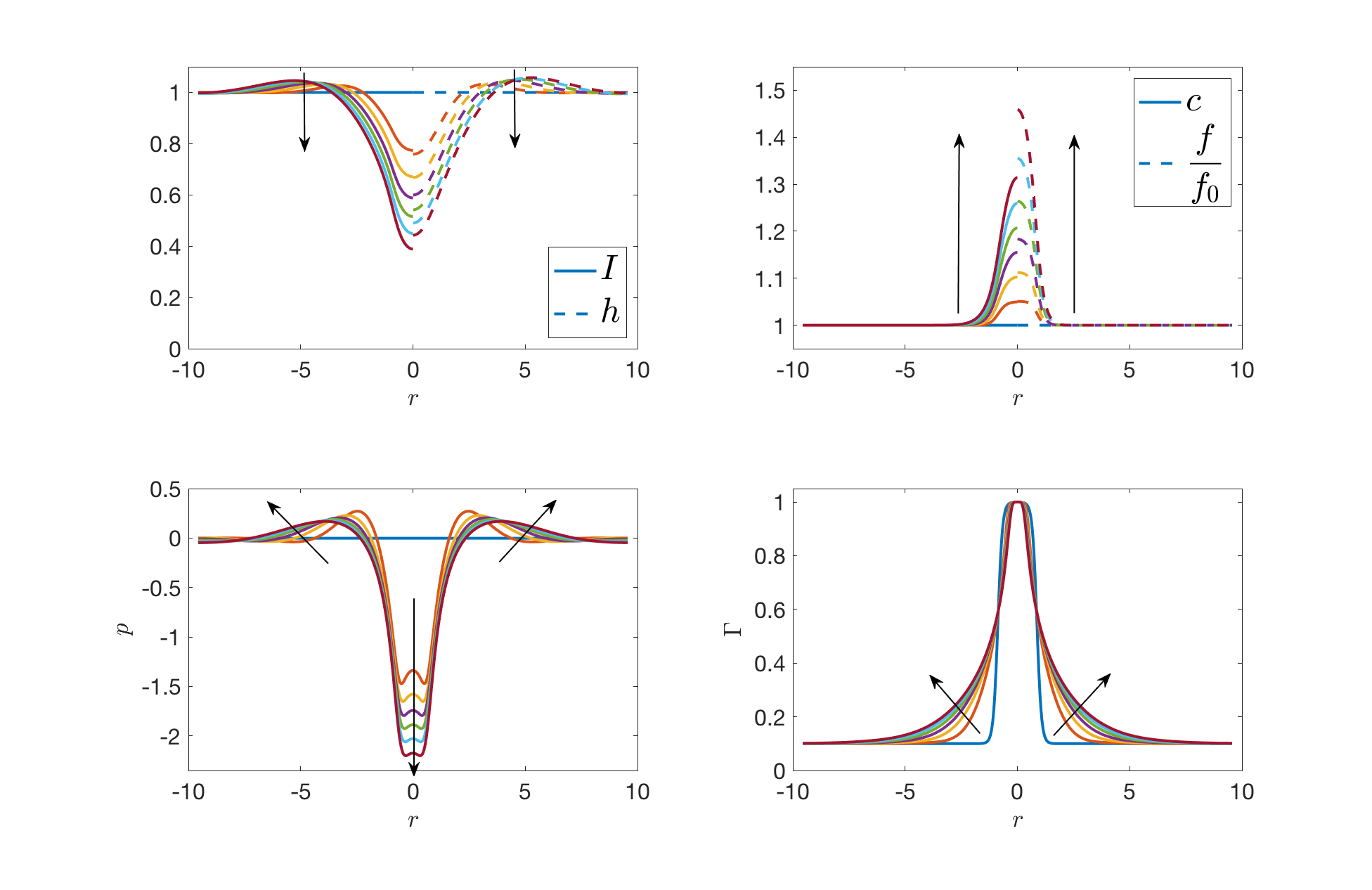}\label{fig:nondim_b}}
\caption{\footnotesize{Nondimensional axisymmetric solutions for $v' = 15 \ \mu$m/min, $R_I' = 0.1$ mm, $(\Delta \sigma)_0 = 20 \ \mu$N/m, $f_0' = 0.3$ \%, and $d = 3 \ \mu$m. The Marangoni number is 2.61. Each curve represents a different time level and arrows indicate increasing time. Intensity has been normalized.}}
\label{fig:nondim}
\end{figure}

In Figure \ref{fig:nondim} we plot the nondimensional axisymmetric solutions for the film thickness $h$, fluid pressure $p$, osmolarity $c$, fluorescein concentration $f$, surfactant concentration $\Gamma$, and the computed theoretical fluorescent intensity $I$. In the top left plot of both subfigures, each theoretical quantity is plotted only on half the domain to allow for more direct comparison: FL intensity is shown as solid lines on the left half, and TF thickness is shown as dashed lines on the right half. The characteristic length along the film for this case is about 0.23 mm, so the nondimensional spot size is a little under 0.5. 

Figure \ref{fig:nondim_a} shows Case (b) evaporation, while Figure \ref{fig:nondim_b} gives the solutions for Case (c). We note the characteristic differences in the solution profiles as a result. For all other conditions held the same, the intensity and TF thickness decrease further in Case (b) as compared to Case (c). Osmolarity and FL concentration increase more in Case (b).

\subsection{Mixed-Mechanism Fitting}

We fit \textit{in vivo} intensity measurements with the mixed-mechanism model discussed in Section \ref{sec:model}. The results are reported in Tables \ref{table:mix_fits} and \ref{table:scalings}. Each FT-TBU instance is labeled by subject, visit, and trial number, the location of the breakup as a clock reading, and the type of breakup (streak or spot). Images showing the FT-TBU instances can be found in Section~\ref{sec:data}. The streaks are fit with our Cartesian model and the spots are fit with our axisymmetric model. Both the experimental and theoretical FL intensities are normalized to the average of the first time level before fitting, and the osmolarity is reported as a multiple of the isotonic concentration. The thinning rate, glob size, and change in surface tension are adjusted to accomplish the fit. Sections \ref{sec:strong}, \ref{sec:mix}, and \ref{sec:weak} show examples of the experimental data, fits, and resulting theoretical solutions using the optimal parameters found by nonlinear least squares minimization. 

We separate the results into three categories: Marangoni effect-dominated, intermediate, and evaporation-dominated. The first is categorized by strong outward flow near the glob edge for the duration of the trial, relatively small evaporation rates, and Marangoni numbers between 2.83 and 5.5. This nondimensional value quantifies the relative strength of the Marangoni effect, and a number above one conveys significance. These large Marangoni numbers coupled with the small evaporation rates signal the relative dominance of the Marangoni effect over evaporation. The first five instances in Table \ref{table:mix_fits} fall into this category.  

The last two instances in Tables \ref{table:mix_fits} and \ref{table:scalings} are evaporation-dominated. Like all other trials, their flow is initially directed outwards, but becomes inward almost immediately. These trials also exhibit relatively large evaporation rates and Marangoni numbers under one, a sign that the Marangoni effect is not important. Taken together and recalling that inward flow is characteristic of the fits seen in \cite{luke2020}, we label these two instances as evaporation-dominated. In fact, the S9v2t5 4:00 spot is also fit well with the evaporation-only model; see Section \ref{sec:evap}.

We label a single instance, the S9v2t1 3:00 streak, as intermediate. This trial exhibits a balance of evaporation and the Marangoni effect, as seen by a relatively large evaporation rate and a Marangoni number of 1.48. Like the evaporation-dominated instances, the depth-averaged flow changes directions early in the trial, but the initial magnitude of the flow is much larger in comparison, and the surface velocity is directed outwards for the duration of the trial. This trial is listed between the two previously mentioned categories in Tables \ref{table:mix_fits} and \ref{table:scalings}.

\begin{table} 
\centering
\tabcolsep=0.14cm
\begin{tabular}{|c|c|c|c|c|c|c|c|c|c|c|}
\hline
\textbf{Trial} & \textbf{\begin{tabular}[c]{@{}c@{}}FT-TBU \\ ID\end{tabular}} & \textbf{\begin{tabular}[c]{@{}c@{}}$\bm{h_0'}$ \\ ($\bm{\mu}$m)\end{tabular}} & \textbf{\begin{tabular}[c]{@{}c@{}}$\bm{f_0'}$ \\(\%)\end{tabular}} & \begin{tabular}[c]{@{}c@{}}$\bm{v'}$\\ \textbf{($\bm{\frac{\mu\text{m}}{\text{min}}}$)} \end{tabular} & \begin{tabular}[c]{@{}c@{}}$\bm{R_I',}$ \\ $\bm{X_I'}$\\ \textbf{(mm)} \end{tabular} & \begin{tabular}[c]{@{}c@{}}$\bm{(\Delta \sigma)_0}$\\ \textbf{($\bm{\frac{\mu\text{N}}{\text{m}}}$)} \end{tabular}  & \textbf{\begin{tabular}[c]{@{}c@{}}Min\\ $\bm{I_{ex}}$\end{tabular}} & \textbf{\begin{tabular}[c]{@{}c@{}}Min\\ $\bm{I_{th}}$\end{tabular}} & \textbf{\begin{tabular}[c]{@{}c@{}}Min\\ $\bm{h'_{th}}$ \\ ($\bm{\mu}$m)\end{tabular}} & \textbf{\begin{tabular}[c]{@{}c@{}}Max\\ $\bm{c_{th}}$ \end{tabular}}\\ \hline
S9v1t4$^+$ & 4:00 \textbf{---} & 3.32 & 0.324 & 6.26 & 0.0750 & 20.3 & 0.106 & 0.398 & 1.32 & 1.23 \\ \hline
S10v1t6$^{++}$ & 12:30 $\bm{\circ}$ & 3.08 & 0.293 & 5.92 & 0.0791 & 60.3 & 0.0635 & 0.0546 & 0.165 & 1.30 \\ \hline
S13v2t10$^+$ & 6:30 \textbf{---} & 3.59 & 0.259 & 13.6 & 0.0786 & 20.0 & 0.116 & 0.286 & 1.15  & 1.55 \\ \hline
S18v2t4$^+$ & 7:30 $\bm{\circ}$ & 2.48 & 0.363 & 15.1 & 0.0808 & 25.6 & 0.143 & 0.115 & 0.410 & 1.86 \\ \hline
S27v2t2$^+$ & 5:00 \textbf{---} & 1.91 & 0.4 & 6.11 & 0.0525 & 25.1 & 0.357 & 0.469 & 0.906 & 1.14 \\ \Xhline{1.5pt} 
S9v2t1 & 3:00 \textbf{---} & 5.01 & 0.292 & 30.3 & 0.140 & 9.85 & 0.0826 & 0.258 & 2.03 & 1.77 \\ \Xhline{1.5pt}
S9v2t5 & 4:00 $\bm{\circ}$ & 2.10 & 0.299 & 26.2 & 0.121 & 4.05 & 0.132 & 0.290 & 1.10 & 1.93 \\ \hline
S9v2t5 & 4:30 $\bm{\circ}$ & 2.33 & 0.299 & 36.9 & 0.123 & 1.74 & 0.0600 & 0.226 & 1.28 & 2.32 \\ \hline
\end{tabular}
\caption{\footnotesize{Results from fitting three parameters. A + denotes using a ``ghost'' first time level. 
The subject (S) number, visit (v) number and (t) trial number are listed, and the FT-TBU location is a clock reading taken from the center of the pupil. A \textbf{---} denotes streak FT-TBU, and a $\bm{\circ}$ is a spot. The initial TF thickness and FL concentration estimates are given. The optimized parameters are the evaporative thinning rate $v'$, the glob radius $R_I'$ or glob half-width $X_I'$, and the change in surface tension $(\Delta \sigma)_0$. The minimum values of both the experimental and theoretical FL intensity and the theoretical thickness are reported. The instances above the thick line are Marangoni effect-dominated, the instance between is a transitional case, and the instances below are evaporation-dominated.
}}
\label{table:mix_fits}
\end{table}

\begin{table} 
\centering
\begin{tabular}{|c|c|c|c|c|c|c|}
\hline
\textbf{Trial} & \textbf{\begin{tabular}[c]{@{}c@{}}FT-TBU \\ ID \end{tabular}} & \textbf{\begin{tabular}[c]{@{}c@{}}Length\\ scale\\ $\bm{\ell}$ (mm)\end{tabular}} & \textbf{\begin{tabular}[c]{@{}c@{}}Time \\ scale\\ $\bm{t_s}$ (s)\end{tabular}} & \textbf{\begin{tabular}[c]{@{}c@{}}Char.\\ velocity\\ $\bm{U}$ ($\bm{\frac{\text{mm}}{\text{s}}}$)\end{tabular}} & \textbf{\begin{tabular}[c]{@{}c@{}}$\bm{M}$ \end{tabular}} & \textbf{\begin{tabular}[c]{@{}c@{}}Res.\\ (norm)\end{tabular}}  \\ \hline
S9v1t4$^+$ & 4:00 \textbf{---} & 0.273 & 4.6 & 0.0593 & 3.06 & 7.49 (2.32) \\ \hline
S10v1t6$^{++}$ & 12:30 $\bm{\circ}$ & 0.198 & 2 & 0.0990 & 5.50 & 4.19 (1.87) \\ \hline
S13v2t10$^+$ &  6:30 \textbf{---} & 0.287 & 4.4 & 0.0652 & 2.83 & 4.16 (1.57)  \\ \hline
S18v2t4$^+$ & 7:30  $\bm{\circ}$ & 0.191 & 2.75 & 0.0695 & 3.36 & 8.13 (2.54)  \\ \hline
S27v2t2$^+$ &  5:00 \textbf{---} & 0.138 & 1.75 & 0.0789 & 2.91 & 3.75 (1.53)   \\ \Xhline{1.5pt}
S9v2t1 &  3:00  \textbf{---} & 0.412 & 6.6 & 0.0624 & 1.48 & 9.19 (2.59)  \\ \Xhline{1.5pt}
S9v2t5 & 4:00 $\bm{\circ}$ & 0.179 & 3.2 & 0.0560 & 0.653 & 2.54 (1.08) \\ \hline
S9v2t5 &  4:30 $\bm{\circ}$ & 0.196 & 3.4 & 0.0577 & 0.275 & 6.14 (2.00) \\ \hline
\end{tabular}
\caption{\footnotesize{Scalings used in the nondimensionalizations of the model in each optimization. A $+$ denotes using a ``ghost'' time level. The instances above the thick line are Marangoni effect-dominated, the instance between is a transitional case, and below are evaporation-dominated.}}
\label{table:scalings}
\end{table}

\subsection{Marangoni Effect-Dominated Thinning}
\label{sec:strong}

\begin{figure} 
\centering
\subfloat[][FL intensity with minima aligned]{\includegraphics[scale=.12]{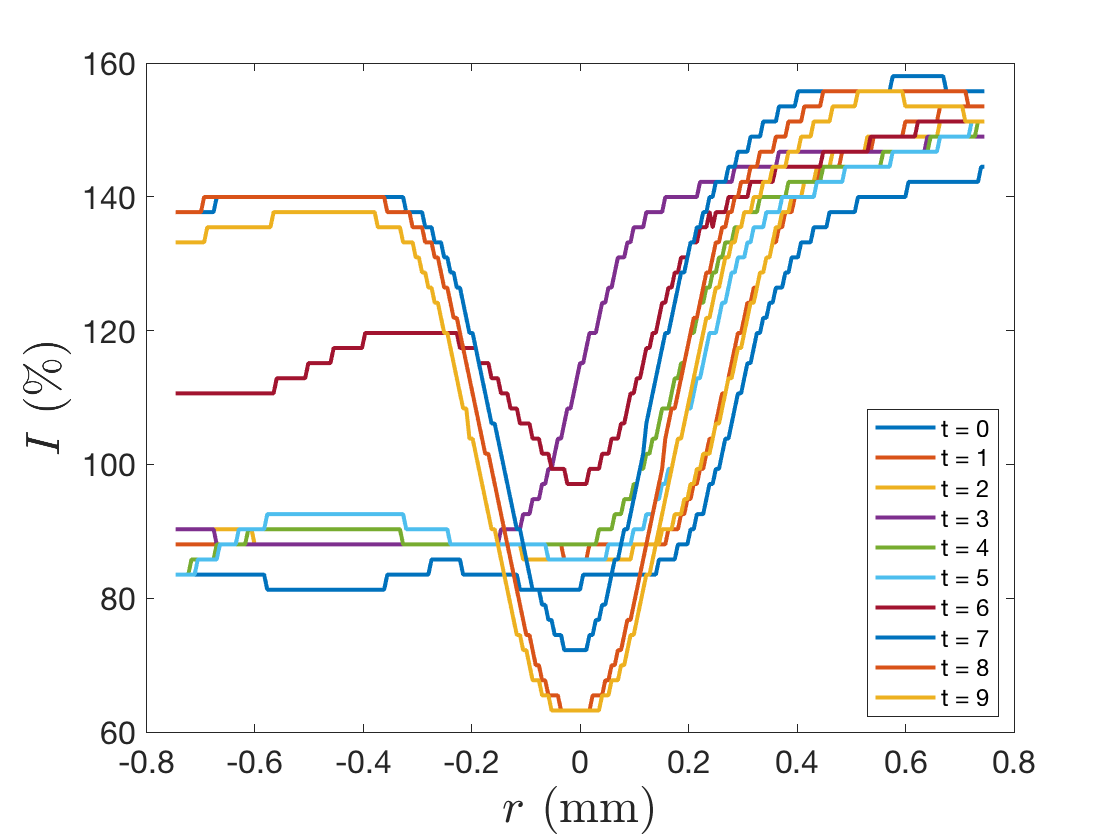}}
\subfloat[][FL intensity decrease]{\includegraphics[scale=.12]{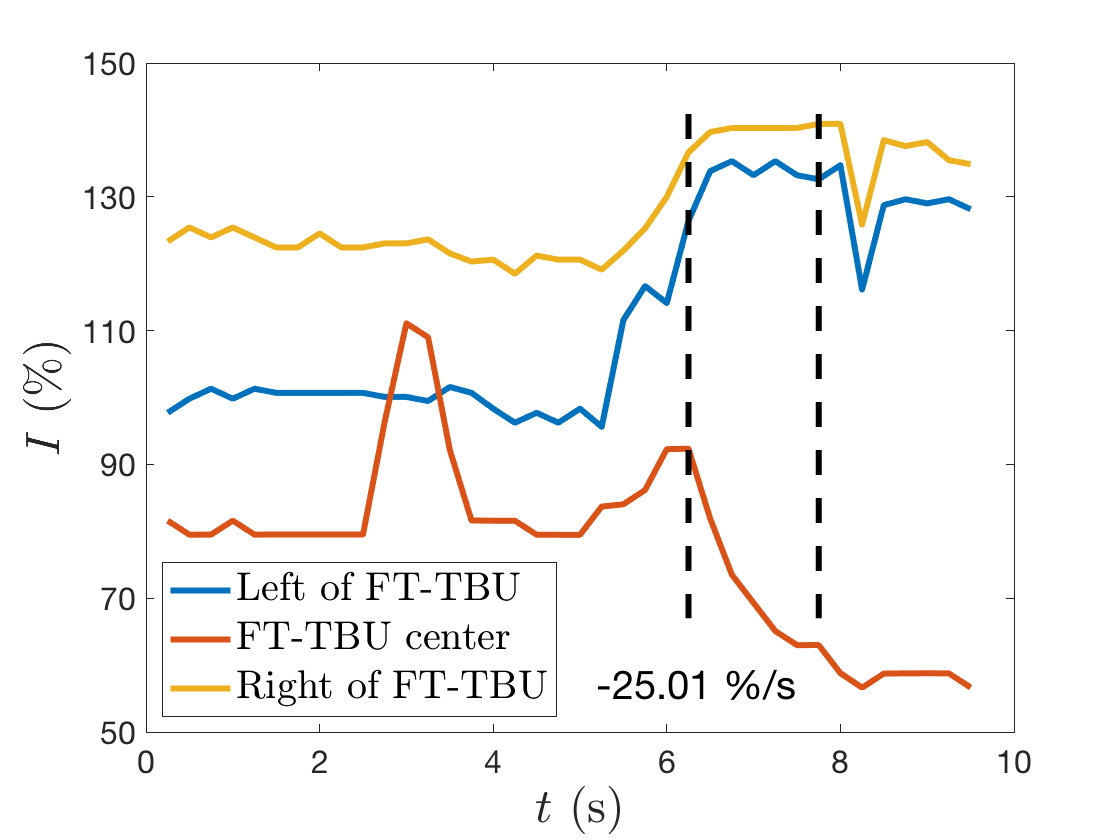}}
\subfloat[][FT-TBU data extraction]{\includegraphics[scale=.08]{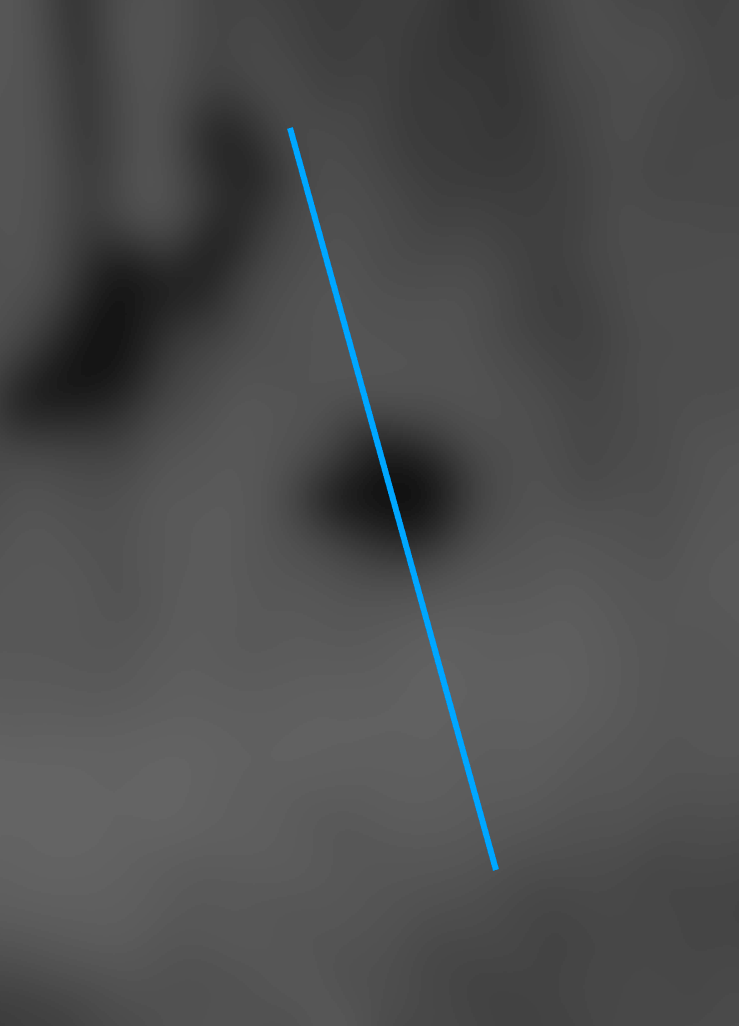}}
\caption{\footnotesize{Extracted data for the S10v1t6 12:30 spot. In (c) the image has been brightened and contrast-enhanced.}}
\label{fig:S10v1t6_1230_data}
\end{figure}

\begin{figure} 
\centering
\subfloat[][Exp. (\textbf{---}) and best fit th. (\textbf{- - -}) FL intensity]{\includegraphics[scale=.15]{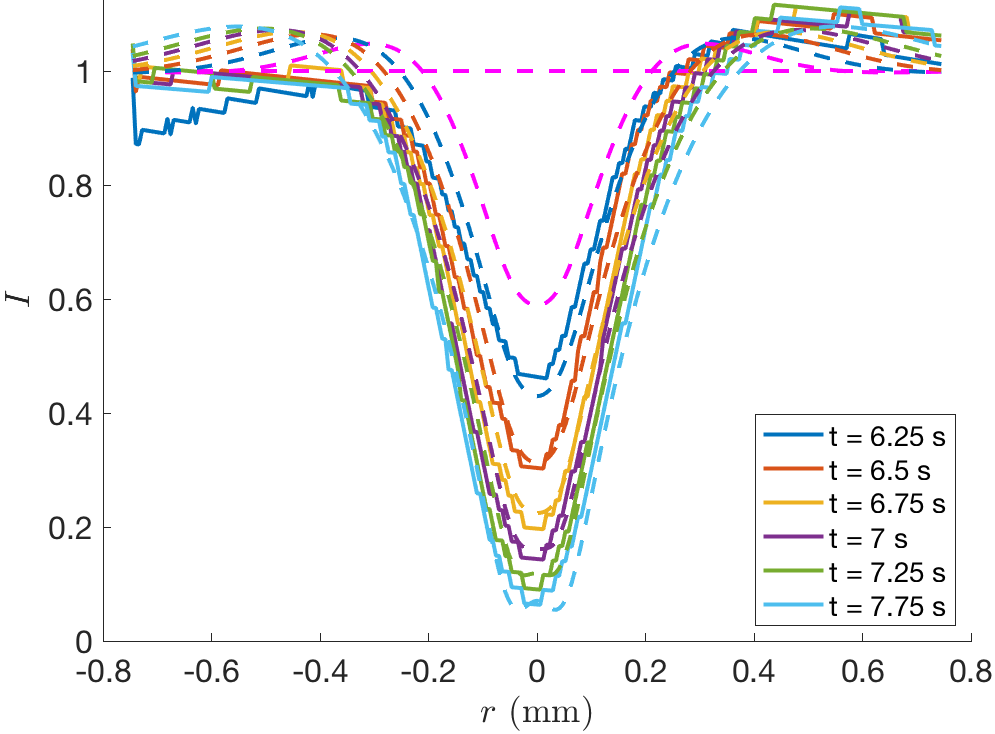}}
\subfloat[][Theoretical TF thickness]{\includegraphics[scale=.15]{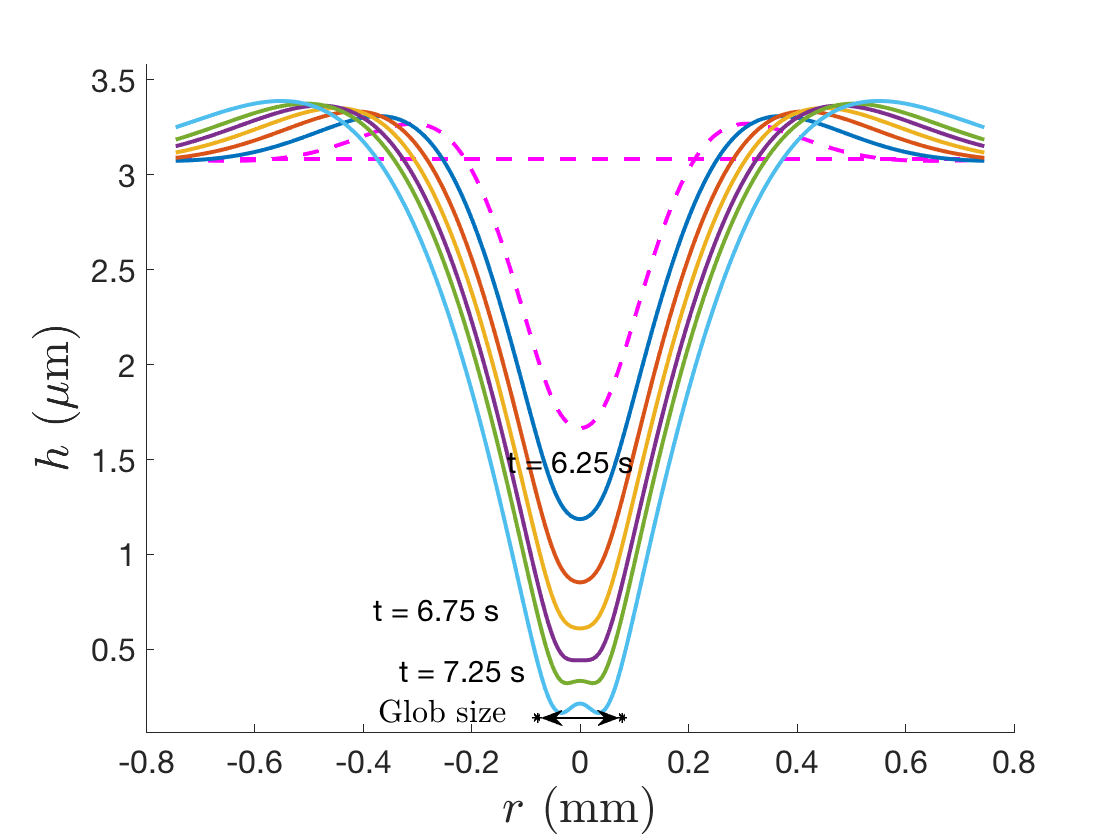}} \\
\subfloat[][Theoretical osmolarity]{\includegraphics[scale=.15]{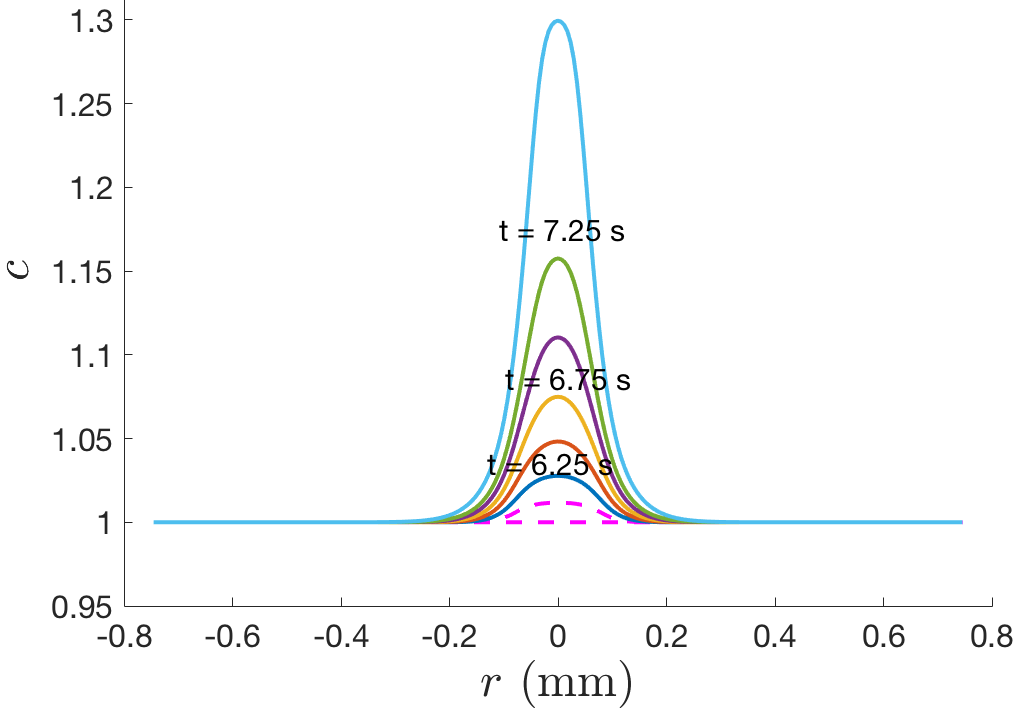}}
\subfloat[][Theoretical fluorescein  concentration]{\includegraphics[scale=.15]{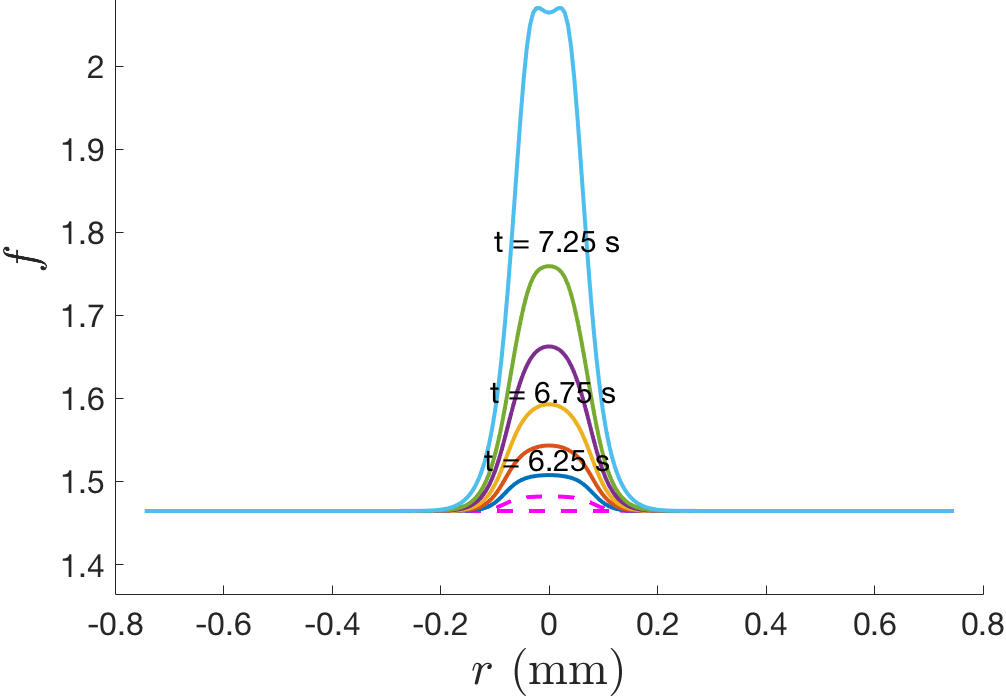}} \\
\subfloat[][Theoretical surfactant  concentration]{\includegraphics[scale=.15]{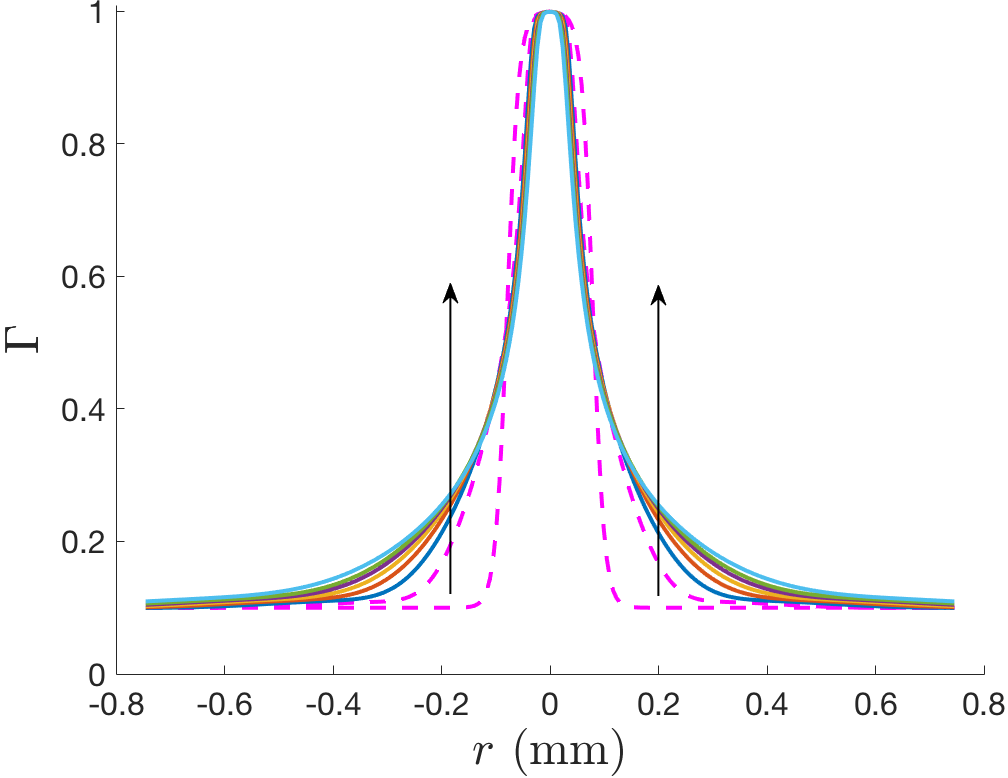}}  
\subfloat[][Theoretical depth-averaged fluid velocity]{\includegraphics[scale=.15]{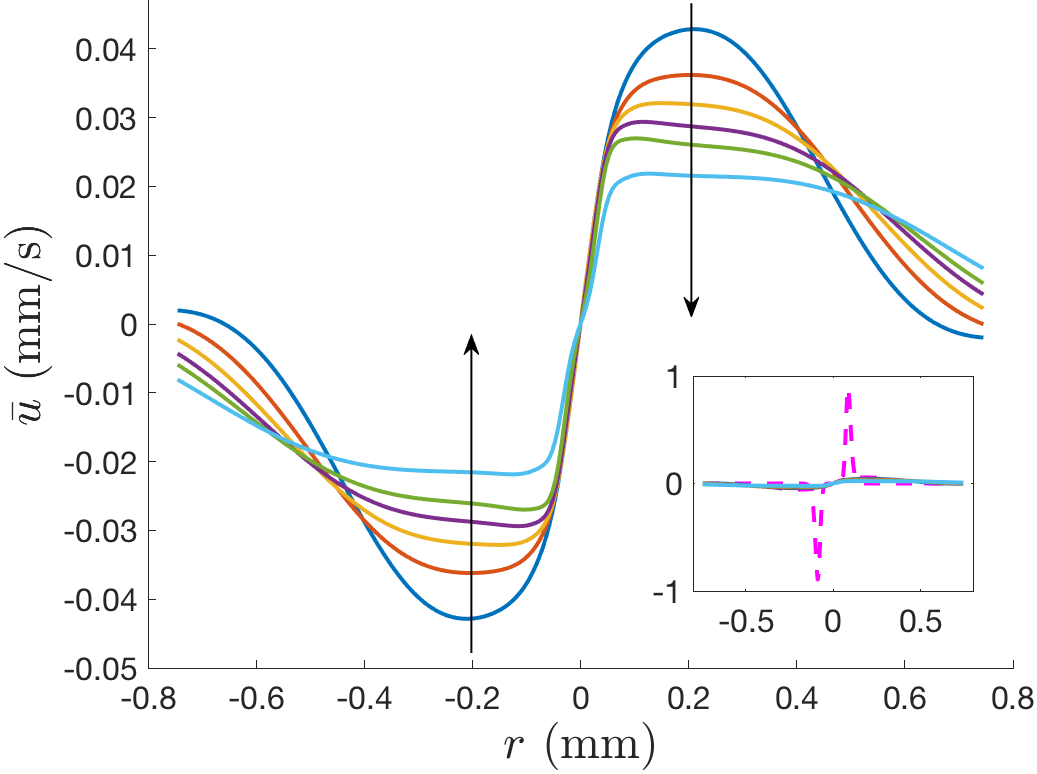}\label{fig:S10v1t6_1230_ubar}} \\
\subfloat[][Theoretical fluid surface velocity]{\includegraphics[scale=.15]{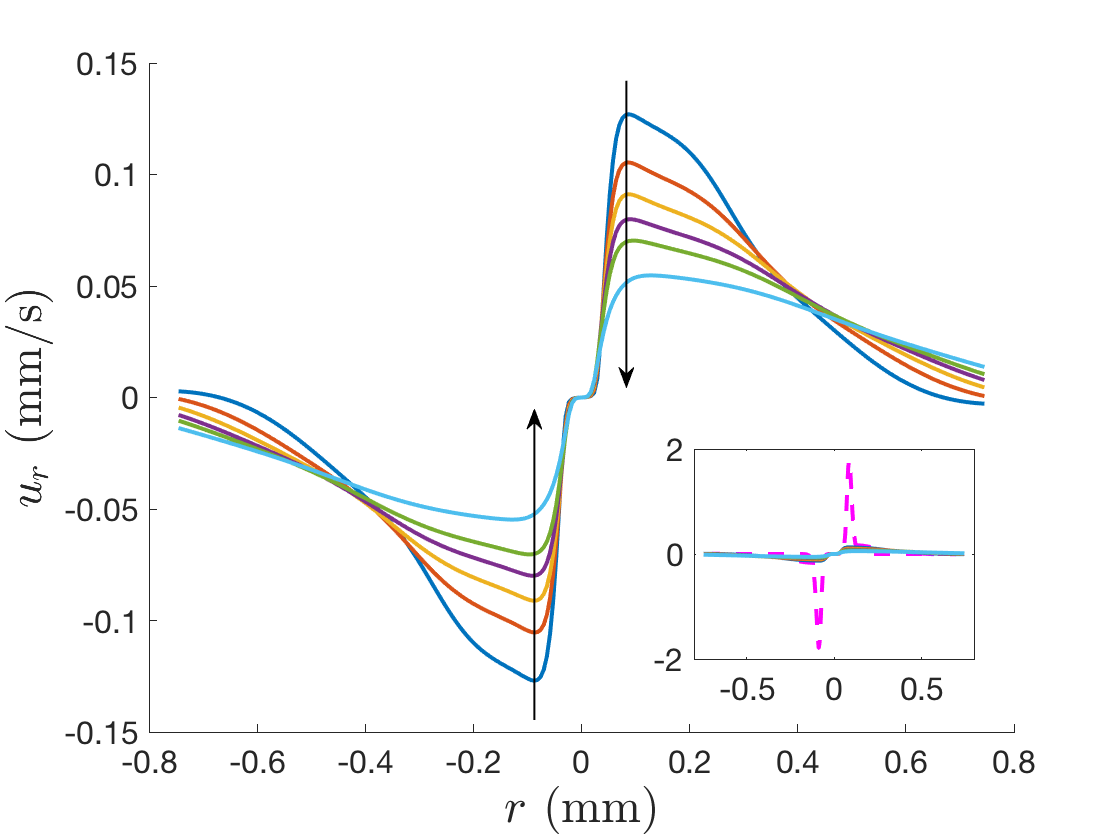}\label{fig:S10v1t6:1230_us}}
\begin{minipage}{18em}
\caption{\footnotesize{S10v1t6 12:30 spot best fit results (Case (c) evaporation). FL intensity has been normalized. Theoretical osmolarity is given as a fraction of the isotonic value. Arrows indicate increasing time.}}
\vspace{2cm}
\label{fig:S10v1t6_1230_fit}
\end{minipage}
\end{figure}

Marangoni effect-dominated thinning is characterized by strong, outward flow and small evaporation rates. Figure \ref{fig:S10v1t6_1230_fit} shows the fit to the S10v1t6 12:30 spot as an example; the data for the fit is shown in Figure \ref{fig:S10v1t6_1230_data}.

The S10v1t6 12:30 spot is a distinctive breakup instance because it is partially hidden by eyelashes and develops very quickly in the later half of the trial, as seen in Figure \ref{fig:S10v1t6_1230_data}b. We fit this instance with two ghost time levels for several reasons. Once the eyelash and lid move so the location where FT-TBU forms is visible, there is already substantial decrease in the measured FL intensity in the center of breakup--a 50\% difference as compared to the edges of the breakup region. If we compare this to the S18v2t4 7:30 spot, which we choose to fit with a single ghost time level, the latter shows only a 20\% difference between the measured FL intensity at the center of breakup and at the edges of the breakup region (see Figure 2 of Online Resource 1). Secondly, fitting the S10v1t6 12:30 spot instance with a single ghost time level results in a 21\% increase in the residual, indicating a significantly worse fit.

The optimal change in surface tension found by the optimization is by far the largest of all instances: 60.3 $\mu$N/m. Correspondingly, the Marangoni number is large, at 5.5. This trial is characterized by significant tangential flow driven by the Marangoni effect that persists throughout the duration of the fit. The magnitude of the flow is more than double any other instance reported in this work. As seen in Table \ref{table:mix_fits}, the S10v1t6 12:30 spot is fit with the smallest evaporation rate of all trials; this fact along with the strength of the flow suggests that the Marangoni effect dwarfs evaporation in terms of importance. The theoretical osmolarity peaks at a dimensional value of 390 mOsmol/L in the center of breakup, which is relatively small in comparison to other instances (see Table \ref{table:mix_fits}), and likely is a result of the short time span of the trial and the slow nature of osmosis. All the Marangoni effect-dominated instances show smaller maximum salt concentration values than the transitional or evaporation-dominated cases; the relatively small amount of evaporation in this first category correlates with a small increase in osmolarity.

All instances recorded in Table \ref{table:mix_fits} are fit well with either evaporation profile Case (b) or (c); we also fit the S18v2t4 7:30 spot with Case (d). The results are shown in Tables 1 and 2 and Figures 1 and 2 of Online Resource 1. By switching from Case (c) to Case (d), the fit was improved by less than 1\%, suggesting that evaporation is not the most important mechanism driving thinning in this instance. 

\subsection{Transitional Thinning}
\label{sec:mix}

Transitional breakup instances are characterized by thinning that is initially dominated by the Marangoni effect, but then becomes evaporation-dominated as the relative importance of the Marangoni effect diminishes as the trial progresses. We show the fit for the S9v2t1 3:00 streak as an example.

\begin{figure} 
\centering
\subfloat[][FL intensity with minima aligned]{\includegraphics[scale=.119]{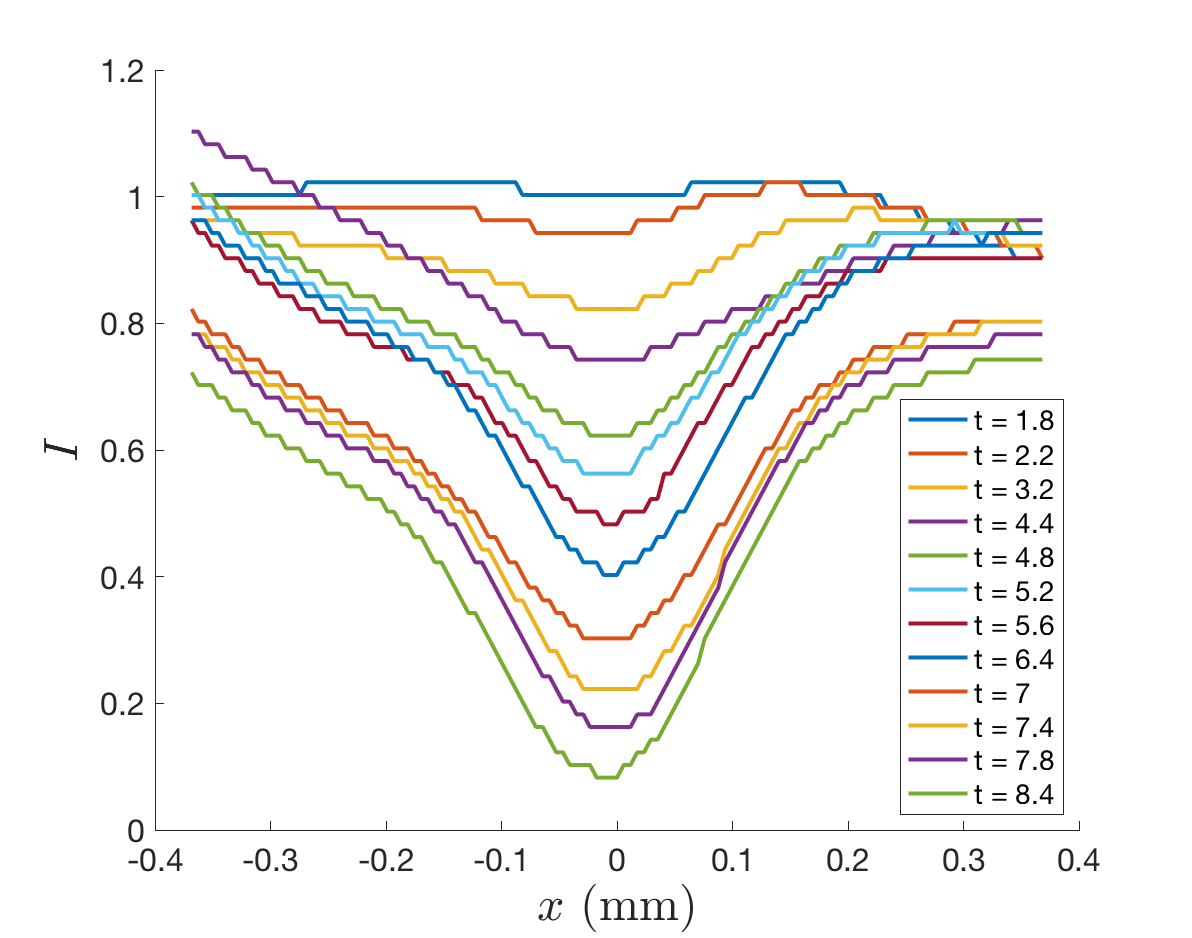}}
\subfloat[][FL intensity decrease]{\includegraphics[scale=.119]{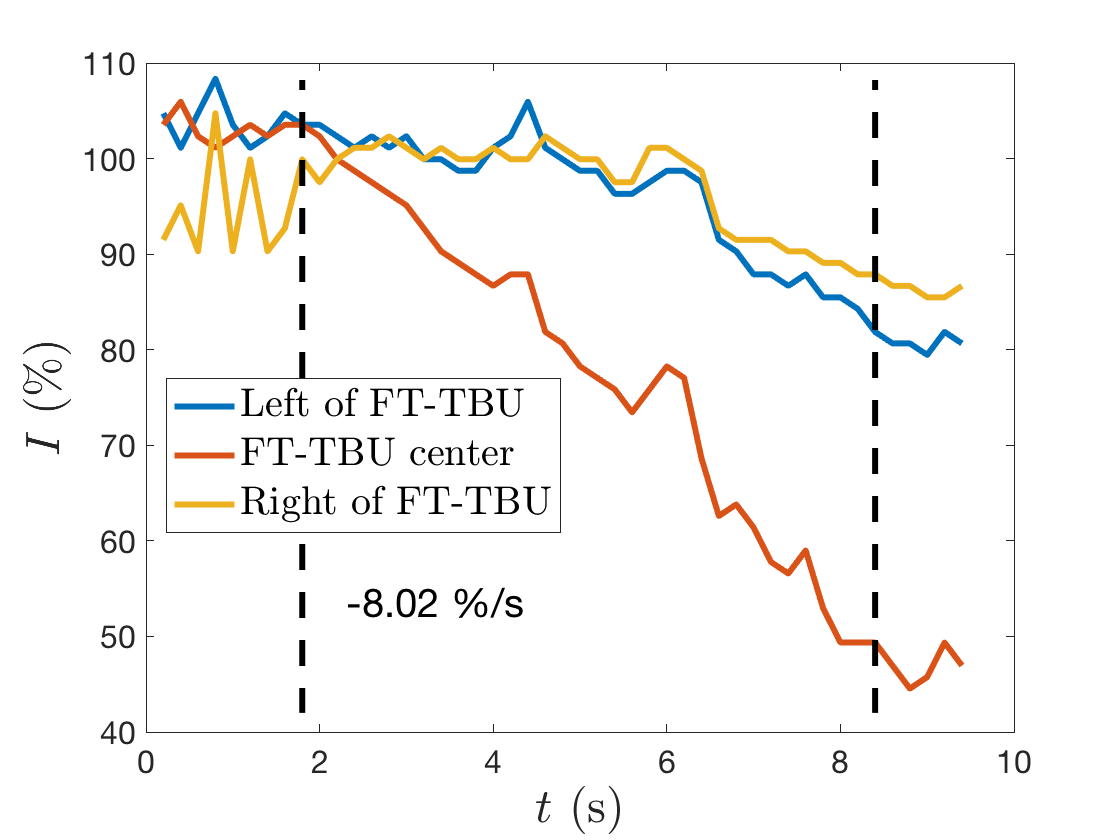}}
\subfloat[][FT-TBU data extraction]{\includegraphics[scale=.08]{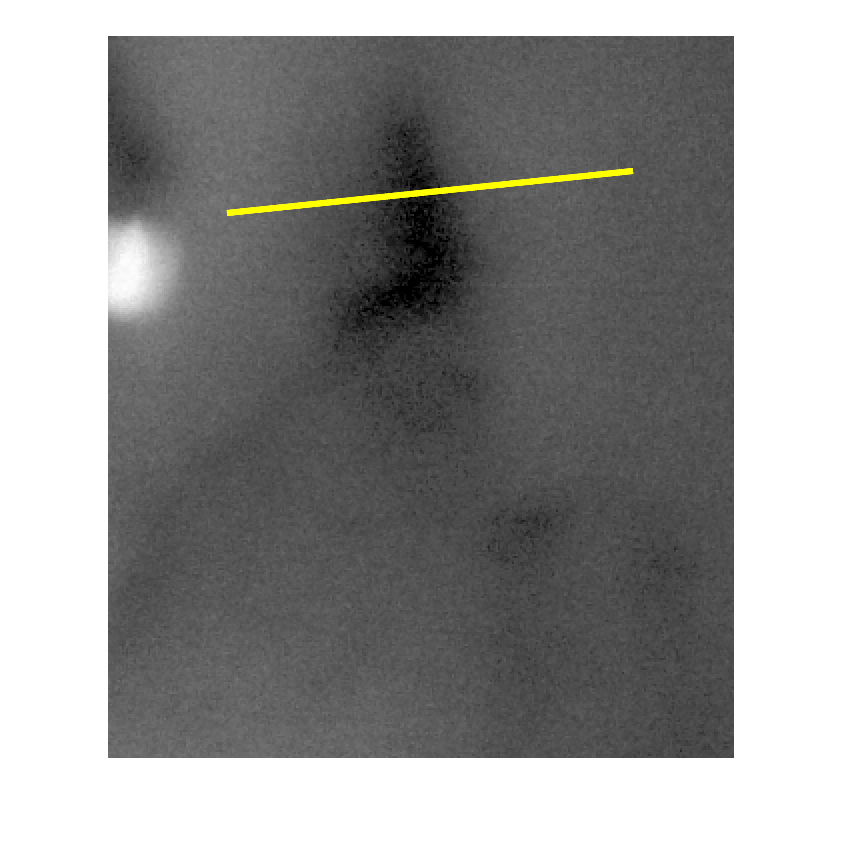}}
\caption{\footnotesize{Extracted data for the S9v2t1 3:00 streak. In (c) the image has been brightened and contrast-enhanced.}}
\label{fig:S9v2t1_3_data}
\end{figure}

\begin{figure} 
\centering
\subfloat[][Exp. (\textbf{---}) and best fit th. (\textbf{- - -}) FL intensity]{\includegraphics[scale=.15]{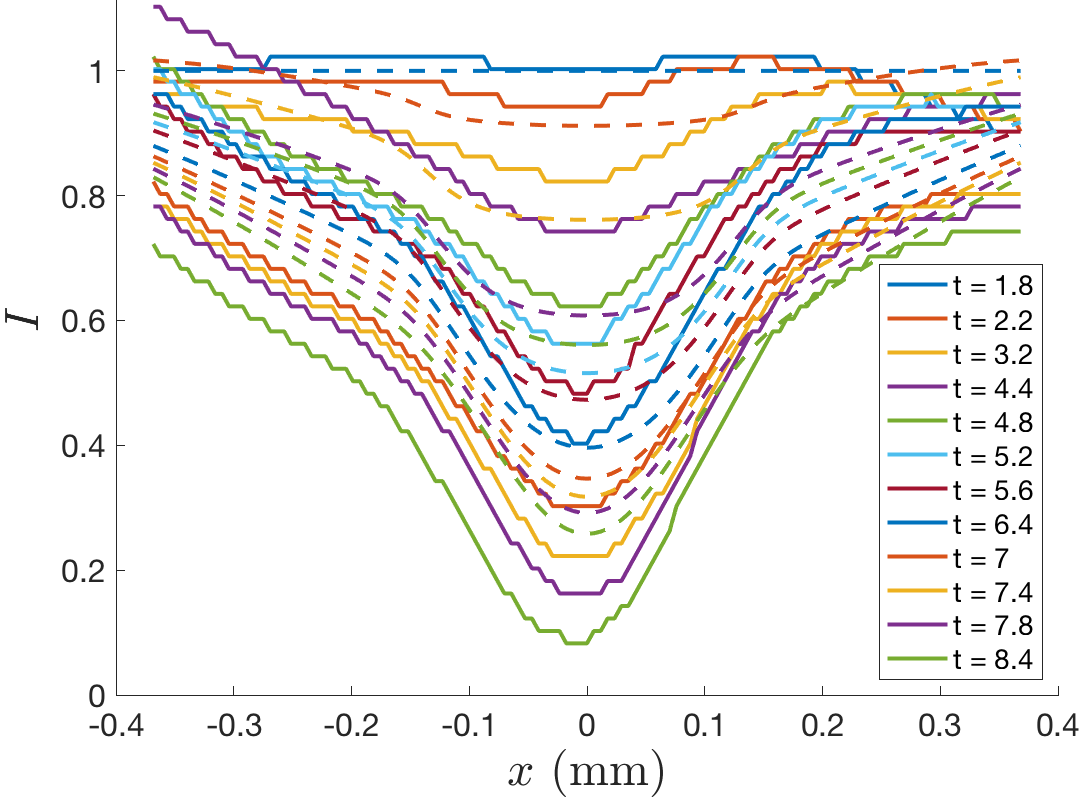}}
\subfloat[][Theoretical TF thickness]{\includegraphics[scale=.15]{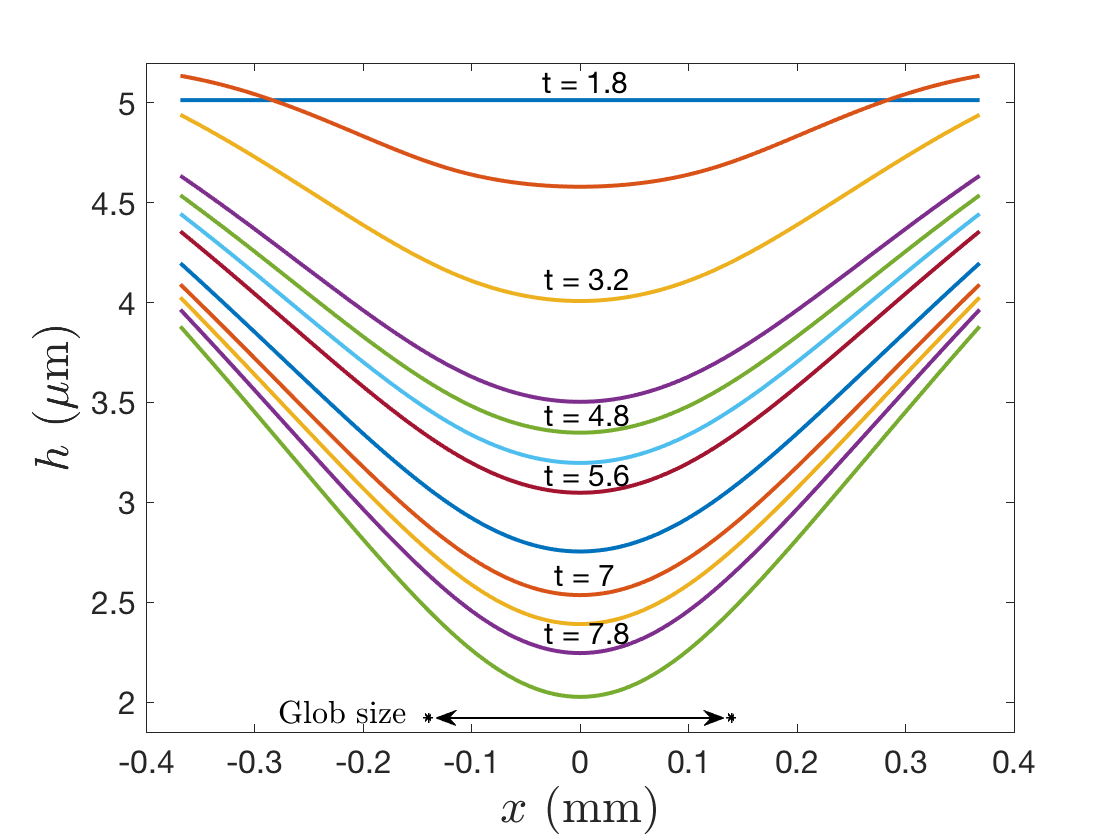}} \\
\subfloat[][Theoretical surfactant \\ concentration]{\includegraphics[scale=.15]{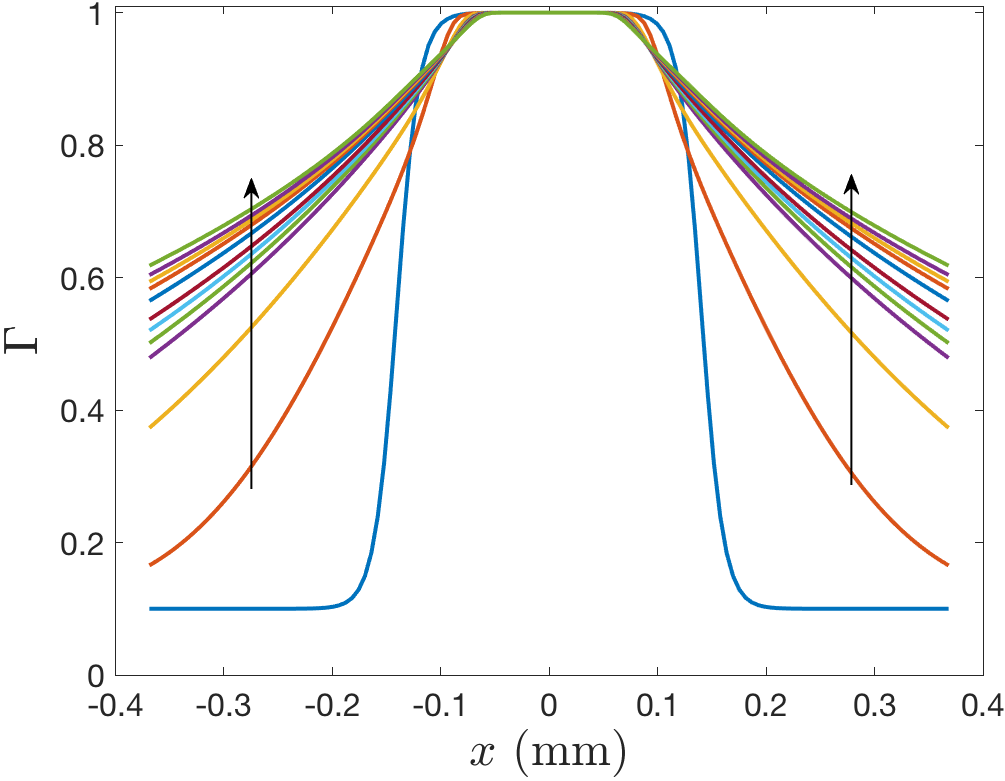}} 
\subfloat[][Theoretical \\ depth-averaged fluid velocity]{\includegraphics[scale=.15]{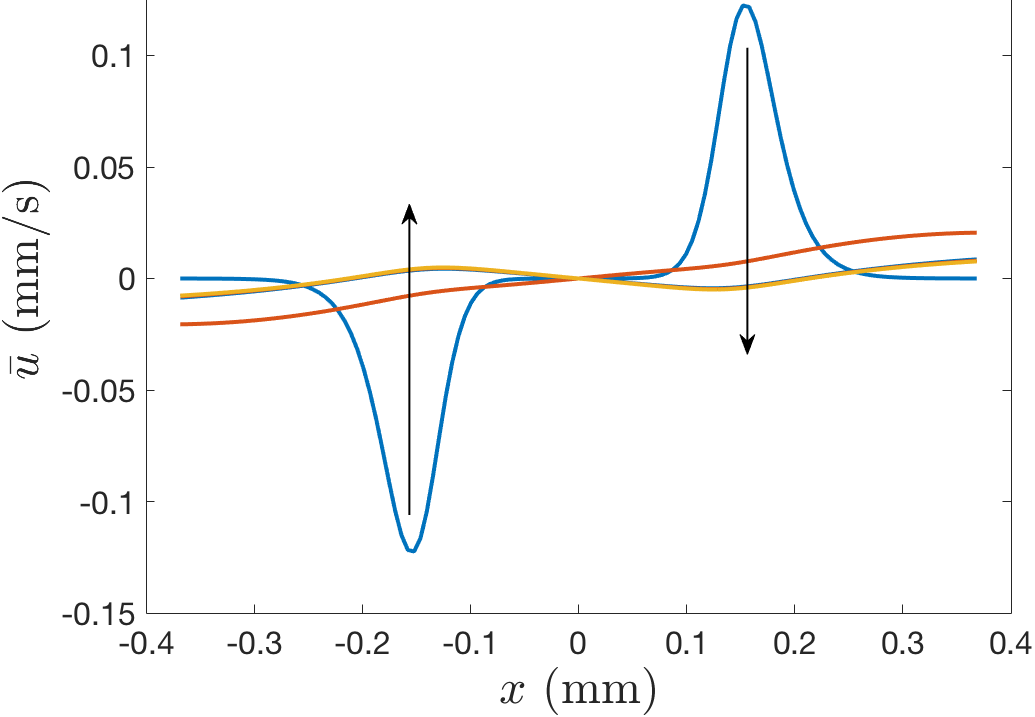}\label{fig:S9v2t1_3_ubar}} \\
\subfloat[][Theoretical fluid surface \\ velocity]{\includegraphics[scale=.15]{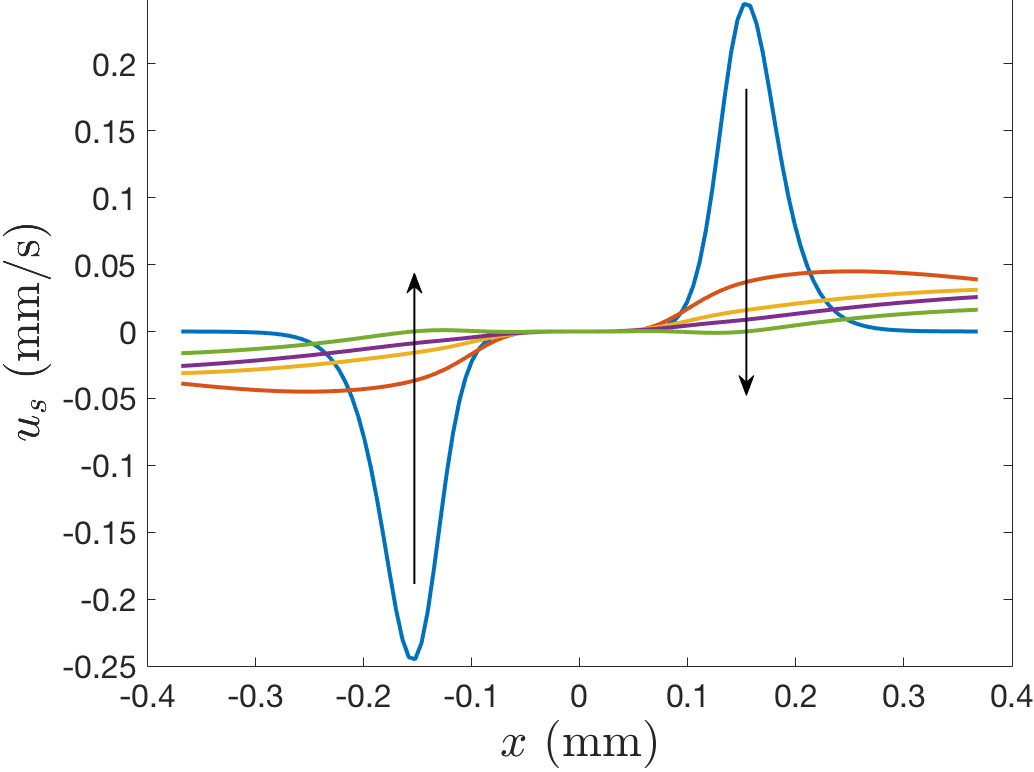}\label{fig:S9v2t1_3_us}}
\begin{minipage}{18em}
\caption{\footnotesize{S9v2t1 3:00 streak best fit results (Case (c) evaporation). FL intensity has been normalized. Theoretical osmolarity is given as a fraction of the isotonic value. Arrows indicate increasing time.}}
\vspace{2cm}
\end{minipage}
\label{fig:S9v2t1_3_fit}
\end{figure}

In Figure \ref{fig:S9v2t1_3_fit}d, we see from the sign of $\bar{u}$ near the glob edge that flow is initially directed outward, indicative of the Marangoni effect, but then quickly reverses direction and becomes healing flow, indicative of evaporation-driven thinning (\citealt{peng2014,luke2020}). This illustrates the fact that Marangoni flow dominates the early thinning of the breakup center, but that evaporation takes over later in the trial. In contrast, the flow is always directed outward far from the center of breakup. This instance has a Marangoni number of 1.48, illustrating the moderate importance of the Marangoni effect. The maximum osmolarity is estimated at a value of 
531 mOsmol/L; this is an intermediate value when compared to the other instances recorded in Table \ref{table:mix_fits}. At 6.6 seconds, this is the longest trial reported in this paper, the duration of which may allow the increase in salt concentration.

\subsection{Evaporation-Dominated Thinning}
\label{sec:weak}

The S9v2t5 4:00 spot is an example of relatively weak Marangoni effect in comparison to evaporation. The change in surface tension is 4.05 $\mu$N/m and the Marangoni number is 0.653. A Marangoni number below one suggests lipid-driven tangential flow plays a relatively weak role in causing thinning.

\begin{figure} 
\centering
\subfloat[][FL intensity with minima aligned]{\includegraphics[scale=.12]{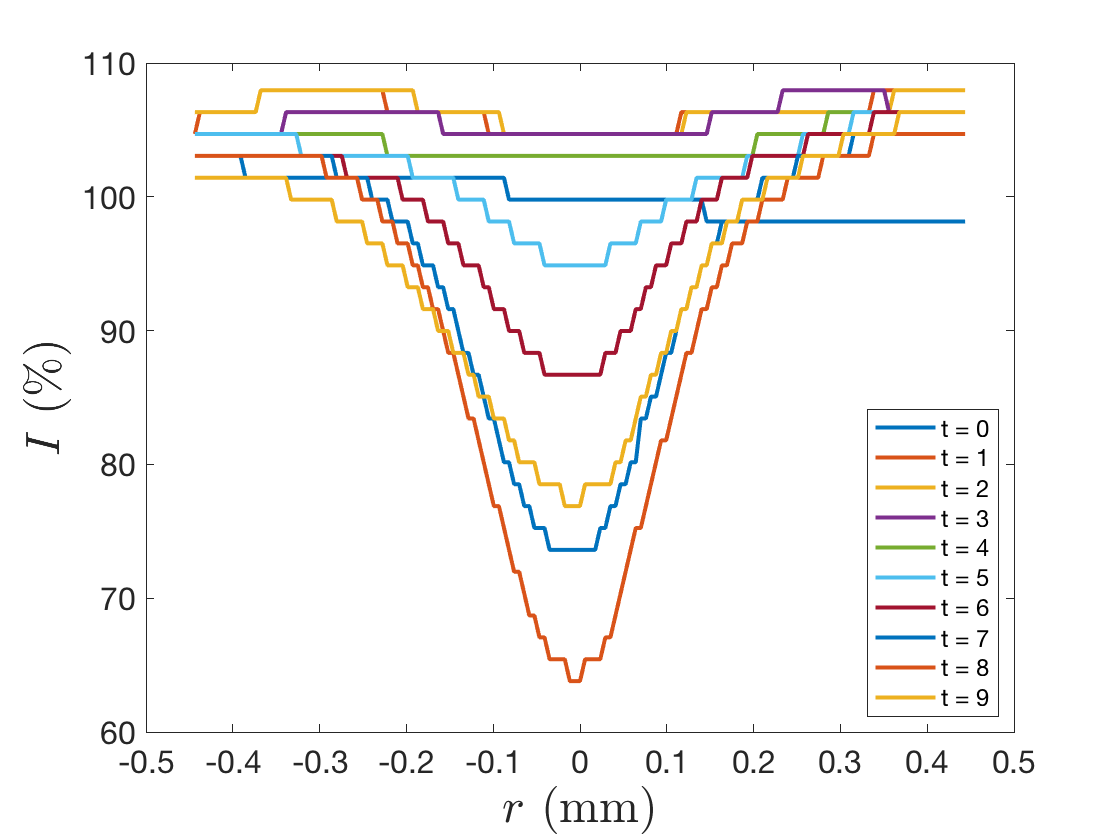}}
\subfloat[][FL intensity decrease]{\includegraphics[scale=.12]{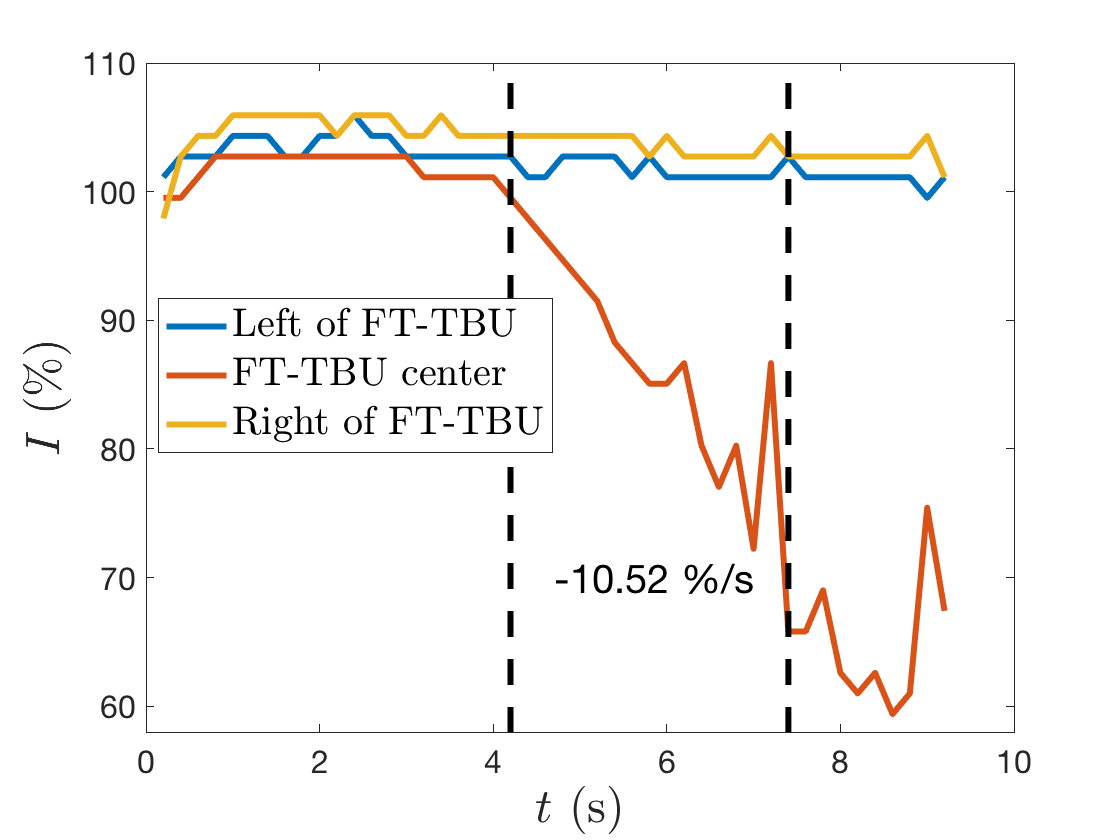}}
\subfloat[][FT-TBU data extraction]{\includegraphics[scale=.09]{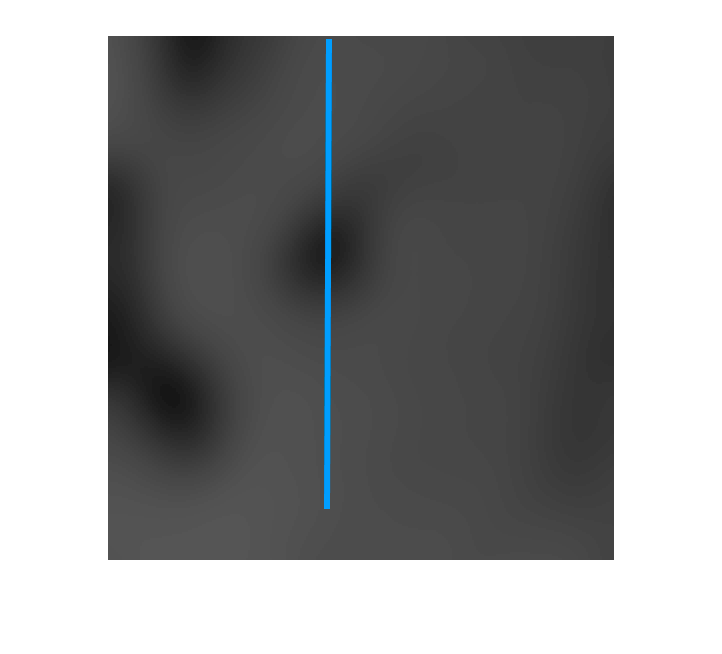}}
\caption{\footnotesize{Extracted data for the S9v2t5 4:00 spot. In (c) the image has been brightened and contrast-enhanced.}}
\label{fig:S9v2t5_4_data}
\end{figure}

\begin{figure} 
\centering
\subfloat[][Experimental (solid) and best fit theoretical (dashed) FL intensity]{\includegraphics[scale=.155]{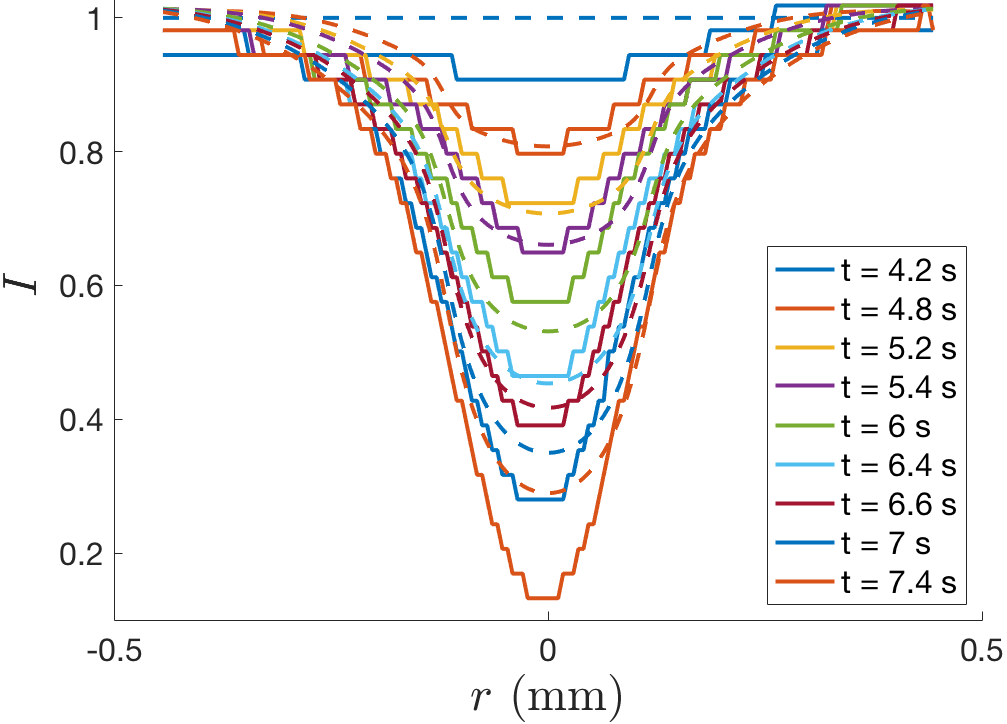}}
\subfloat[][Theoretical TF thickness]{\includegraphics[scale=.155]{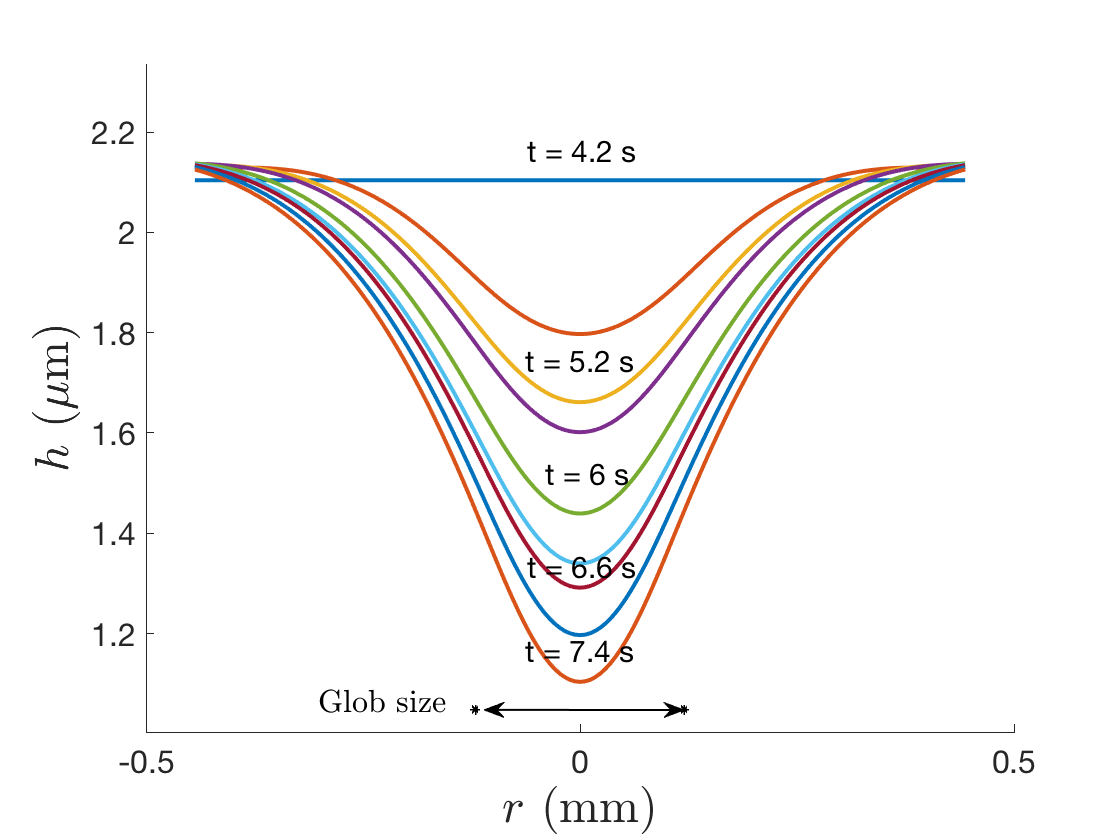}} \\
\subfloat[][Theoretical depth-averaged fluid velocity]{\includegraphics[scale=.155]{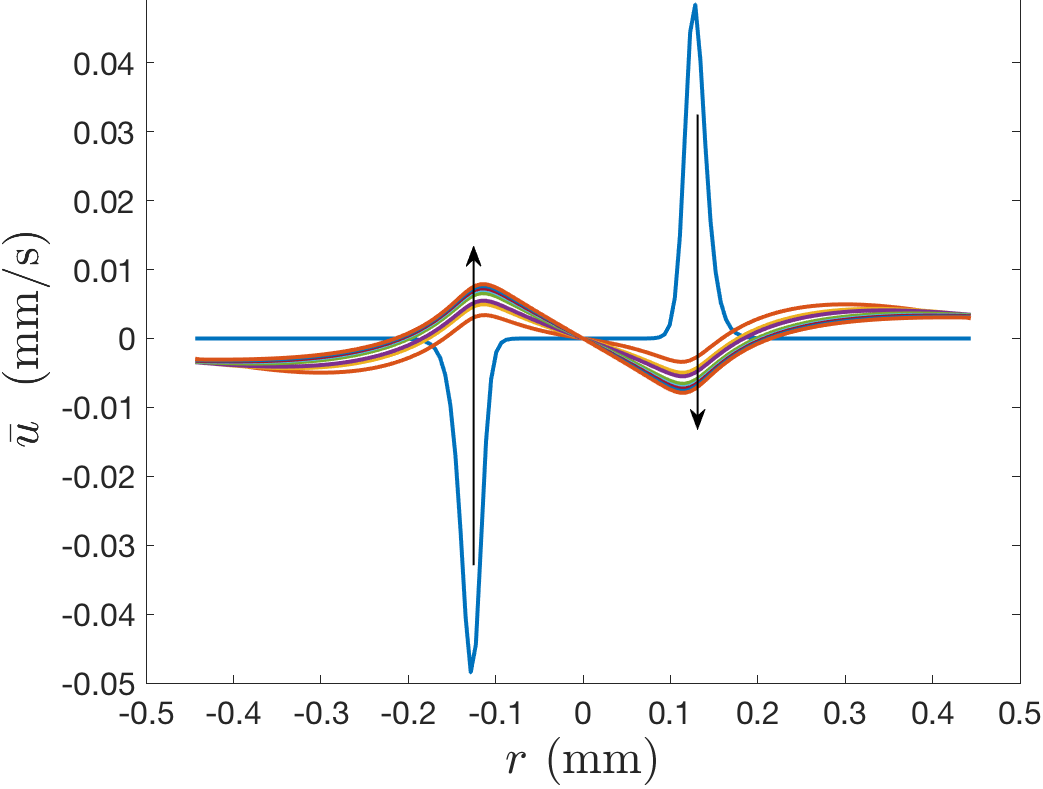}\label{fig:S9v2t5_4_ubar}}
\subfloat[][Theoretical fluid surface velocity]{\includegraphics[scale=.155]{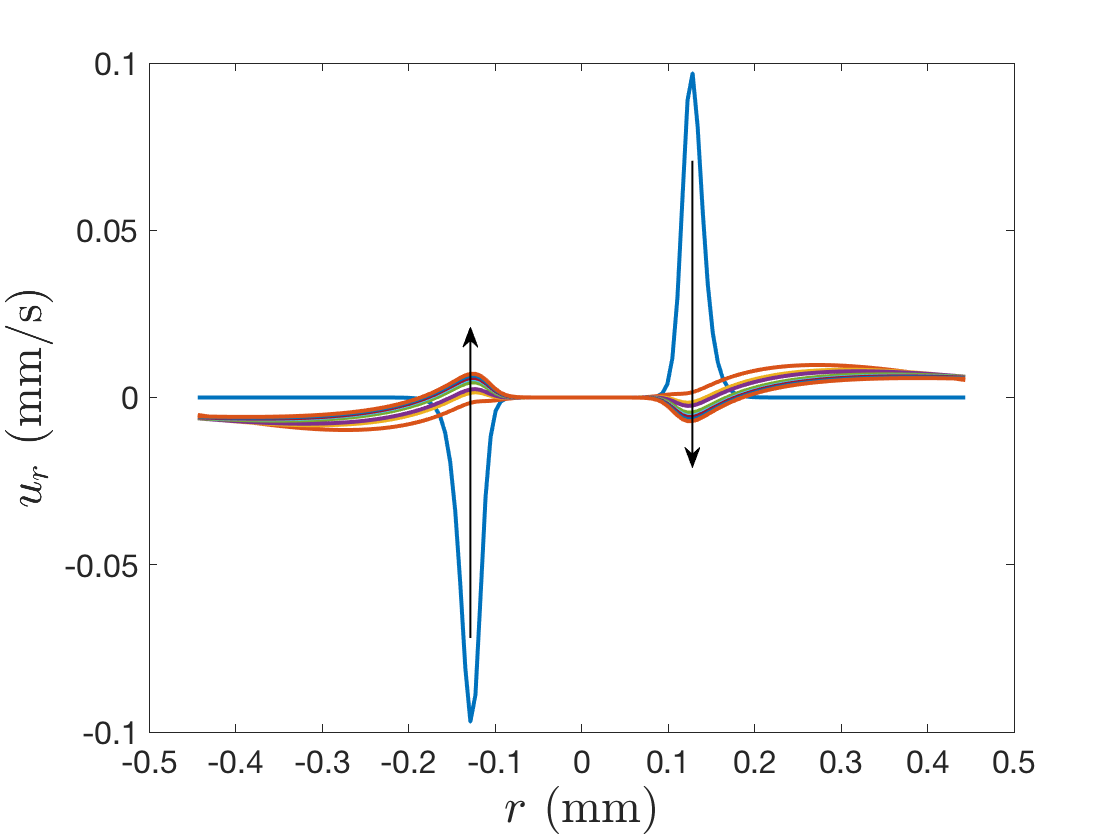}\label{fig:S9v2t5_4_us}}
\caption{\footnotesize{S9v2t5 4:00 spot best fit results (Case (c) evaporation). FL intensity has been normalized. Theoretical osmolarity is given as a fraction of the isotonic value. Arrows indicate increasing time.}}
\label{fig:S9v2t5_4_fit}
\end{figure}

Both $\bar{u}$ and $u_r$ in Figures \ref{fig:S9v2t5_4_fit}c,d show outward flow near the edge of the glob in the first time level that quickly changes to inward, healing flow. This healing flow is stronger in magnitude than the weak, outward tangential flow at the edges of the domain. The evaporation rate of this instance is higher compared to others, as evaporation must overcome the inward flow to create the spot. The scale of the initial outward flow is much smaller than that described in Sections \ref{sec:strong} and \ref{sec:mix}, further evidence that the Marangoni effect plays a weak role in thinning the TF in this instance. The maximum theoretical osmolarity, 579 mOsmol/L, is nearly twice the isotonic value. The two instances we categorize as evaporation-dominated show the highest maximum osmolarity values; this suggests that significant evaporation is related to a large increase in osmolarity.

\subsection{Evaporation-Only Model}
\label{sec:evap}

We fit the breakup instances recorded in Tables \ref{table:mix_fits} and \ref{table:scalings} with an evaporation-only model when possible (some fits were not successful). This model uses a Gaussian distribution for evaporation; the parameters that are adjusted are the peak thinning rate, $v_{\max}'$, background thinning rate, $v_{\min}'$, and Gaussian distribution width, $r_w'/x_w'$. We record the results in Table \ref{table:evap}. Most values of the optimal evaporation rates are at the top end of what may be considered realistic (\citealt{nichols2005}), and some are above what we think is possible (over 40 $\mu$m/min). This is strong evidence that evaporation alone cannot cause thinning occurring in this short time. It is important to note that \cite{nichols2005} recorded overall thinning rates, which may underestimate the evaporation rate if inward healing flow slows thinning. We compare an average of the overall thinning rates from our models, $\partial h'/\partial t'$, with this experimental data in Section \ref{sec:disc}.

An example fit can be seen for the S18v2t4 7:30 spot in Figure 4 of Online Resource 1. The optimal evaporation rate is 53.3 $\mu$m/min (see Table \ref{table:evap}), which is well above what has been recorded for evaporation. In contrast, this instance is fit well with the mixed-mechanism model (see Table \ref{table:mix_fits}).

The S10v1t2 8:00 streak is an instance where the optimal parameters from a fit using the mixed-mechanism model suggest the Marangoni effect plays essentially no role in causing the thinning. The optimal change in surface tension is 0.0380 $\mu$N/m, and the corresponding Marangoni number is 9.25$\times 10^{-3}$. In comparison, fitting this instance with the evaporation-only model results in a realistic thinning rate of 19.3 $\mu$m/min and a better fit (see Table \ref{table:evap}). This suggests evaporation alone dominates the thinning in this instance. The corresponding data and fit can be seen in Figures 5 and 6 of Online Resource 1.

\begin{table} 
\centering
\tabcolsep=0.15cm
\begin{tabular}{|c|c|c|c|c|c|c|c|c|c|c|}
\hline
\textbf{Trial} & \textbf{\begin{tabular}[c]{@{}c@{}}FT-TBU \\ ID\end{tabular}}  & \textbf{\begin{tabular}[c]{@{}c@{}}$\bm{h_0'}$ \\ ($\bm{\mu}$m)\end{tabular}} & \textbf{\begin{tabular}[c]{@{}c@{}}$\bm{f_0'}$ \\(\%)\end{tabular}} & \begin{tabular}[c]{@{}c@{}}$\bm{v'_{\max}}$\\ \textbf{($\bm{\frac{\mu\text{m}}{\text{min}}}$)} \end{tabular} & \begin{tabular}[c]{@{}c@{}}$\bm{v'_{\min}}$\\ \textbf{($\bm{\frac{\mu\text{m}}{\text{min}}}$)} \end{tabular} & \begin{tabular}[c]{@{}c@{}}$\bm{r_w'/x_w'}$\\ \textbf{(mm)} \end{tabular}   & \textbf{\begin{tabular}[c]{@{}c@{}}Min\\ $\bm{I_{ex}}$\end{tabular}} & \textbf{\begin{tabular}[c]{@{}c@{}}Min\\ $\bm{I_{th}}$\end{tabular}} & \textbf{\begin{tabular}[c]{@{}c@{}}Min\\ $\bm{h_{th}'}$\end{tabular}} & \textbf{\begin{tabular}[c]{@{}c@{}}Max\\ $\bm{c_{th}}$\end{tabular}} \\ \hline
S9v1t4$^+$ & 4:00 \textbf{---} & 3.32 &0.324 & 33.5 & 4.32 & 0.110 & 0.119 & 0.206 & 1.49 & 2.39 \\ \hline
S9v2t1 & 3:00 \textbf{---} & 5.01 & 0.292 & 45.3 & 5.08 & 0.0858 & 0.0826 & 0.215 & 2.70 & 2.40 \\ \hline
S9v2t5 & 4:00 $\bm{\circ}$ & 2.10 & 0.299 & 37.7 & -0.584 & 0.0840 & 0.138 & 0.281 & 1.21 & 1.96 \\ \hline
S10v1t2 & 8:00 \textbf{---} & 2.17 & 0.255 & 19.3 & 1.69 & 0.0629 & 0.133 & 0.231 & 0.979 & 2.14 \\ \hline
S13v2t10$^+$ & 6:30 \textbf{---} & 3.59 & 0.259 & 40.0 & 8.70 & 0.117 & 0.116 & 0.151 & 1.27 & 2.99 \\ \hline
S18v2t4$^+$ & 7:30 $\bm{\circ}$ & 2.48 & 0.363 & 53.3 & 11.9 & 0.100 & 0.143 & 0.0756 & 0.704 & 3.49 \\ \hline
S27v2t2$^+$ & 5:00 \textbf{---} & 1.91 & 0.4 & 35.3 & 2.31 & 0.0925 & 0.340 & 0.340 & 1.07 & 1.80 \\ \hline
\end{tabular}
\caption{\footnotesize{Results from fitting three parameters with an evaporation-only model with a Gaussian evaporation profile. A  + denotes a ``ghost'' time level was used. The subject (S) number, visit (v) number and (t) trial number are listed, and the FT-TBU location is a clock reading taken from the center of the pupil. A \textbf{---} denotes streak FT-TBU, and a $\bm{\circ}$ is a spot. The initial TF thickness and FL concentration estimates are given. The optimized parameters are the peak evaporative thinning rate $v_{\max}'$, the background evaporative thinning rate $v_{\min}'$, and the Gaussian evaporation profile radius or half-width $r_w'/x_w'$. The minimum values of both the experimental and theoretical FL intensity and the theoretical thickness are reported.
}}
\label{table:evap}
\end{table}

\subsection{Zero Evaporation Model}
\label{sec:mar}

In order to determine whether evaporation is necessary in the instances we study, we fit the data with a model that excludes evaporation. Successful fits are recorded in Table \ref{table:mar}. While the parameter values are reasonable, the residuals of the fits are far higher than those given by the mixed-mechanism model. There is rapid change in the theretical FL intensity and TF thickness in the beginning, but the decrease slows and cannot capture the behavior of the later experimental data. Two example fits are shown in Figures 7 and 8 of Online Resource 1 respectively: the S10v1t6 12:30 spot and the S18v2t4 7:30 spot. Comparing Tables \ref{table:mix_fits} and \ref{table:mar}, we see that the optimal values for both $(\Delta \sigma)_0$ and $R_I'$ for the S10v1t6 12:30 spot are fairly similar. However, the fit shown in Section \ref{sec:strong} captures the qualitative nature of the data in the last few time levels better, as the theoretical intensity for the Marangoni effect-only model shown in Online Resource 1 exhibits an upturn near the center of breakup. Further, the residual is 6\% smaller in the mixed-mechanism model case as compared with the Marangoni effect-only model fit. The mixed-mechanism fit for the S10v1t6 12:30 spot exhibits Marangoni effect-dominated flow; the relatively successful zero-evaporation fit to this instance is strong support of this interpretation.  In contrast, the S18v2t4 7:30 spot is an intermediate case where both evaporation and the Marangoni-effect play important roles in causing the thinning; this is seen in the relatively poor fit to the data when evaporation is turned off.

\begin{table} 
\centering
\tabcolsep=0.205cm
\begin{tabular}{|c|c|c|c|c|c|c|c|c|c|}
\hline
\textbf{Trial} & \textbf{\begin{tabular}[c]{@{}c@{}}FT-TBU \\ ID \end{tabular}} & \textbf{\begin{tabular}[c]{@{}c@{}}$\bm{h_0'}$ \\ ($\bm{\mu}$m)\end{tabular}} & \textbf{\begin{tabular}[c]{@{}c@{}}$\bm{f_0'}$ \\(\%)\end{tabular}} & \begin{tabular}[c]{@{}c@{}}$\bm{(\Delta \sigma)_0}$\\ \textbf{($\bm{\frac{\mu\text{N}}{\text{m}}}$)} \end{tabular}  & \begin{tabular}[c]{@{}c@{}}$\bm{R_I', X_I'}$\\ \textbf{(mm)} \end{tabular}   & \textbf{\begin{tabular}[c]{@{}c@{}}Min\\ $\bm{I_{ex}}$\end{tabular}} & \textbf{\begin{tabular}[c]{@{}c@{}}Min\\ $\bm{I_{th}}$\end{tabular}} & \textbf{\begin{tabular}[c]{@{}c@{}}Min\\ $\bm{h_{th}'}$\end{tabular}} & \textbf{Ma}  \\ \hline
S10v1t6$^*$ & 12:30 $\bm{\circ}$ & 3.08 & 0.293 & 60.7 & 0.0624 & 0.0635 & 0.0714 & 0.182 & 5.54 \\  \hline
S13v2t10$^*$ & 6:30 \textbf{---} & 3.59 & 0.259 & 43.0 & 0.119 & 0.116 & 0.371 & 1.16 & 6.07 \\ \hline
S18v2t4$^*$ & 7:30 $\bm{\circ}$ & 2.48 & 0.363 & 37.2 & 0.125 & 0.143 & 0.316 & 0.677 & 4.88 \\ \hline
S27v2t2$^*$ & 5:00 \textbf{---} & 1.91 & 0.4 & 25.2 & 0.0647 & 0.340 & 0.625 & 1.11 & 2.92 \\ \hline
\end{tabular}
\caption{\small{Results from fitting two parameters with a zero evaporation model. A $+$ denotes a ``ghost'' time level was used. The subject (S) number, visit (v) number and (t) trial number are listed, and the FT-TBU location is a clock reading taken from the center of the pupil. A \textbf{---} denotes streak FT-TBU, and a $\bm{\circ}$ is a spot. The initial TF thickness and FL concentration estimates are given. The optimized parameters are the change in surface tension $(\Delta \sigma)_0$ and the glob radius $R_I'$ or half-width $X_I'$. The minimum values of both the experimental and theoretical FL intensity and the theoretical thickness are reported as well as the Marangoni number.}}
\label{table:mar}
\end{table}

\subsection{Fluid Flow Profiles}

\begin{figure} 
\centering
\subfloat[][Local extrema of $\bar{u}$]{\includegraphics[scale=.155]{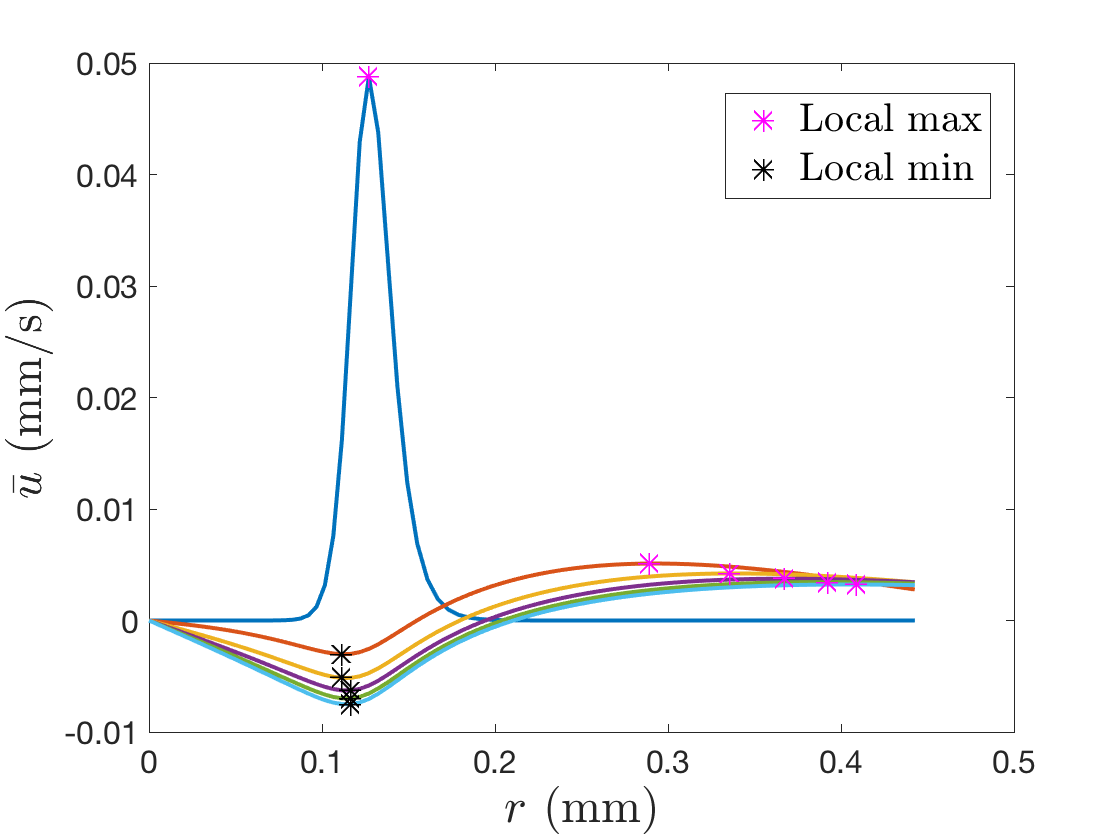}}
\subfloat[][Local extrema of $u_r$]{\includegraphics[scale=.155]{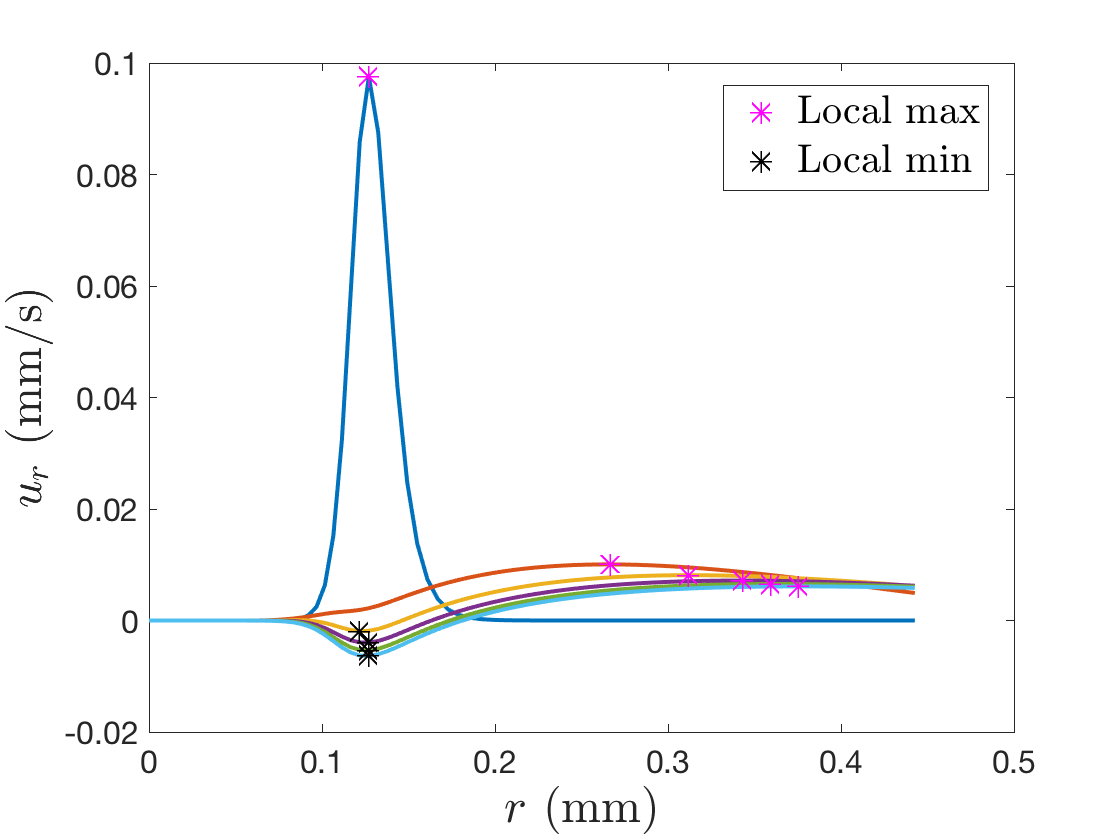}}
\caption{\footnotesize{Local extrema for the S9v2t5 4:00 spot.}}
\label{fig:ubar_us}
\end{figure}

The theoretical fluid profiles $\bar{u}$ and $u_r$ or $u_s$ that result from an optimization both illustrate the dynamic nature of the thinning and reveal the relative importance of the Marangoni effect and evaporation. Figures \ref{fig:S10v1t6_1230_ubar},g and \ref{fig:S9v2t1_3_ubar},e show the fluid profiles for the S10v1t6 12:30 spot and S9v2t1 3:00 streak, respectively. As previously discussed, the S10v1t6 12:30 spot exhibits strong tangential flow for the duration of the fit, while $\bar{u}$ for the S9v2t1 3:00 streak transitions from strong outward to weakly inward flow near the glob edge with slightly stronger outward flow near the edges of the domain. Figure \ref{fig:ubar_us} shows an example of transitional thinning in which the inward flow near the glob rises in importance and overtakes the outward flow away from the glob by the end of the trial. We mark the spatial locations of the relative extrema; note that the maxima move significantly to the right over time, indicating that the strongest outward flow moves to the edge of the domain as the spot forms and widens slightly. Near the glob edge, healing flow forms and acts in a narrow spatial region for the majority of the trial. Looking at $\bar{u}$ and $u_r$ or $u_s$ helps us categorize the three example instances in Sections \ref{sec:strong}, \ref{sec:mix}, and \ref{sec:weak} as Marangoni effect-dominated, transitional, and evaporation-dominated thinning, respectively.

\begin{figure} 
\centering
\subfloat[][Depth-averaged fluid velocity]{\includegraphics[scale=.155]{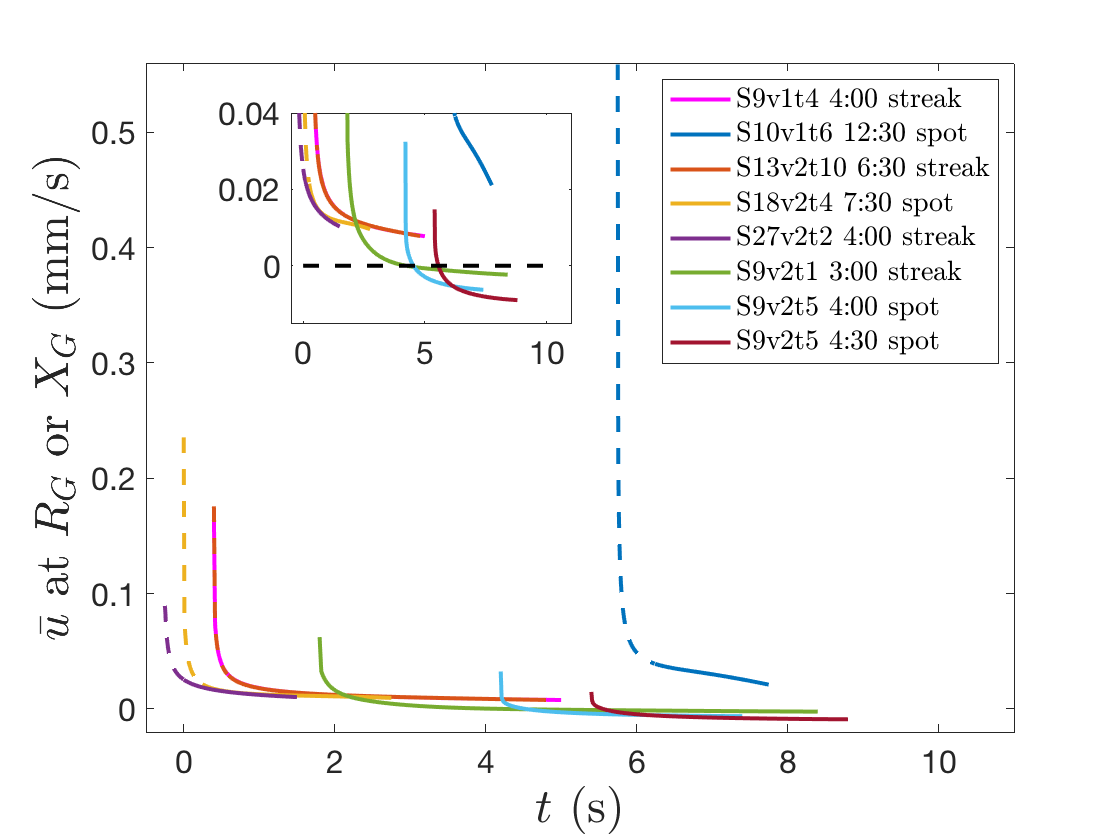}}
\subfloat[][Fluid surface velocity]{\includegraphics[scale=.155]{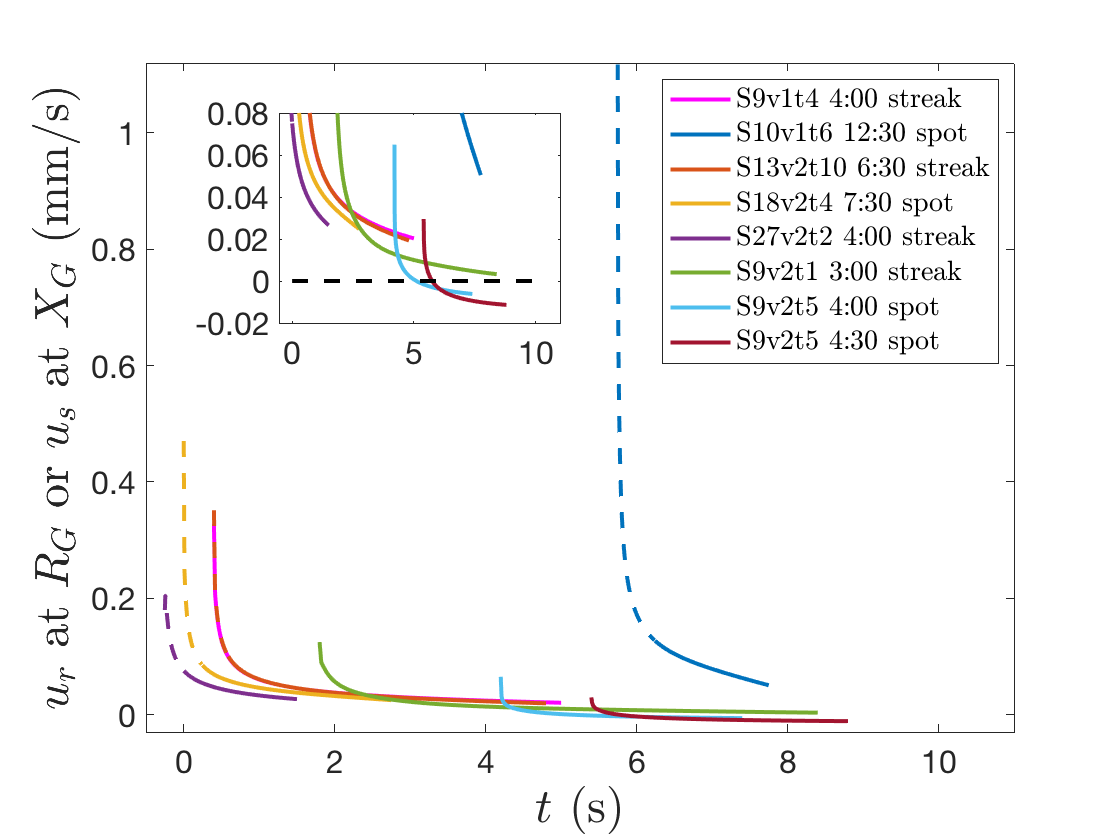}}
\caption{$\bar{u}$ and $u_r$ at $R_G = R_I + R_W$ or $u_s$ at $X_G = X_I + X_W$. Dashed lines indicate ghost time levels and solid lines indicate fit times. Insets: a close-up near the origin, with a dashed line at zero.}
\label{fig:ubar_us_all}
\end{figure}

We investigate and compare the fluid flow profiles of all eight mixed-mechanism instances at the edge of the glob, which is given by $R_G = R_I + R_W$, the glob radius + the transition width, or $X_G = X_I + X_W$, the half-width + the transition width (depending on spot or streak). By recording $\bar{u}$ and $u_r$ or $u_s$ at $R_G$ or $X_G$, we ensure that our measurement is outside of the glob. As previously mentioned, the S10v1t6 12:30 spot exhibits the strongest flow of any instance by more than a factor of two. Both $\bar{u}$ and $u_r$ or $u_s$ of each instance decrease significantly in magnitude in less than a second; this is evidence of the rapidly-acting Marangoni effect that wanes in importance as the glob spreads out. The inset of Figure \ref{fig:ubar_us_all}a shows that three trials exhibit flow at the glob edge that begins outward and then turns inward. These are the S9v2t1 3:00 streak, which we categorize as transitional thinning, and the S9v2t5 4:00 and 4:30 spots, which we designate as evaporation-dominated thinning. This inward flow at the glob edge indicates that capillary flow has overtaken tangential flow, and thus evaporation has become or is the dominant mechanism.

\subsection{Effect of Initial FL Concentration on Fitting}

\begin{figure} 
\centering
\subfloat[][Case (b) evaporation]{\includegraphics[scale=.2]{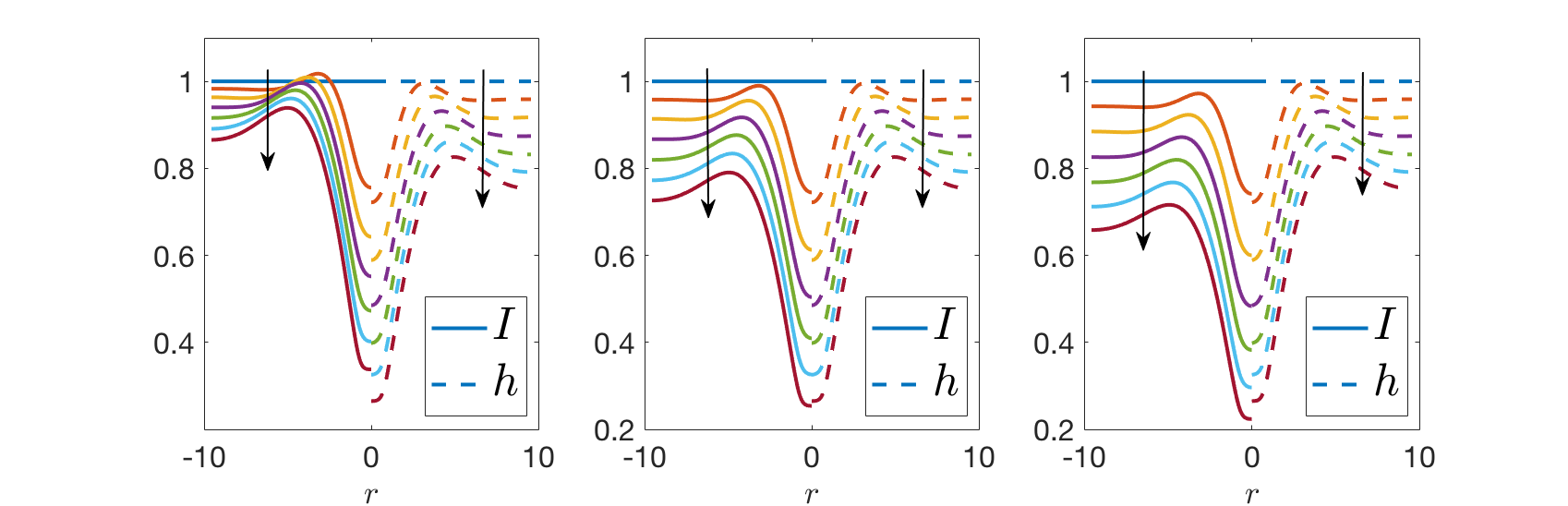}
\label{fig:nd_sols_dif_f0_cE}} \\
\subfloat[][Case (c) evaporation]{\includegraphics[scale=.2]{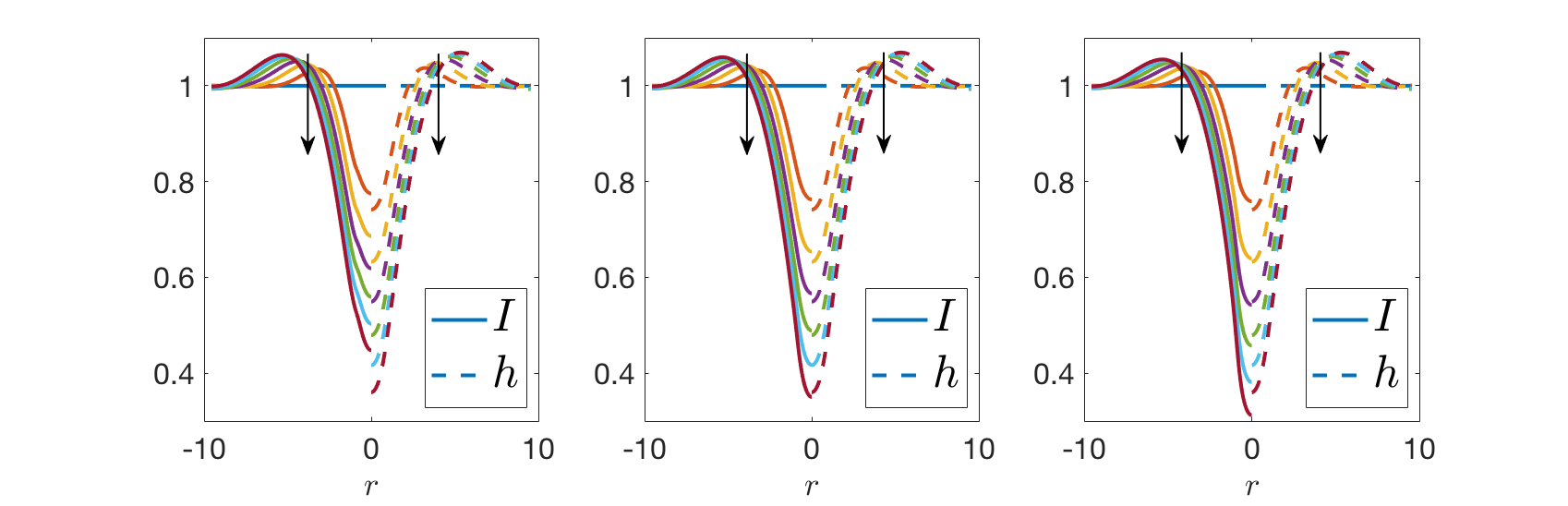}
\label{fig:nd_sols_dif_f0_ncE}}
\caption{\footnotesize{Nondimensional solutions for $I$ and $h$ for three different choices of $f_0'$: from left to right, $f_0' = 0.1$\%, $f_0' = 0.2$\%, and $f_0' = 0.3$\%. Arrows indicate increasing time. The parameters are $v' = 15 \ \mu$m/min, $R_I' = 0.1$ mm, $(\Delta \sigma)_0 = 20 \ \mu$N/m, $f_0' = 0.2$ \%, and $d = 3 \ \mu$m. The Marangoni number is 2.61.}}
\label{fig:nd_sols_dif_f0}
\end{figure}

As has been discussed elsewhere (\citealt{nichols2012,braun2014,braun2015,luke2020}), the FL concentration affects the relationship between FL intensity and TF thickness and can complicate interpretation of results. In particular, we investigate how the initial FL concentration estimate we obtain and fix during our optimization procedure can affect our fits. Figure \ref{fig:nd_sols_dif_f0} shows the qualitative and quantitative similarities of normalized theoretical TF thickness and FL intensity.  Each subfigure shows plots for three different initial FL concentration estimates: 0.1\%, 0.2\%, and 0.3\%. The initial FL concentration estimates for the results shown in Table \ref{table:mix_fits} have a mean and standard deviation of $0.316 \pm 0.0451$\%. Thus, our fits should exhibit dynamics most like the right-most plot in both Figures \ref{fig:nd_sols_dif_f0_cE} and \ref{fig:nd_sols_dif_f0_ncE}. We see that for initial FL concentrations near 0.3\%, TF thickness is initially ahead of FL intensity at the origin, but falls behind at later time levels.

\begin{figure} 
\centering
\includegraphics[scale=.2]{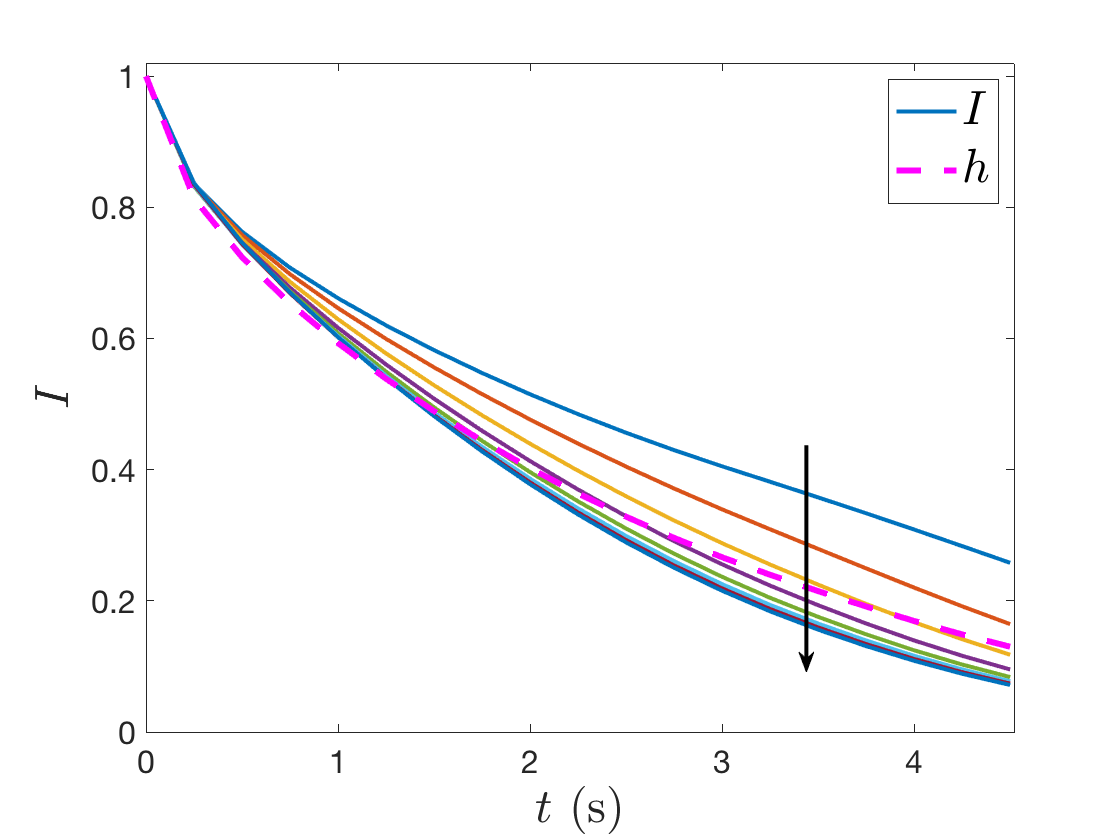}
\caption{\footnotesize{Minimum FL intensity value (normalized) plotted against time for various $f_0$. The arrow indicates increasing initial FL concentration, from 0.05\% to 0.4\% in increments of 0.05\%. Minimum TF thickness (normalized) has been plotted as a dashed line for comparison. The parameters are $v' = 15 \ \mu$m/min, $R_I' = 0.1$ mm, $(\Delta \sigma)_0 = 20 \ \mu$N/m, $f_0' = 0.2$ \%, and $d = 3 \ \mu$m. The Marangoni number is 2.61.}}
\label{fig:min_I_vs_t_f0}
\end{figure}

Figure \ref{fig:min_I_vs_t_f0} shows normalized minimum theoretical FL intensity plotted against time for varying initial FL concentration values. A dashed line indicates normalized minimum theoretical TF thickness, which correlates most closely with an initial FL concentration value of 0.15\% for parameter values that are characteristic for the fits reported in Table \ref{table:mix_fits}. Thus, the average initial FL concentration of our trials is above the ideal value to draw conclusions about TF thickness from measuring and fitting FL intensity. 

We explore the effect of varying the initial FL concentration $f_0'$ on all subsequent computations, including the determination of the optimal parameters. We report the results from examining the S9v2t5 4:00 spot, discussed in Section \ref{sec:weak}, and the S10v1t6 12:30 spot, discussed in Section \ref{sec:strong}. Ten different $f_0'$ values normally distributed around the initial FL concentration estimates recorded in Table \ref{table:mix_fits} were used with a standard deviation of $s = 0.05$\%.  Only results from the runs with residuals less than 10\% above the original value are reported in Tables \ref{tab:f0_strong} and \ref{tab:f0_weak}. For the S9v2t5 4:00 spot, the statistics for the initial FL estimates used are 0.284 $\pm \ 0.0561$\%, and for the resulting initial TF thickness estimates, 2.08 $\pm$ 0.353 $\mu$m, and for the S10v1t6 12:30 spot, the statistics are $f_0'$: 0.315 $\pm 0.0214$\% and $h_0'$: 3.14 $\mu$m $\pm$ 0.285 $\mu$m.

 The parameters are reported in Tables \ref{tab:f0_weak} and \ref{tab:f0_strong}. We denote the mean and standard deviation as $m$ and $s$, respectively. The values were on average 16.0 \% and 16.9 \% away from the optimal parameters recorded in Table \ref{table:mix_fits}, for the S9v2t5 4:00 and S10v1t6 12:30 spots, respectively. For comparison, the $f_0'$ values generated were 16.9\% and 5.03\% away from the mean on average for the S9v2t5 4:00 and S10v1t6 12:30 spots, respectively. For the S9v2t5 4:00 spot, the change in surface tension $(\Delta \sigma)_0$ showed significantly more variance from the optimal value as compared to both $R_I'$ and $v'$. This may be further evidence that the Marangoni effect is not very important to the thinning in this instance. In contrast, for the S10v1t6 12:30 spot, the thinning rate $v'$ showed the most variance from the optimal value, suggesting evaporation is not an important factor in causing the thinning in this case. 

\begin{table} 
\centering
\begin{tabular}{|c|c|c|c|c|}
\hline
\textbf{Quantity} & \textbf{$ \bm{m \pm s}$} & \textbf{Range} & \textbf{\begin{tabular}[c]{@{}c@{}}Range in\\ \%\end{tabular}} & \textbf{\begin{tabular}[c]{@{}c@{}}Avg. \% change\\ from opt\end{tabular}} \\ \hline
$v'$ ($\mu$m/min) & 27.5 $\pm$ 3.78 & 22.2--33.5 & 3.38--27.7 & 12.7 \\ \hline
$R_I'$ (mm) & 0.126 $\pm$ 0.00843 & 0.118--0.147 & 0.373--21.2 & 5.59 \\ \hline
$(\Delta \sigma)_0$ ($\mu$N/m) & 3.60 $\pm$ 1.51 & 0.325--5.66 & 2.88--92.0 & 29.8 \\ \hline
\end{tabular}
\caption{\footnotesize{S9v2t5 4:00 spot: statistics from varying $f_0'$ on the optimal parameters.}}
\label{tab:f0_weak}
\end{table}

\begin{table} 
\centering
\begin{tabular}{|c|c|c|c|c|}
\hline
\textbf{Quantity} & \textbf{$ \bm{ m \pm s}$} & \textbf{Range} & \textbf{\begin{tabular}[c]{@{}c@{}}Range in\\ \% \end{tabular}} & \textbf{\begin{tabular}[c]{@{}c@{}}Avg. \% change\\ from opt\end{tabular}} \\ \hline
$v'$ ($\mu$m/min) & 5.47 $\pm$ 2.82 & 1.92-9.41 & 17.4-67.5 & 39.1 \\ \hline
$R_I'$ (mm) & 0.0790 $\pm$ 0.00315 & 0.0740-0.0828 & 0.539-6.39 & 3.09 \\ \hline
$(\Delta \sigma)_0$ ($\mu$N/m) & 62.1 $\pm$ 7.01 & 54.9-74.4 & 2.36-23.3 & 8.40 \\ \hline
\end{tabular}
\caption{\footnotesize{S10v1t6 12:30 spot: statistics from varying $f_0'$ on the optimal parameters.}}
\label{tab:f0_strong}
\end{table}

\section{Discussion}
\label{sec:disc}

We fit PDE models to experimental FL data by optimizing several clinically-relevant parameters as model inputs. In comparison to conducting evaporation-only fits, the mixed-mechanism model poses more challenges when fitting to data. We successfully explain several of the varying situations we observe with different evaporation profiles and the inclusion of one or two ghost time levels. There is no unifying theme of the instances we report here, in contrast to the pure-evaporation fits recorded in \cite{luke2020}. There seem to be far fewer instances of intermediate and rapid, hypothesized glob-driven (\citealt{Cho1992,yokoi2019}) thinning than evaporation-driven thinning in the FL data that we studied. Regardless, we are successful in obtaining highly detailed information about the breakup instances we study. Our optimizations are robust as they are insensitive to initial guesses and noise.

\cite{zhong2019} varied nondimensional glob sizes $R_I$ between 0.25 and 3 to examine the effect of FT-TBUT (referred to as TBUT in their paper) on $R_I$.
The length scale used was 0.0742 mm, which is about 4 times smaller than what is found for the fits shown here. The Marangoni-driven instance that the authors used had a trial length of 2.5 seconds. They noted that capillary pressure driven by increased curvature in the TF shape results in a longer FT-TBUT for a nondimensional $R_I < 0.025$. All of our nondimensional optimal glob sizes are above this value, ranging from 0.254 to 0.628. The authors also plotted the location of the minimum thickness versus glob size and showed the existence of a crossover point above which the location of the FT-TBU is outside the glob radius. FT-TBU happens under globs smaller than $R_I = 0.9$ nondimensionally (0.067 mm dimensionally when $d = 3.5 \ \mu$m), and at or outside the edge when the glob is larger. Marangoni effect-driven shear stress extracts fluid from underneath small globs, but cannot affect the TF near the center of a larger glob. While we use a different length scaling than \cite{zhong2019} did, all of our glob sizes are significantly smaller than the cutoff value of 0.9 nondimensionally. Thus, we expect that breakup is taking place under the globs.

\begin{figure} 
\centering
\subfloat[][Evaporation rates]{\includegraphics[scale=.2]{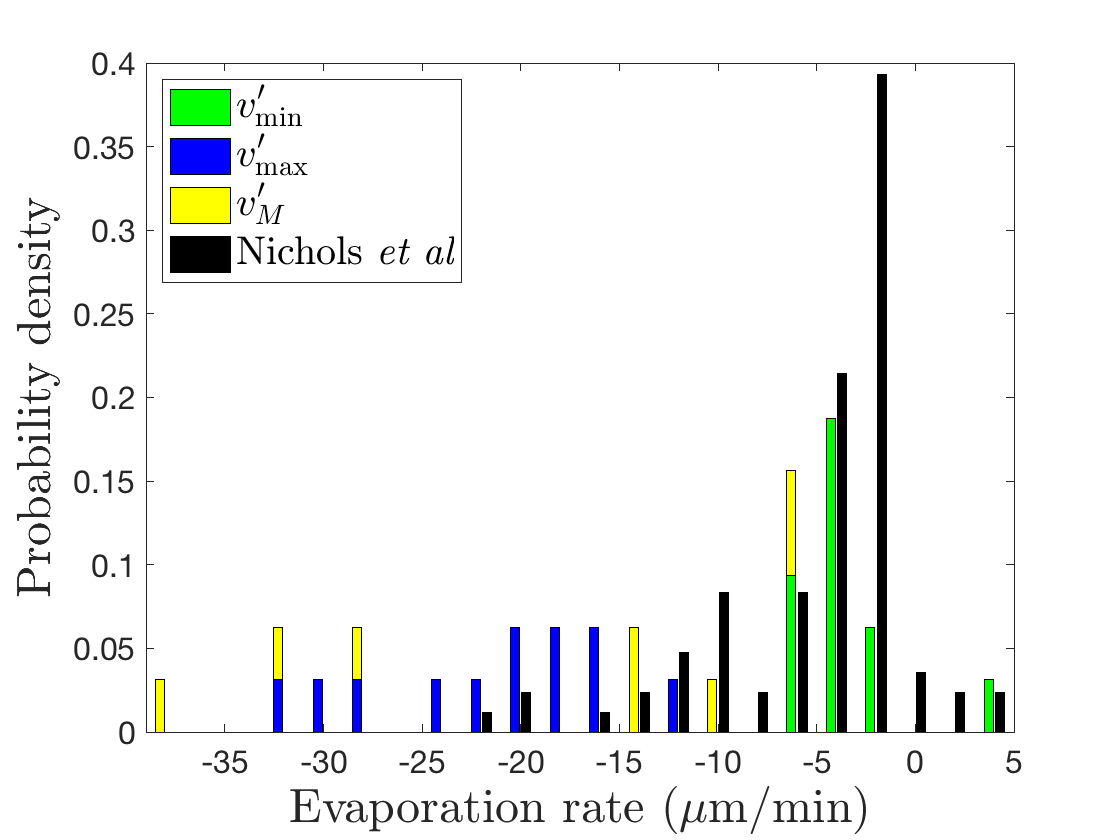}\label{fig:thin_FL_hist_a}} \\
 \subfloat[][Overall thinning rates]{\includegraphics[scale=.2]{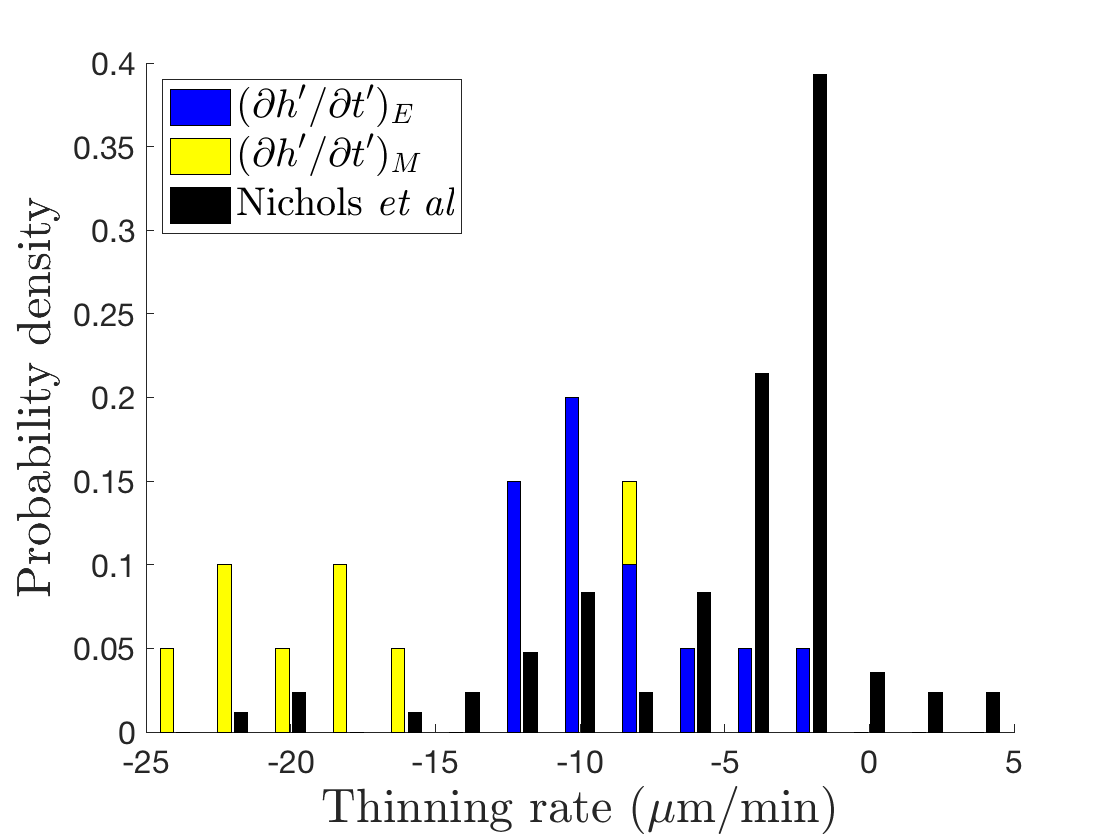}\label{fig:thin_FL_hist_b}}
\caption{\footnotesize{Histograms of rates of change plotted against experimental point measurements from \cite{nichols2005}. The background and peak evaporation rates $v_{\min}'$ and $v_{\max}'$, respectively, are for the evaporation-only model fits shown in \cite{luke2020}, and the single evaporation rate $v_M'$ is for the mixed-mechanism model fits reported in Tables \ref{table:mix_fits} and \ref{table:scalings}.}}
\label{fig:thin_FL_hist}
\end{figure}

Figure \ref{fig:thin_FL_hist_a} compares data from \cite{nichols2005} with our optimal evaporation rates from fitting with the evaporation-only model and the mixed-mechanism model. Figure \ref{fig:thin_FL_hist_b} compares the data with our overall thinning rate $\partial h'/\partial t'$. Both include evaporation values reported in \cite{luke2020}. The values recorded in \cite{nichols2005} compare most closely with our values in Figure \ref{fig:thin_FL_hist_b} because the authors were unable to separate the effects of evaporation and tangential flow. Further, their point measurements did not target breakup, and as such, the distribution has a smaller mean than our optimizations, since we specifially fit regions of significant FL intensity decrease. Our evaporation-only model thinning rates fall within this experimental range, whereas our mixed-mechanism model results have more variation and some values fall just outside their range. As expected, the evaporation-only thinning rates are smaller on average than the mixed-mechanism cases. The overall thinning rates $\partial h'/\partial t'$ are smaller than their corresponding evaporation rates $v'$ for all the evaporation-only cases, as well as the transitional or evaporation-dominated mixed-mechanism cases where the fluid flow is directed inwards for the majority of the trial. This is because the inward flow, characteristic for the evaporation-only model, combats evaporation and retards overall thinning. In contrast, the Marangoni effect-dominated instances have overall thinning rates that are larger than their respective evaporation rates. The strong, outward flow that defines these breakup cases augments evaporation and creates even faster thinning.

\begin{figure} 
\centering
\subfloat[][gradual streak FT-TBU]{
\includegraphics[scale=.153]{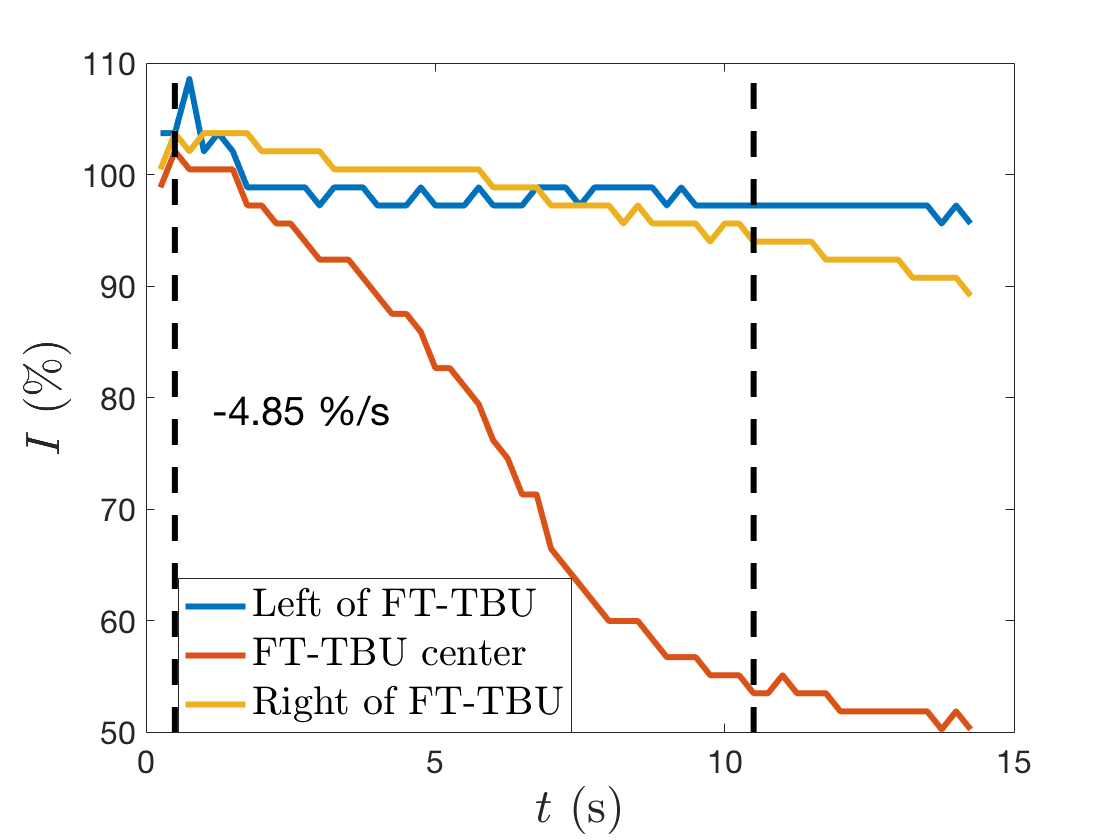}\label{fig:min_max_a}} 
\subfloat[rapid spot FT-TBU]{
\includegraphics[scale=.153]{7_24_20_S10v1t6_1230_dec.png}\label{fig:min_max_b}} 
\caption{\footnotesize{Comparison of FL intensity change over time of FT-TBU from the same subject and visit in different trials, shown with average percentage decrease per second for various time regions during the trial.}}
\label{fig:min_max}
\end{figure}

\begin{figure} 
\centering
\includegraphics[scale=.22]{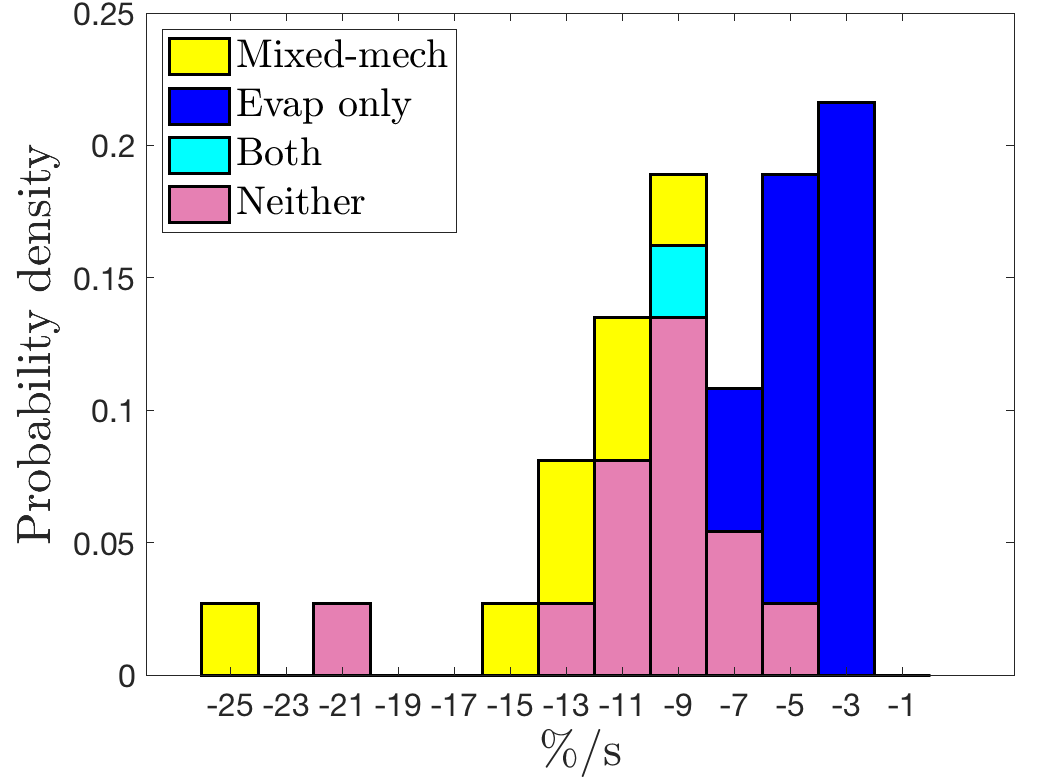}
\caption{\footnotesize{Histogram of percent FL intensity change per second categorized by best model fit.}}
\label{fig:FL_dec}
\end{figure}

In Figures \ref{fig:min_max_a},b, we compare two breakup instances from the same subject and visit. The left instance (S10v1t2 8:00 streak) is fit by the evaporation-driven thinning model (see Table \ref{table:evap}); the right instance is fit with our mixed-mechanism model (S10v1t6 12:30 spot, see Table \ref{table:mix_fits}). The qualitative and quantitative differences in intensity decrease over time from Figure \ref{fig:min_max_a} to Figure \ref{fig:min_max_b} suggest the possibility of different mechanisms driving FT-TBU. We report the percent FL intensity decrease per second for all breakup instances studied (not necessarily reported in this paper or \cite{luke2020}). This is shown in Figure \ref{fig:FL_dec}. The data points are categorized by which model produced a successful fit: mixed-mechanism, evaporation-only, or neither. Neither also includes instances for which an evaporation-only fit was not attempted. Faster instances are fit well with the mixed-mechanism model and slower instances are fit well with the evaporation-only model. We were unable to fit some instances with either model. The approximate delineation by percent FL intensity rate of decrease of which mechanisms are important is further evidence that the time scale on which breakup forms is important (\citealt{awisi19}). 

\begin{figure} 
\centering
\subfloat[][Maximum osmolarity]{\includegraphics[scale=.155]{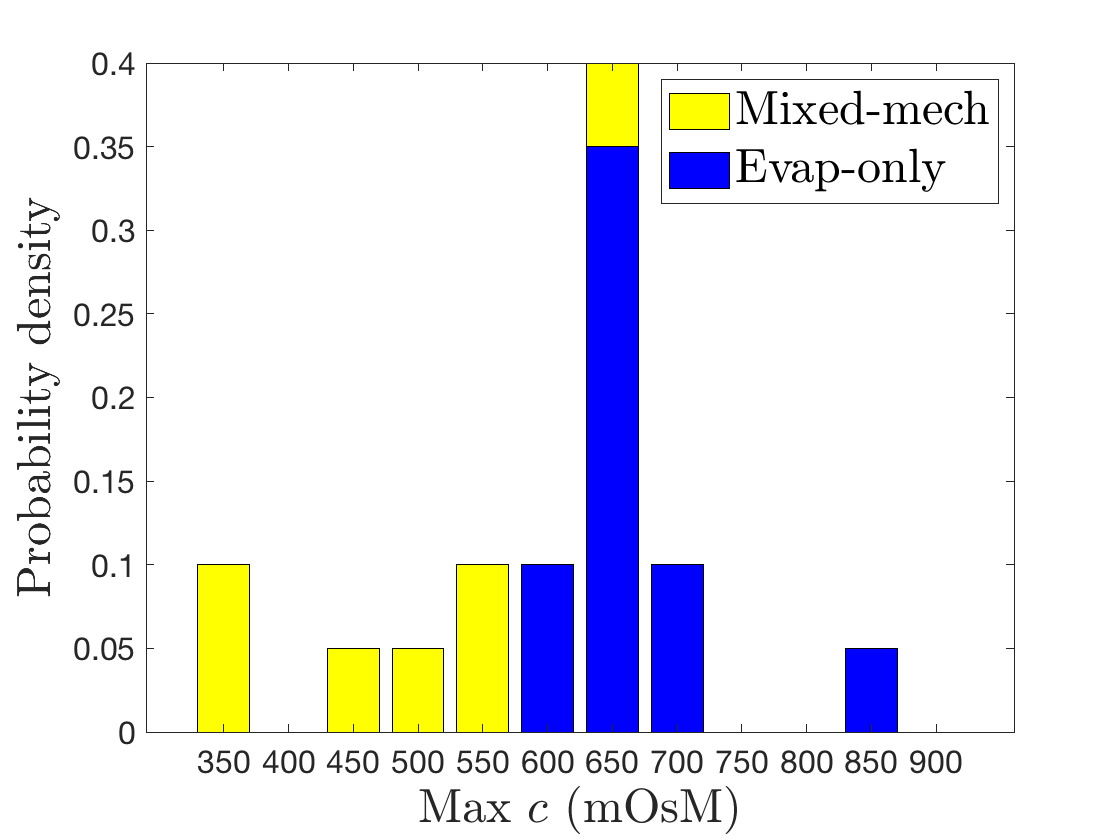}\label{fig:c_hist}}
\subfloat[][Minimum thickness]{\includegraphics[scale=.155]{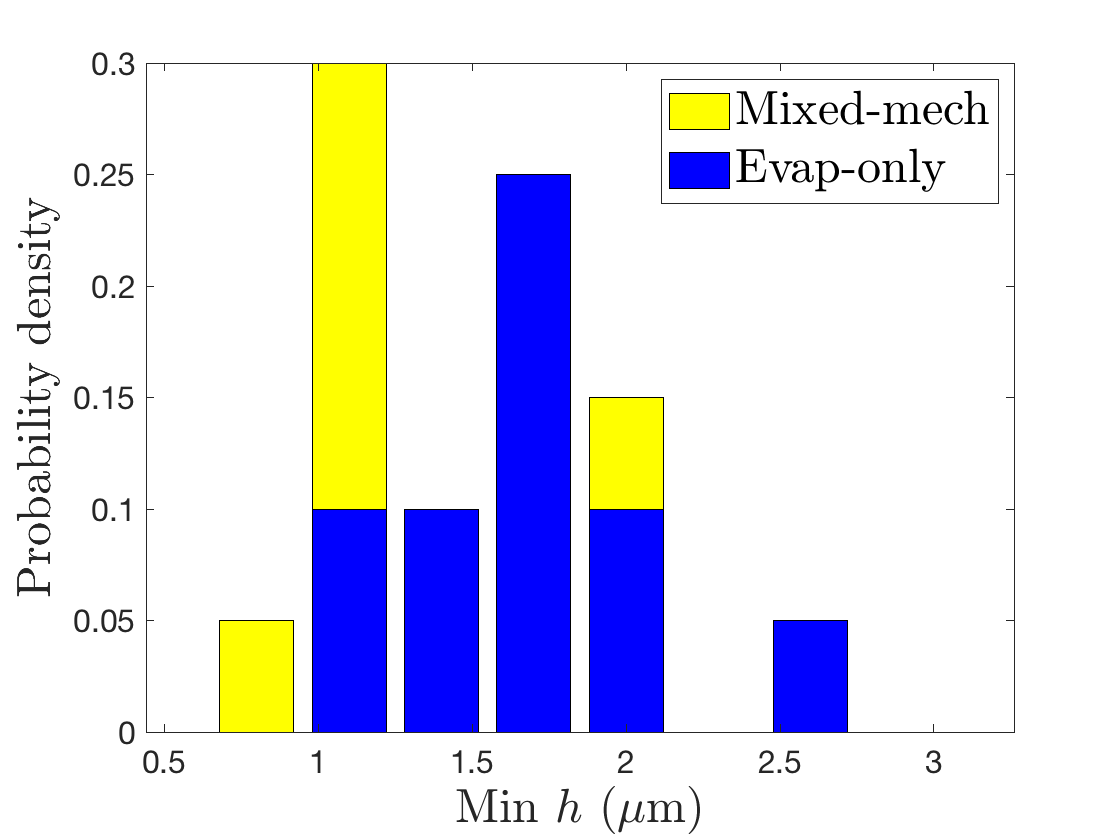} \label{fig:h_hist}} \\
\subfloat[][Max $c$ vs. trial length]{\includegraphics[scale=.155]{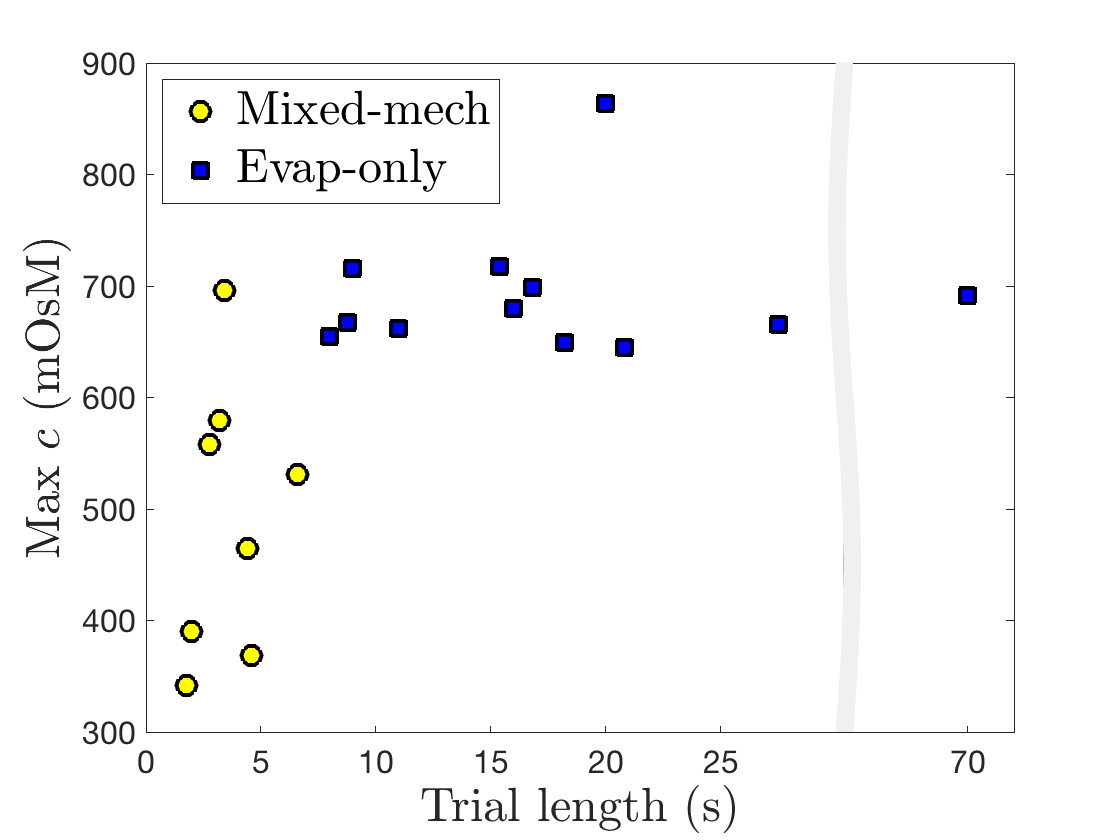}\label{fig:maxc_vs_t}} 
\subfloat[][Max $h$/min $h$ vs. max $c$/min $c$]{\includegraphics[scale=.155]{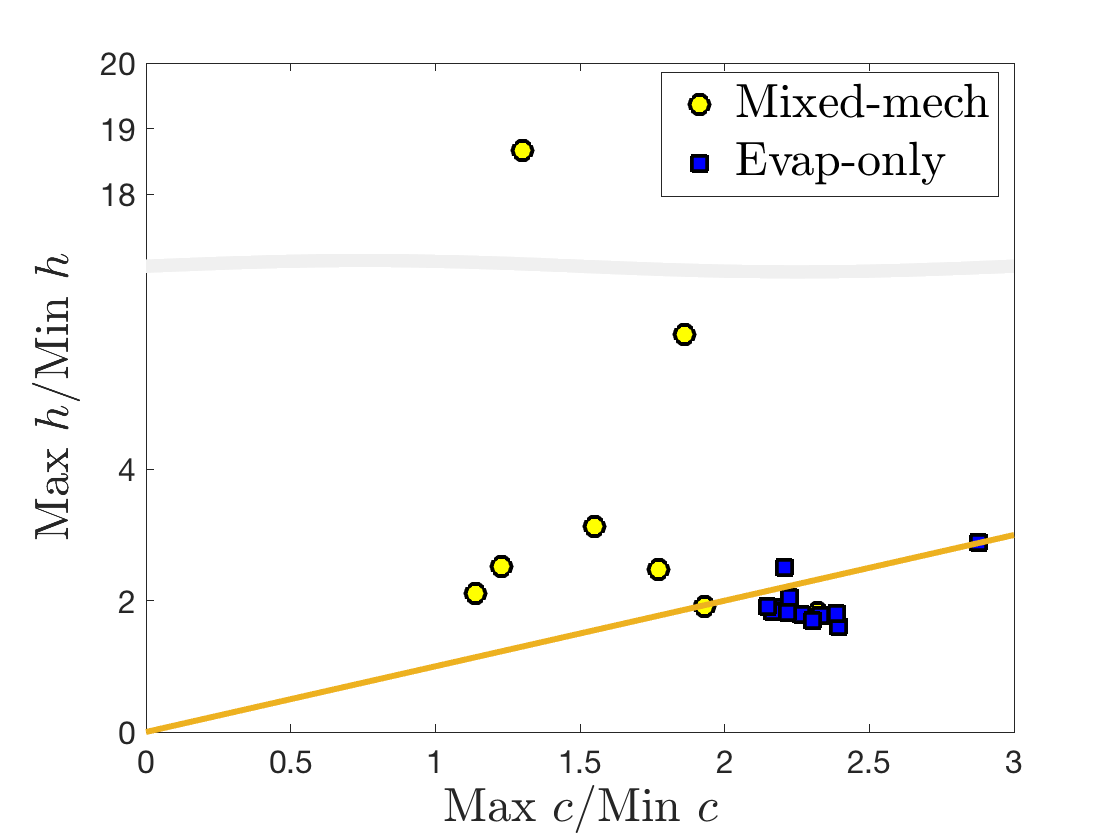}\label{fig:h_c_ratio}}
\caption{\footnotesize{(a), (b): Histograms of maximum osmolarity and minimum thickness (final times of fit). (c): Maximum osmolarity versus trial fitting time. Mixed-mech denotes the mixed-mechanism model fits and evap-only denotes the evaporation model fits in \cite{luke2020}. (d): Relative change in theoretical TF thickness and osmolarity. A line of slope one has been added to show the flat film approximation. Note that in (c) and (d) axes breaks have been used. Maximum thickness and minimum osmolarity are initial conditions.}}
\label{fig:h_and_c}
\end{figure}

Figure \ref{fig:h_and_c} displays various quantities for TF thickness and osmolarity and compares the results from this report with those given in \cite{luke2020}. The maximum osmolarity and minimum thickness of the theoretical solution of each fit is shown in the histogram in Figures \ref{fig:c_hist} and \ref{fig:h_hist}. Referring to Table \ref{table:mix_fits}, there is a direct, apparently linear relationship between the optimal rate of evaporation and the maximum osmolarity. Further, the mixed-mechanism fits display lower maximum osmolarity values on average than the evaporation-only fits. This is largely due to the opposing flow directions of the models. Solutes are advected into the breakup region for the duration of the trial in evaporative cases, increasing osmolarity in that region. In contrast, outward-directed flow carries salt and fluorescein ions away from the dry spot in mixed-mechanism instances, lowering the central concentration. Minimum thickness values are on average lower for the mixed-mechanism cases.

The maximum osmolarity for each fit in Table \ref{table:mix_fits} and the values from \cite{luke2020} are plotted against the time interval of the fit in Figure \ref{fig:maxc_vs_t}. The result gives evidence that osmolarity has a large range in as short a time frame as ten seconds.
In the evaporation-only cases, osmolarity tends to level off around a similar value regardless of trial length. This supports the notion that the TF thickness has reached the height of the glycocalyx at the end of the fit interval. Therefore, the salt concentration cannot increase beyond this point as its movement is tied to the fluid dynamics, which have essentially halted. 
The time resolution of the data inhibits our ability to resolve the rapid dynamics that occur in the shorter trials; this needs to be taken into consideration when drawing conclusions about our results.

The ratio of maximum to minimum theoretical TF thickness has been plotted against the ratio of maximum to minimum theoretical osmolarity in Figure \ref{fig:h_c_ratio}. Mass conservation in the flat film approximation without flow given in \cite{braun2014} satisfies
\begin{equation}
hc = \text{constant},
\end{equation}
and therefore gives a straight line for relative change. Most mixed-mechanism fits recorded in Table \ref{table:mix_fits} fall above this line. The rapid, outward flow that characterizes the mixed-mechanism fits aids thinning and thus the thickness ratio is higher than osmolarity in most cases. The mixed-mechanism outlier above the axis break is evidence that evaporation, which increases osmolarity, did not have time to act within this trial. Most evaporation-only fits from \cite{luke2020} fall just below the straight line from the flat-film approximation. The osmolarity ratio is larger than that of thickness in most evaporation-only cases because inward flow sweeps salt into the breakup region. The inclusion of spatial variation in our PDE models allows salt ions to leave the breakup region by diffusion, whereas the osmolarity in the ODE flat-film model can become large enough to stop thinning by inducing vertical flow from the cornea. Osmosis never overcomes thinning in the PDE model, as seen previously (\citealt{peng2014,braun2015}).

 All of our indicators point to breakup that is not terribly severe for normal subjects individuals that we fit. This is evidenced by minimum thickness estimates that rarely approach zero and maximum osmolarity values below those estimated experimentally or modeled elsewhere (\citealt{liu09, braun2015, peng2014, li2016}). This may be a limitation of our data, both in its imaging modality and time resolution, or it may indicate that our model needs to include other mechanisms not yet considered. \cite{braun2017} showed that tear film models with spatial variation produce smaller peak osmolarity values than the theoretical limit of the flat film result. \cite{king2018} found evidence that evaporation continues after FT-TBUT, causing the appearance of ``hollows'' in the corneal surface (see their Figure 1b). We miss these dynamics since we halt our fitting procedure at FT-TBUT; incorporating this into our model could yield simulations that more closely match severe breakup. In a few of the instances recorded in Tables \ref{table:mix_fits} and \ref{table:scalings} it is possible to fit slightly longer in time; as a result we are fairly confident that we underestimate the maximum osmolarity and overestimate the minimum TF thickness. 

\section{Conclusions and Future Work}

The mixed-mechanism model gives the best fit to the intermediate thinning instances we study, as compared to evaporation-only or zero evaporation models as measured by smaller residuals and realistic parameter values. We take this as strong evidence that both evaporation and the Marangoni effect affect the formation of some cases of FT-TBU that occurs in 1-8 seconds. While the relative importance of each mechanism may vary between instances, leaving out one or the other significantly decreases the quality of the fit and/or the feasibility of the optimal parameters. The Marangoni effect dominates the dynamics early on in the trial as evidenced by the significant outward flow that characterizes each instance we report, and evaporation plays a supporting role that becomes increasingly important as time increases and the Marangoni effect diminishes in magnitude. Capillary flow may overtake the initial outward tangential flow and inward flow may attempt to fill the forming spot or streak.

Our results are differentiated by the optimal parameter values; we categorize various ranges by which mechanism, if any, dominates the thinning seen in the trial. We obtain estimates from our optimizations for parameters that cannot currently be measured directly \textit{in vivo}.
Some optimal parameter values fall within published ranges of experimental point measurements (\citealt{nichols2005}), while others lie above them. This discrepancy is likely due to our ability to target breakup, as the experimental data was taken from the center of the cornea regardless of whether breakup occurs. 

Improvements could be made to this model. The glycocalyx could be modeled as a porous medium instead of using a no slip condition at the ocular surface; this could promote breakup at shorter times for smaller spots (\citealt{nong2010}).   A two layer model for the TF system could also be used (see \citealt{BruBrew14,stapf2017}), and the fit could be conducted over two spatial dimensions.

A local ODE model that approximates the fluid flow dynamics near the center of FT-TBU has been created; fitting, analysis, and comparison to the PDE results are underway. We aim to use this simplified version of the model to elucidate mechanistic information from the data and automate the process of identifying a wider range of breakup regions and fitting them with a model in order to estimate relevant TF quantities in FT-TBU. This approach would give a representative statistical view of the dataset, rather than the complex and detailed information from a handful of instances of specific shapes.

\section{Funding}

This work was supported by National Science Foundation grant DMS 1909846 and National Institutes of Health grant NEI R01EY021794. The content is solely the responsibility of the authors and does not necessarily represent the official views of the funding sources.

\appendix

\section{Appendix}
\label{sec:app}

\setcounter{equation}{0}

\subsection{Governing Dimensional Equations}
\label{sec:app_gov}

 For the circular case we use the dimensional axisymmetric coordinates $(r',z')$ to denote the position and $\bm{u}' = (u',w')$ to denote the fluid velocity. The TF is modeled as an incompressible Newtonian fluid on $0 < r' < R_0$ and $0 < z' < h'(r',t')$, where $h'(r',t')$ denotes the thickness of the film. Conservation of mass and momentum of the TF fluid and transport of solutes within the fluid are given, respectively, by
\begin{equation}
\nabla' \cdot \bm{u}' = 0,
\end{equation}
\begin{equation}
\rho (\bm{u}'_{t'} + (\bm{u}' \cdot \nabla') \bm{u}') = -\nabla' p' + \mu \nabla'^2 \bm{u}',
\end{equation}
\begin{equation}
\p_{t'} S' + \nabla' \cdot( \bm{u}' S') = D_S \nabla'^2 S',
\end{equation}
where $p'$ is the fluid pressure and $S'$ represents either $c'$, the osmolarity, or $f'$, the FL concentration, with diffusivities $D_o$ and $D_f$, respectively. The fluid density is $\rho$ and the dynamic viscosity is $\mu$.

 At the film/cornea interface $z' = 0$, we require no slip and osmosis across a perfect semipermeable membrane:
\begin{equation}
u' = 0, \quad w' = P_o V_w (c'-c_0).
\end{equation}
The membrane permeability is given by $P_o$, the molar volume of water is $V_w$, and $c_0$ is the isotonic osmolarity.

 We enforce no flux of solutes across the film/cornea interface:
\begin{equation}
D_o \partial_{z'} c' - w'c' = 0, \quad D_f \partial_{z'} f' - w' f' = 0.
\end{equation}
At $z' = 0'$, the outward normal is given by $\bm{n} = (0,-1)$, and at $z = h'(r',t')$,
\begin{equation}
\bm{n} = \ds \frac{(-\nabla'_{II} h',1)}{(1 + |\nabla'_{II}h'|^2)^{1/2}},
\end{equation}
where $\nabla'_{II}$ is the gradient in the plane of the substrate parallel to $z' = 0$.

The kinematic condition implies that the balance of the material derivative of the TF thickness and the fluid velocity in the $z'-$ direction is controlled by the evaporative mass flux, $J'$:
\begin{equation}
\rho \left[ \frac{ \p_{t'} h' + u' \nabla'_{II} h' - w'}{(1 + |\nabla'_{II} h'|^2)^{1/2}} \right] = -J',
\end{equation}
where
\begin{equation}
J' = \rho v'
\end{equation}
gives the rate of change of mass per unit area for uniform evaporation, which is Case (b) from Section \ref{sec:evap_dist}. Here, $v'$ is the uniform thinning rate. 

 The normal stress condition at $z' = h'(r',t')$ is given by
\begin{equation}
-p'_v - \bm{n}' \cdot \bm{T}' \cdot \bm{n}'= - \l \sigma_0 \nabla'_s \cdot \bm{n}' + \frac{A^*}{h'^3}\r,
\end{equation}
where $p'_v$ is atmospheric pressure, the Newtonian stress tensor is $T' = -p' \bm{I} + \mu (\nabla' \bm{u}' + \nabla' \bm{u}'^T)$, $\sigma_0$ is the surface tension, $\nabla'_s = (I - \bm{n}' \bm{n}') \cdot \nabla$ is the surface Laplacian (\citealt{stone1990}), and $A^*$ is the Hamaker constant.

No flux of solutes across the free surface is enforced by

\begin{equation}
D_S \bm{n}' \cdot \nabla S' - \bm{n}' \cdot (\bm{u}' - \bm{u}_I') S' = 0,
\end{equation}
where $D_S$ is again either $D_O$ or $D_f$ and $S'$ is either $c$ or $f$, and $\bm{u}_I'$ is the interface velocity.

We model the lipid on the TF surface $z' = h'(r', t')$ as a tangentially immobile aqueous/glob interface with higher concentration of lipid on $0 < r' < R_I'$ and with a fixed size and concentration. Outside the glob on $R_I' < r' < R_0'$, the aqueous/air interface is mobile. Conservation of mass for the insoluble polar lipid is given by the following transport equation:

\begin{equation}
\Gamma' = \Gamma_0 \text{ on } [0, R_I'], \quad \text{ and } \quad \partial_{t'} \Gamma' + \nabla_s' \cdot (\Gamma' \bm{u}') = D_s \nabla_s'^2 \Gamma' \text{ on } [R_I', R_0'].
\end{equation}
Surfactant mass is conserved in the separate regions inside and outside the glob, but the glob acts as a source of surfactant for the mobile region when these regions are considered together.

The tangential stress boundary condition varies based on spatial location: under or outside the glob. The no-slip boundary condition replaces the tangential stress boundary condition at $z' = h'$ on $(0, R_I')$:

\begin{equation}
u' = 0 \text{ in } (0, R_I').
\end{equation} 
On $(R_I', R_0')$ at $z' = h'$, the tangential stress balance is given by:

\begin{equation}
\bm{t}' \cdot \bm{T}' \cdot \bm{n}' = \partial_s \sigma = \bm{t}' \cdot \nabla' [\sigma_0 - (\partial_{\Gamma} \sigma)_0 (\Gamma' - \Gamma_0)]
 \text{ in } (R_I', R_0').
\end{equation}
Here, the unit tangent vector to the interface is $\bm{t}' = \ds \frac{(1, \nabla'_{II} h')}{(1 + |\nabla'_{II}h'|^2)^{1/2}}.$

 \subsection{Derivation of TF Equations, Spot Case}
\label{s:spot_eqns}

Using the scalings (\ref{scale1}), (\ref{scale2}), we nondimensionalize the governing equations as in \cite{zhong2019}. At leading order, conservation of mass and momentum of the fluid on $0 < z < h(r,t)$ and transport of solutes are given by
\begin{equation}
\frac{1}{r} \p_r(ru) + \p_z w = 0,
\label{nd_mass}
\end{equation}
\begin{equation}
\p_z^2 u = \p_r p, \quad \p_z p = 0,
\label{nd_mom}
\end{equation}
\begin{equation}
\partial_t S + u \partial_r S + w \partial_z S = \text{Pe}_S^{-1} \left( \frac{1}{r} \partial_r(r \partial_r S) + \epsilon^{-2} \partial_z^2 S \right),
\label{solute}
\end{equation}
where now $S = c$ or $f$ nondimensionally. We keep all powers of $\epsilon$  before assuming an expansion in this small parameter. The P\'{e}clet number, Pe$_S = \ds \frac{U \ell}{ D_S}$, is Pe$_c$ or Pe$_f$ for osmolarity and fluorescein, respectively.
The leading order boundary conditions at $z = 0$ are
\begin{equation}
u = 0, \quad w = P_c(c-1).
\label{bc_0}
\end{equation}
We enforce zero flux of solutes by
\begin{equation}
\epsilon^{-2} \text{Pe}_S^{-1} \partial_z S = w S.
\end{equation}
The leading order boundary conditions at $z = h(r,t)$ are
\begin{equation}
\p_t h + u \p_r h - w = -J,
\label{kin}
\end{equation}
\begin{equation}
p = - \frac{1}{r} \p_r (r \p_r h) - \frac{A}{h^3},
\end{equation}
\begin{equation}
\partial_z S = \epsilon^2 (\text{Pe}_S JS + \partial_r h \partial_r S).
\end{equation}
For the lipid, under the glob on $[0, R_I]$ at $z = h$, we have no slip and a fixed concentration:

\begin{equation}
u = 0, \quad \text{and} \quad \Gamma = 1.
\end{equation}
Outside the glob on $[R_I, R_L]$ at $z = h$, surfactant transport and the tangential stress boundary condition are given, respectively, by

\begin{equation}
\partial_t \Gamma + \frac{1}{r} \partial_r (r u_r \Gamma) = \text{Pe}_s^{-1} \frac{1}{r}\partial_r(r \partial_r \Gamma) \quad \text{and}  \quad \partial_z u = -M \partial_r \Gamma.
\end{equation}
Here, $u_r$ denotes the surface velocity at leading order and $M$ is the Marangoni number.

Equation \ref{nd_mom} implies that $p = p(r,t)$, and thus

\begin{equation}
u = \frac{\partial_r p}{2} z^2 + C(r,t)z + D(r,t).
\end{equation}
We use the transition function $B(r)$ described in the text to write the boundary equation for $u$ at $z =h$ in the following way, and give the boundary condition at $z = 0$:

\begin{equation}
u = 0 \quad \text{at} \quad z = 0, \quad u(1-B(r)) + (\partial_z u + M \partial_r \Gamma) B(r) = 0 \quad \text{at} \quad z = h.
\end{equation}
This implies that

\begin{equation}
u = \frac{\partial_r p}{2}z^2 + C(r,t) z, \quad C(r,t) = - \frac{\frac{\partial_r p}{2}h [h(1-B) + 2B]+ MB \partial_r \Gamma}{h(1-B) + B}.
\end{equation}
 Integrating (\ref{nd_mass}) over the vertical domain, applying the Leibniz rule, and using (\ref{kin}) and (\ref{bc_0}) to substitute in for the first three resulting terms gives
\begin{equation}
\p_t h  + J - P_c(c(r, 0, t)-1) + \frac{1}{r } \p_r (r h \bar{u}) = 0,
\end{equation}
where
\begin{equation}
\ds \bar{u}(r,t) = \frac{1}{h} \int_0^h u(r,z,t) dz
\end{equation}
is the depth-averaged fluid velocity.

For solutes, we continue the derivation for the osmolarity $c$ following \cite{JenGrot93}.

The solute boundary condition at $z = 0$ is
\begin{equation}
\epsilon^{-2} \text{Pe}_c^{-1} \p_z c = wc,
\end{equation}
and the boundary condition at $z = h(r,t)$ is
\begin{equation}
   \p_z c = \text{Pe}_c\epsilon ( u \p_r h - w) c + \epsilon^2 \p_r h \p_r c .
\end{equation}
Assume that $c(r,z,t)$ can be expanded as:
\begin{equation}
c = c_0(r,z,t) + \epsilon^2 c_1(r,z,t) + O(\epsilon^4).
\end{equation}
After substituting this expression for $c$ into (\ref{solute}), the leading order equation is given by
\begin{equation}
\p_z^2 c_0 = 0,
\end{equation}
and thus $c_0 = c_0(r,t)$. The next order in $\epsilon$ results in
\begin{equation}
\p_z^2 c_1 = \text{Pe}_c (\p_t c_0 + u \p_r c_0) - \frac{1}{r} \p_r (r \p_r c_0).
\label{next}
\end{equation}
Integrating (\ref{next}) over the vertical domain gives
\begin{equation}
\p_z c_1(r,h,t) - \p_z c_1(r,0,t) = \text{Pe}_c (h \p_t c_0 + h \bar{u} \p_r c_0) - \frac{1}{r} h \p_r (r \p_r c_0).
\end{equation}
The terms involving $c_1$ can be eliminated by identifying the boundary conditions at $O(\epsilon^2)$; these result in an equation for $c_0$. We drop the subscript to give our leading order PDE for osmolarity:
\begin{equation}
h(\p_t c  + \bar{u} \p_r c) = \frac{1}{r}\text{Pe}_c^{-1} \p_r (r \p_r c) + Jc - P_c(c-1)c.
\end{equation}
The evolution equation for $f$ may be obtained similarly:
\begin{equation}
h(\p_t f  + \bar{u} \p_r f) = \frac{1}{r}\text{Pe}_f^{-1} \p_r (r \p_r f) + Jf - P_c(c-1)f.
\end{equation}
The collected equations, initial and boundary conditions are given in the text.

\subsection{TF Equations, Streak Case}
\label{s:streak_eqns}

The derivation of the problem in the linear case for streaks is similar to the axisymmetric case, and more details may be found in \cite{zhong2019}. The non-dimensionalization is the same in both cases.

The problem is solved on $0 < x < X_0$ and $0  < z < h(x,t)$, where $h$ is the TF thickness. Homogeneous Neumann boundary conditions are applied at $x = 0$ and $x = X_0$. The initial conditions are the same as for the spot case with $r$ and $R_0$ replaced by $x$ and $X_0$. The fluid velocity coordinates in the $(x,z)$ directions are given by $(u,w)$. Nondimensionally, the system is given by

\begin{equation}
\p_t h + J  - P_c(c-1) + \p_x(h \bar{u}) = 0,
\label{eq:dhdt_x}
\end{equation}
\begin{equation}
p = - \p_x^2 h - A h^{-3},
\end{equation}
\begin{equation}
\partial_t \Gamma = B(x)[\text{Pe}_s^{-1} \partial_x^2 \Gamma - \partial_x(u_s \Gamma)],
\end{equation}
\begin{equation}
h(\p_t c + \bar{u} \p_x c) = \text{Pe}_c^{-1} \p_x^2 c + \partial_x h \partial_x c - P_c(c-1) c + Jc,
\end{equation}
\begin{equation}
h(\p_t f + \bar{u} \p_x f) = \text{Pe}_f^{-1} \p_x^2 f + \partial_x h \partial_x f - P_c(c-1) f + Jf.
\label{eq:dfdt_x}
\end{equation}
Here, 

\begin{equation}
\bar{u} =  - \frac{\frac{1}{3} \partial_x p h^2 [ \frac{1}{4} (1 - B)h + B] - \frac{1}{2} M \partial_x \Gamma B h}{(1-B)h + B},
\end{equation}
\begin{equation}
u_s = - \frac{B[\frac{1}{2}h^2 \partial_x p + M h \partial_x \Gamma]}{(1-B)h + B}.
\end{equation}
The nondimensional choices for the evaporation profile are the same as for the spot case given in Section \ref{sec:evap_dist}.

\subsection{FL Intensity}
\label{sec:app_intensity}

The following equation gives the FL intensity $I$ (\citealt{webber86}; \citealt{nichols2012}):

\begin{equation}
I' = I_0 \frac{1 - \exp(-\epsilon_f h' f')}{1 + (f'/f_{\text{cr}})^2}
\label{eq:I_dim}
\end{equation}
Here, $h'$ is the TF thickness, $f'$ is the FL concentration, $\epsilon_f$ is the Napierian extinction coefficient, and $I_0$ is a normalization factor calculated using model eye measurements. Once we solve the system of nondimensional equations given either in (\ref{eq:dhdt})-(\ref{eq:dfdt}) for spots or (\ref{eq:dhdt_x})-(\ref{eq:dfdt_x}) for streaks, we then compute the nondimensional FL intensity $I$ given in Equation \ref{eq:I_nd}.

Our \emph{in vivo} observations (\citealt{awisigyau2020}) typically operate in a regime near or slightly above the peak of the $I$ vs. $f'$ curve. Asymptotic expansions for fixed $h'$ show that $I$ decreases quadratically with increasing $f'$ in the self-quenching regime (\citealt{braun2014}).

\subsection{FL Concentration}
\label{sec:app_conc}

FL concentration is typically reported as a percentage in the ocular literature. For a particular FL concentration $f$ given as a percentage, this quantity is converted to molar as $f_M$ by

\begin{equation}
f_M = \frac{\rho}{M_w}\frac{f}{100},
\end{equation}
where $\rho$ is the density of water (Table \ref{table:dim}) and $M_w$ is the molecular weight of sodium fluorescein (approximately 376 g/mol). Critical FL concentration $f_{\text{cr}}$, 0.2\%, makes an 0.0053 M solution when dissolved in water. This conversion of $f_{\text{cr}}$ to molar is necessary to compute the dimensionless Napierian extinction coefficient $\phi$ (Table \ref{table:nondim}).

 \section*{Conflict of Interest}

 The authors declare that they have no conflict of interest.
 
  \clearpage

 \bibliographystyle{spbasic}
\bibliography{Evap_paper_bib}

\clearpage
 
 \section{Supplementary Material}
 
 \setcounter{equation}{0}
 \setcounter{figure}{0}
 \setcounter{table}{0}

\subsection{Mixed-mechanism modifications}
\label{sec:mod}

\begin{figure}[H]
\centering
\subfloat[][FL intensity with minima aligned]{\includegraphics[scale=.12]{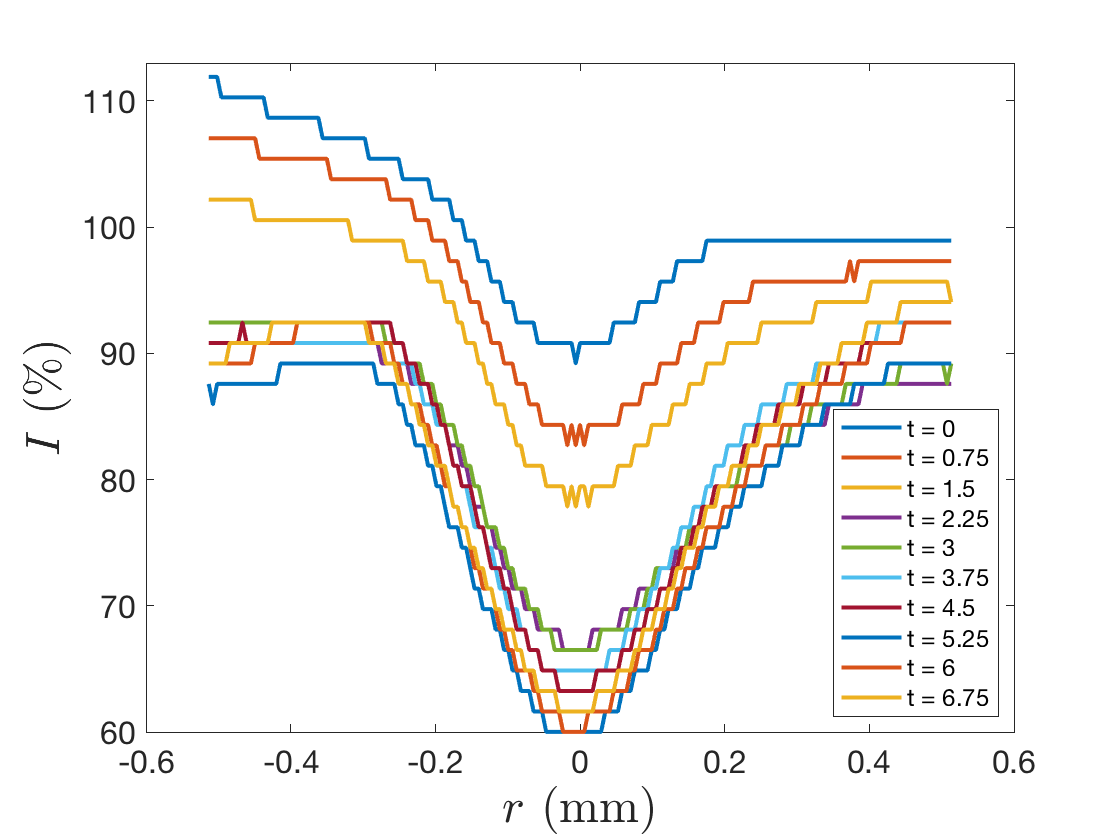}}
\subfloat[][FL intensity decrease]{\includegraphics[scale=.12]{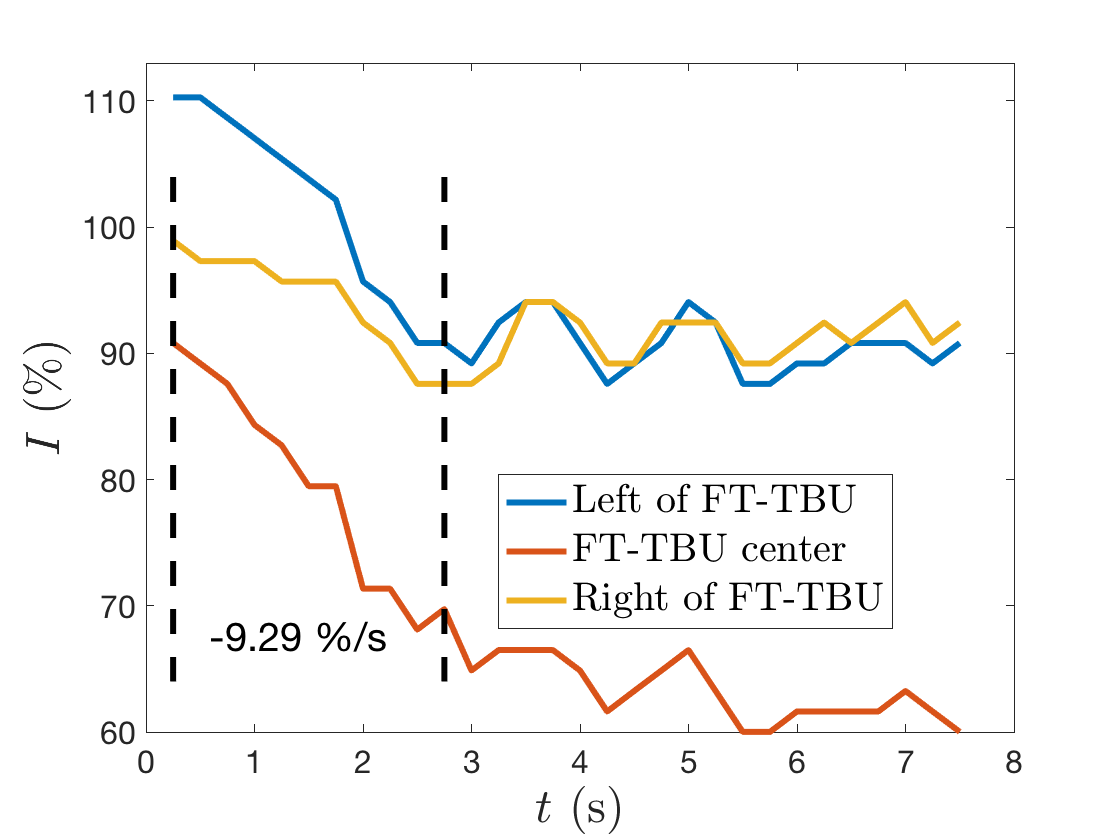}}
\subfloat[][FT-TBU data extraction]{\includegraphics[scale=.04]{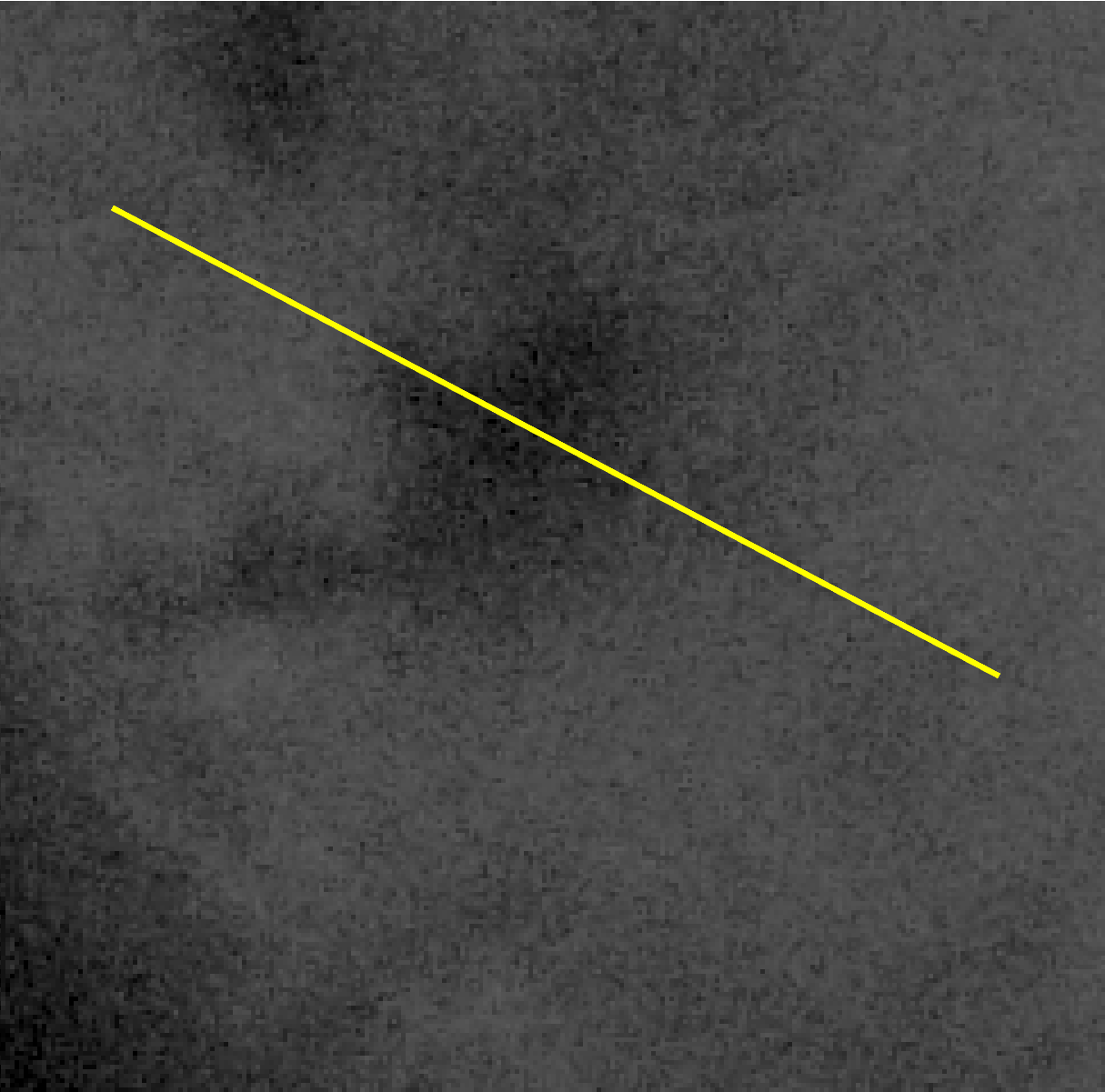}}
\caption{\footnotesize{Extracted data for the S18v2t4 7:30 spot. In (c) the image has been brightened and contrast-enhanced.}}
\label{fig:S18v2t4_730_data}
\end{figure}

\begin{figure}[H]
\centering
\subfloat[][Exp. (---) and best fit th. (- - -) FL intensity]{\includegraphics[scale=.15]{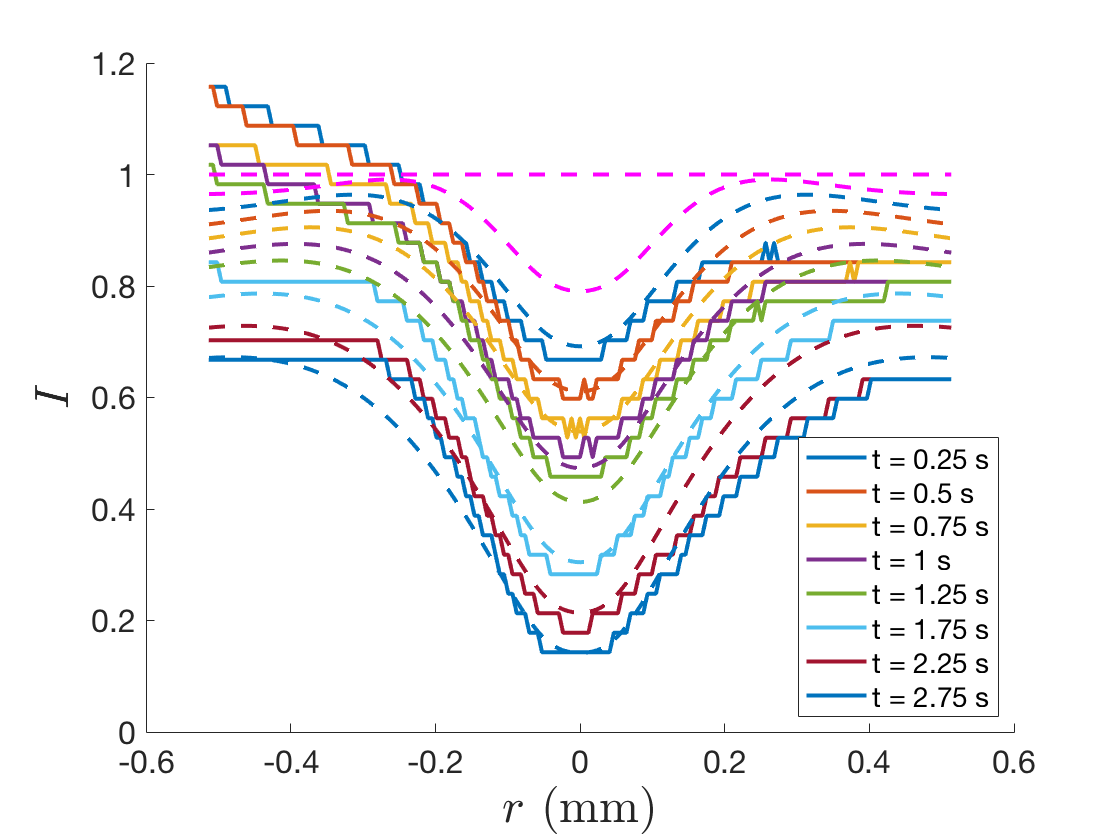}}
\subfloat[][Theoretical TF thickness]{\includegraphics[scale=.15]{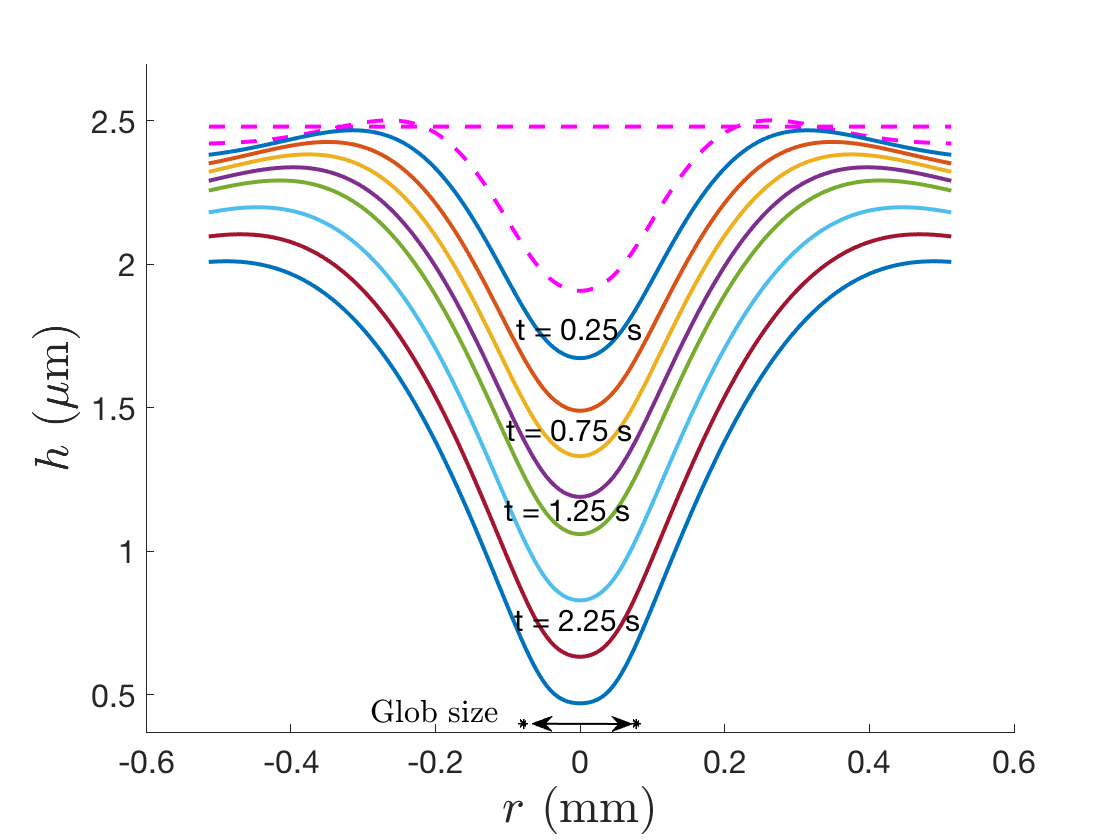}} \\
\subfloat[][Theoretical osmolarity]{\includegraphics[scale=.15]{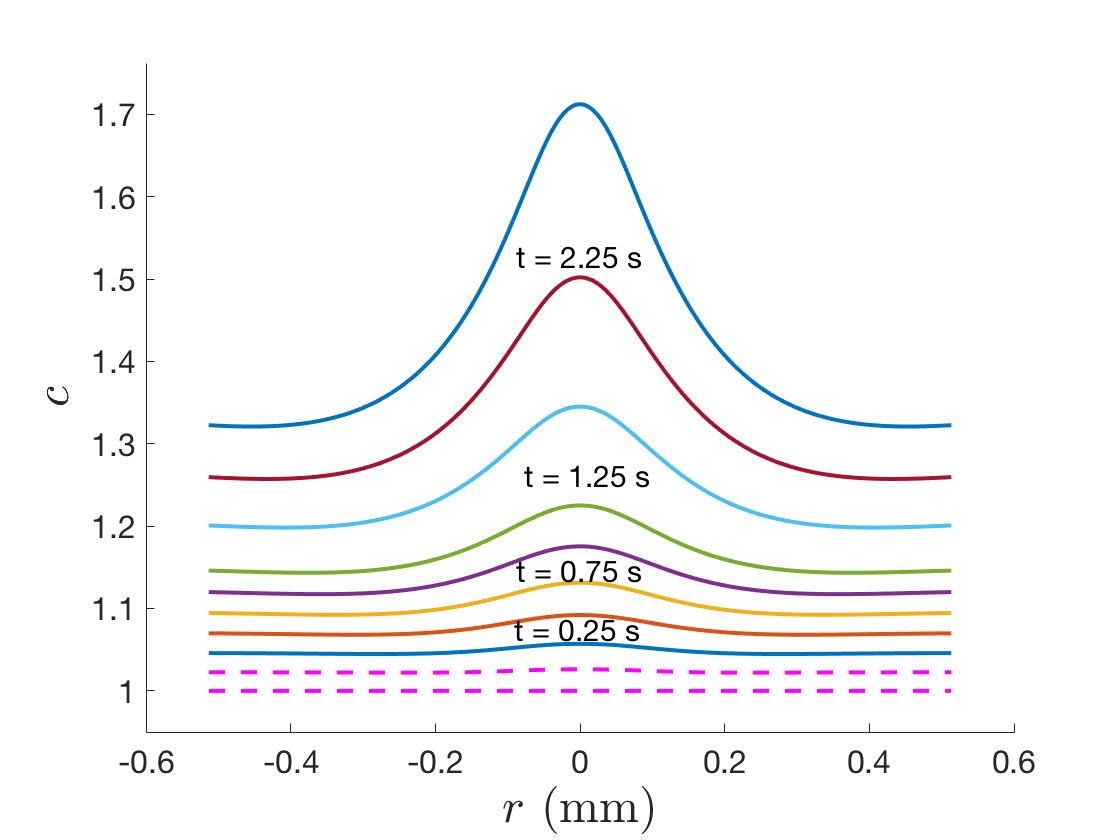}}
\subfloat[][Theoretical depth-averaged fluid velocity]{\includegraphics[scale=.15]{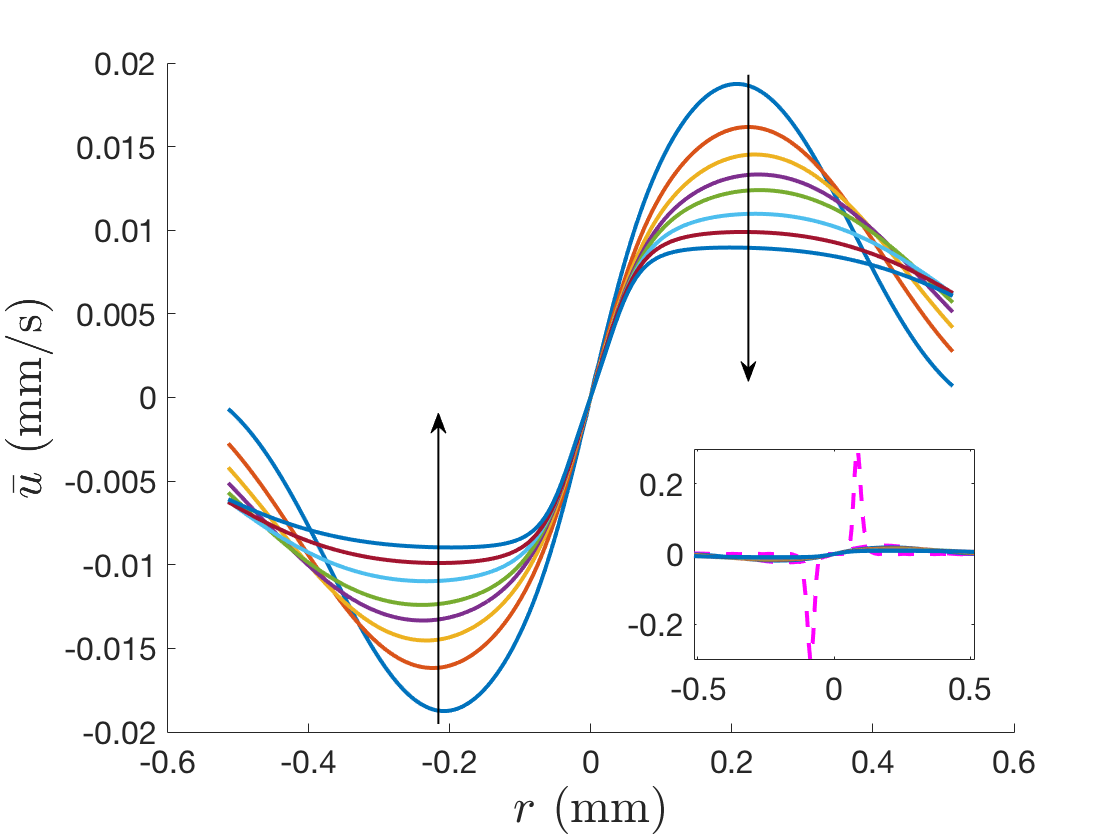}}
\caption{\footnotesize{S18v2t4 7:30 spot best fit results (Case (b) evaporation). FL intensity has been normalized. Theoretical osmolarity is given as a fraction of the isotonic value. Arrows indicate increasing time.}}
\label{fig:S18v2t4_730_fit_2}
\end{figure}

\begin{figure}[H]
\centering
\subfloat[][Exp. (\textbf{---}) and best fit th. (- - -) FL intensity]{\includegraphics[scale=.15]{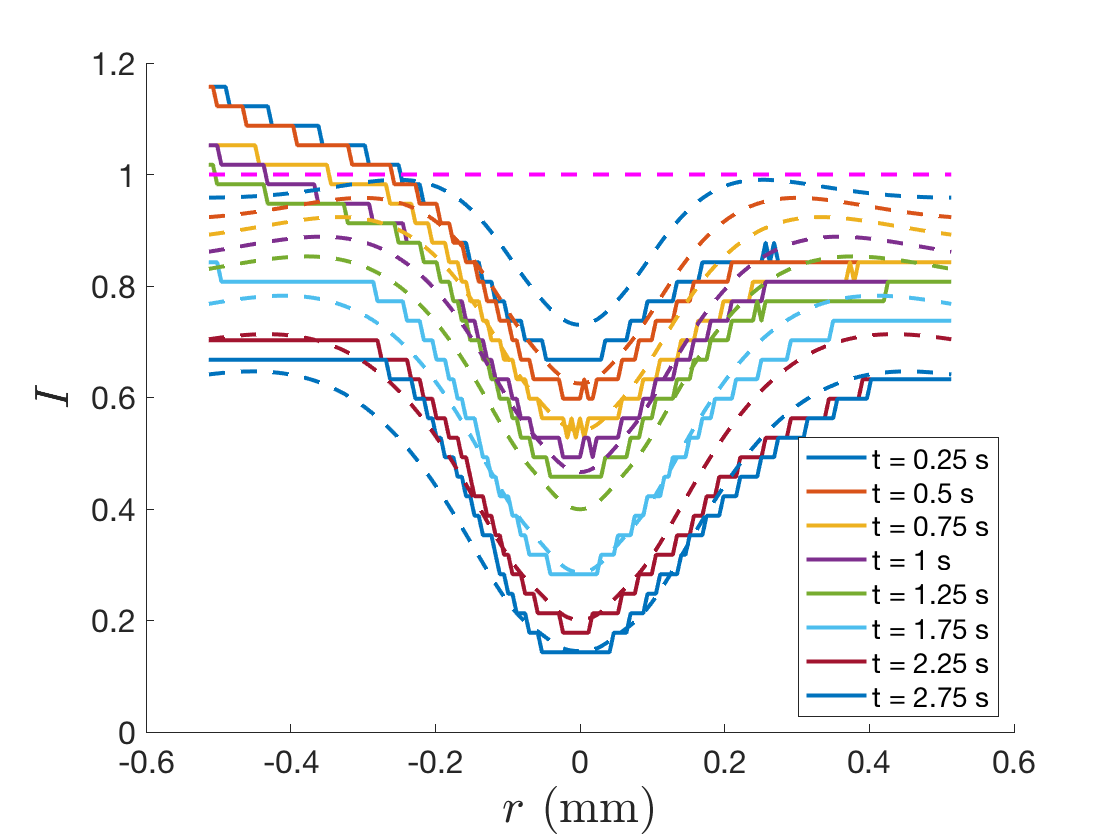}}
\subfloat[][Theoretical TF thickness]{\includegraphics[scale=.15]{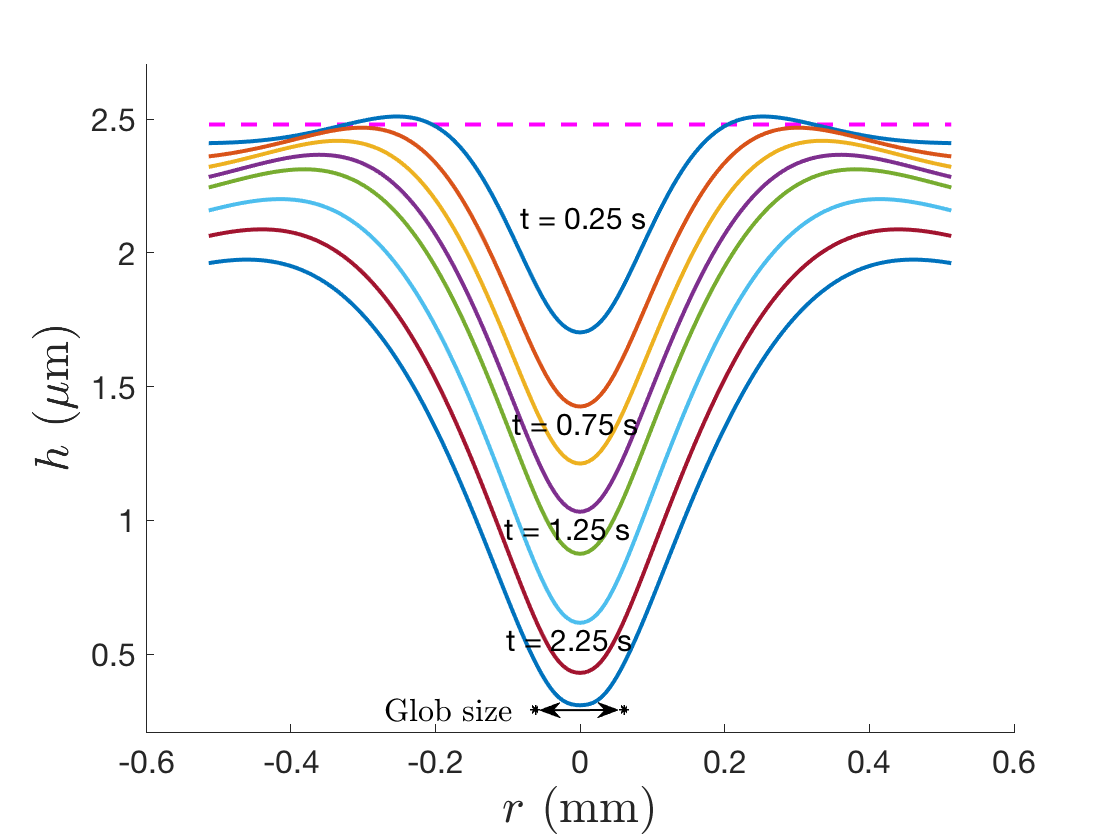}} \\
\subfloat[][Theoretical osmolarity]{\includegraphics[scale=.15]{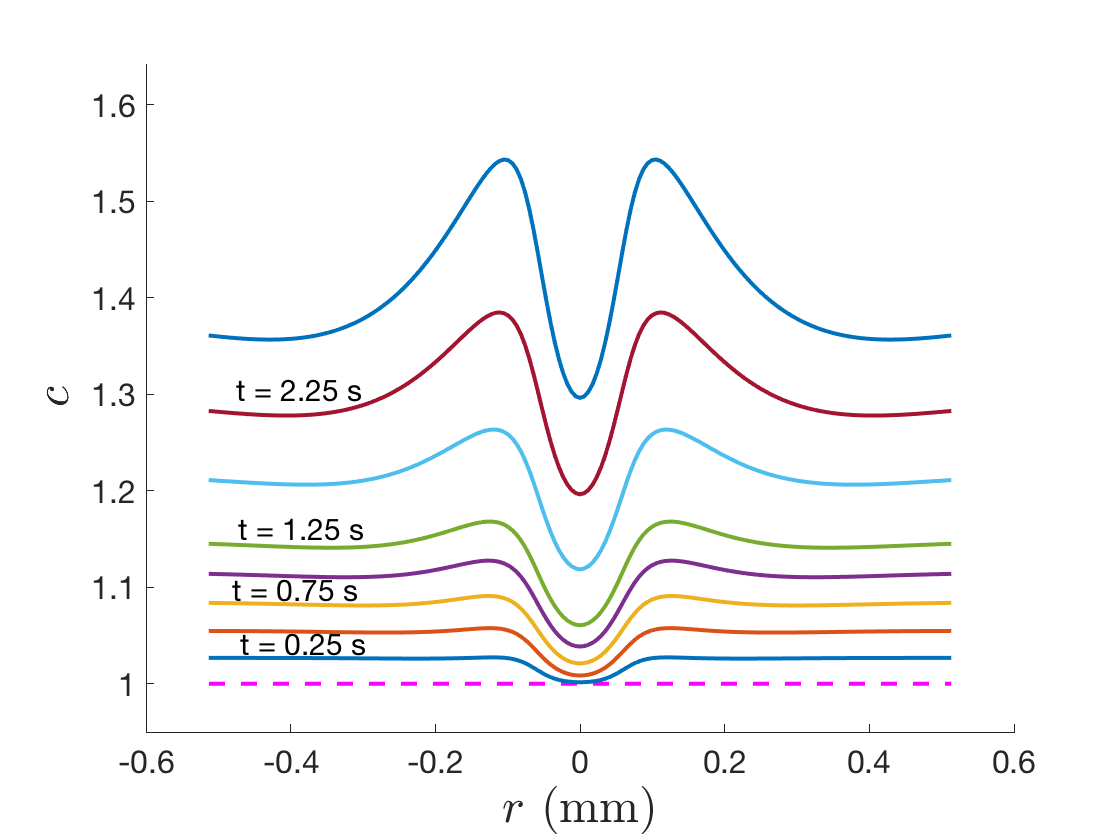}} 
\subfloat[][Theoretical fluorescein concentration]{\includegraphics[scale=.15]{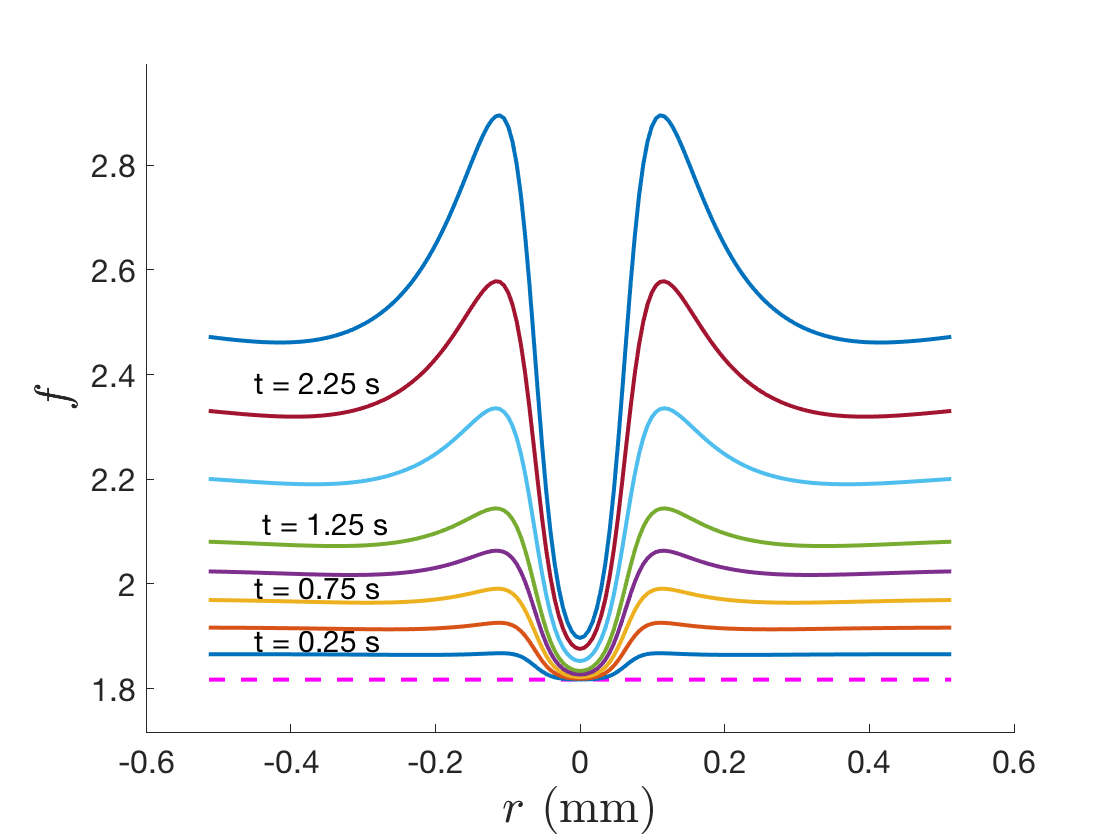}} 
\caption{\footnotesize{S18v2t4 7:30 spot best fit results (Case (d) evaporation). FL intensity has been normalized. Theoretical osmolarity is given as a fraction of the isotonic value.}}
\label{fig:S18v2t4_730_fit_d}
\end{figure}

We explore two modifications to our mixed-mechanism fitting scheme, using the S18v2t4 7:30 spot as an example (see Figure \ref{fig:S18v2t4_730_data}). First, we fit the S18v2t4 7:30 spot with two ghost time levels in Figure \ref{fig:S18v2t4_730_fit_2}. Second, we fit the S18v2t4 7:30 spot with Case (d) evaporation. The fit is shown in Figure \ref{fig:S18v2t4_730_fit_d}. The results for both fits are summarized in Table \ref{table:results_sup}. The results from the paper are listed in the first row for comparison.

Fitting the S18v2t4 7:30 spot with two ghost time levels produces a 2\% smaller residual than using a single time level. The optimal evaporation rate decreases by 13\%, the glob width decreases by 3.7\%, and the change in surface tension decreases by 6.6\%, resulting in a Marangoni number that is 6.5\% smaller. Since the addition of a second ghost time level did not significantly improve the fit, we report the results of using a single ghost time level in the paper. This is in contrast to the S10v1t6 12:30 spot, whose fit benefits greatly from using two ghost time levels.

Fitting the S18v2t4 7:30 spot with Case (d) evaporation gave a similar looking fit to that in the paper. The optimal evaporation rate is a mere 2.7\% larger than using Case (b) evaporation, the glob width is 23\% smaller, and the change in surface tension is 42.5\% larger, resulting in a Marangoni number that is also 42.5\% larger. The residual is 1\% smaller. This suggests that evaporation plays a relatively small role in causing the thinning in this instance as compared to the Marangoni effect.

\begin{table}[H]
\centering
\tabcolsep=0.16cm
\begin{tabular}{|c|c|c|c|c|c|c|c|c|c|c|}
\hline
\textbf{\begin{tabular}[c]{@{}c@{}}Modifi-\\ cation\end{tabular}}  & \begin{tabular}[c]{@{}c@{}}$\bm{v'}$\\ \textbf{($\bm{\frac{\mu\text{m}}{\text{min}}}$)} \end{tabular} & \begin{tabular}[c]{@{}c@{}}$\bm{R_I'}$ \\ \textbf{(mm)} \end{tabular} & \begin{tabular}[c]{@{}c@{}}$\bm{(\Delta \sigma)_0}$\\ \textbf{($\bm{\frac{\mu\text{N}}{\text{m}}}$)} \end{tabular} & \textbf{\begin{tabular}[c]{@{}c@{}}Min\\ $\bm{I_{th}}$\end{tabular}} & \textbf{\begin{tabular}[c]{@{}c@{}}Min\\ $\bm{h_{th}'}$ \\ ($\bm{\mu}$m)\end{tabular}} & \textbf{\begin{tabular}[c]{@{}c@{}}Max\\ $\bm{c_{th}}$ \end{tabular}} & \textbf{\begin{tabular}[c]{@{}c@{}}Time \\ scale\\ $\bm{t_s}$ (s)\end{tabular}} & \textbf{\begin{tabular}[c]{@{}c@{}}Char.\\ velocity\\ $\bm{U}$ ($\bm{\frac{\text{mm}}{\text{s}}}$)\end{tabular}} & \textbf{\begin{tabular}[c]{@{}c@{}}$\bm{M}$ \end{tabular}} & \textbf{\begin{tabular}[c]{@{}c@{}}Res.\\ (norm)\end{tabular}}\\ \hline
$+$  & 15.1 & 0.0808 & 25.6  & 0.115 & 0.410 & 1.86 & 2.75 & 0.0695 & 3.36 & 8.13 (2.54) \\ \hline
${++}$  & 13.1 & 0.0778 & 23.9  & 0.142 & 0.469 & 1.71 & 3 & 0.0637 & 3.14 & 7.60 (2.49) \\ \hline
$+$ (d)  & 15.5 & 0.0619 & 36.5  & 0.145 & 0.309 & 1.54 & 2.75 & 0.0695 & 4.49 & 8.05 (2.54) \\ \hline
\end{tabular}
\caption{\footnotesize{Results from fitting the S18v2t4 7:30 spot with three parameters. The initial TF thickness estimate is 2.48 $\mu$m and the initial FL concentration estimate is 0.363\%. The minimum experimental FL intensity is 0.143. The optimized parameters are the evaporative thinning rate $v'$, the the glob radius or glob half-width $R_I'/X_I'$, and the change in surface tension $(\Delta \sigma)_0$. The minimum values of both the experimental and theoretical FL intensity and the theoretical thickness are reported. Case (d) evaporation is denoted with a (d). Scalings used in the nondimensionalizations of the model in each optimization are given.
}}
\label{table:results_sup}
\end{table}

\subsection{Evaporation-only results}
\label{sec:evap}

\begin{figure}[H]
\centering
\subfloat[][Exp. (---) and best fit th. (- - -) FL intensity] {\includegraphics[scale=.15]{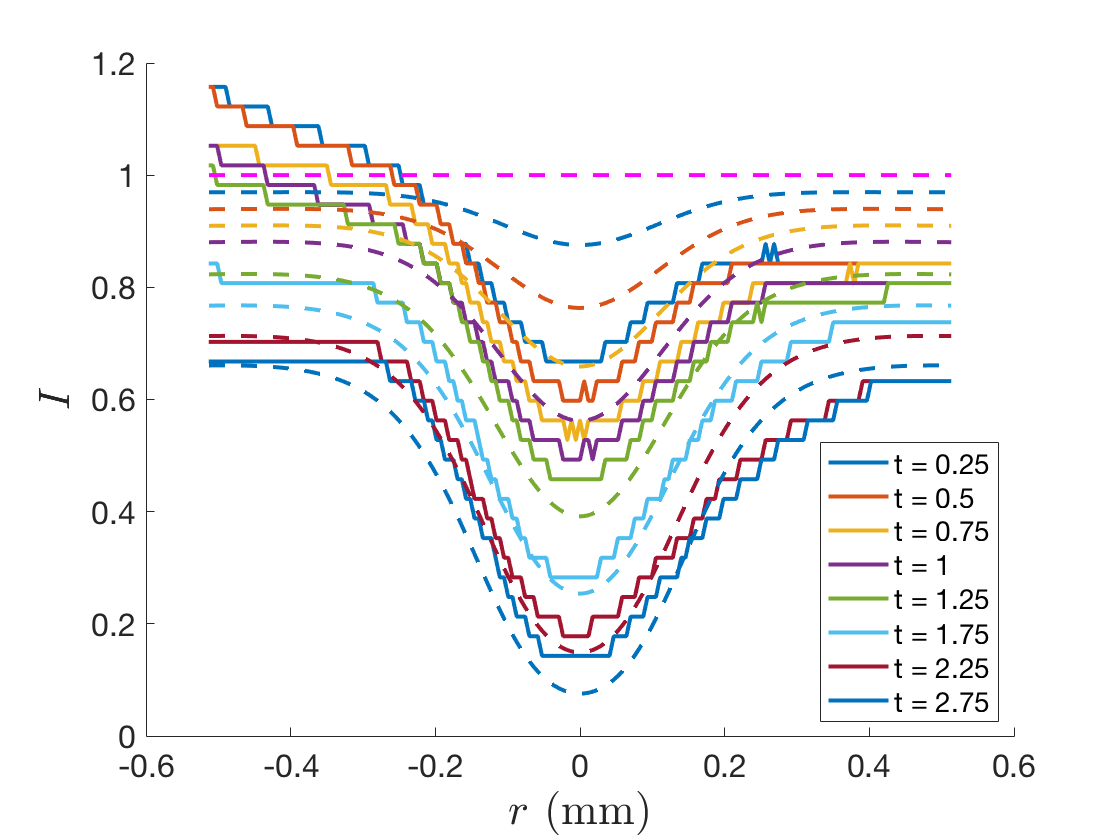}}
\subfloat[][Theoretical TF thickness]{\includegraphics[scale=.15]{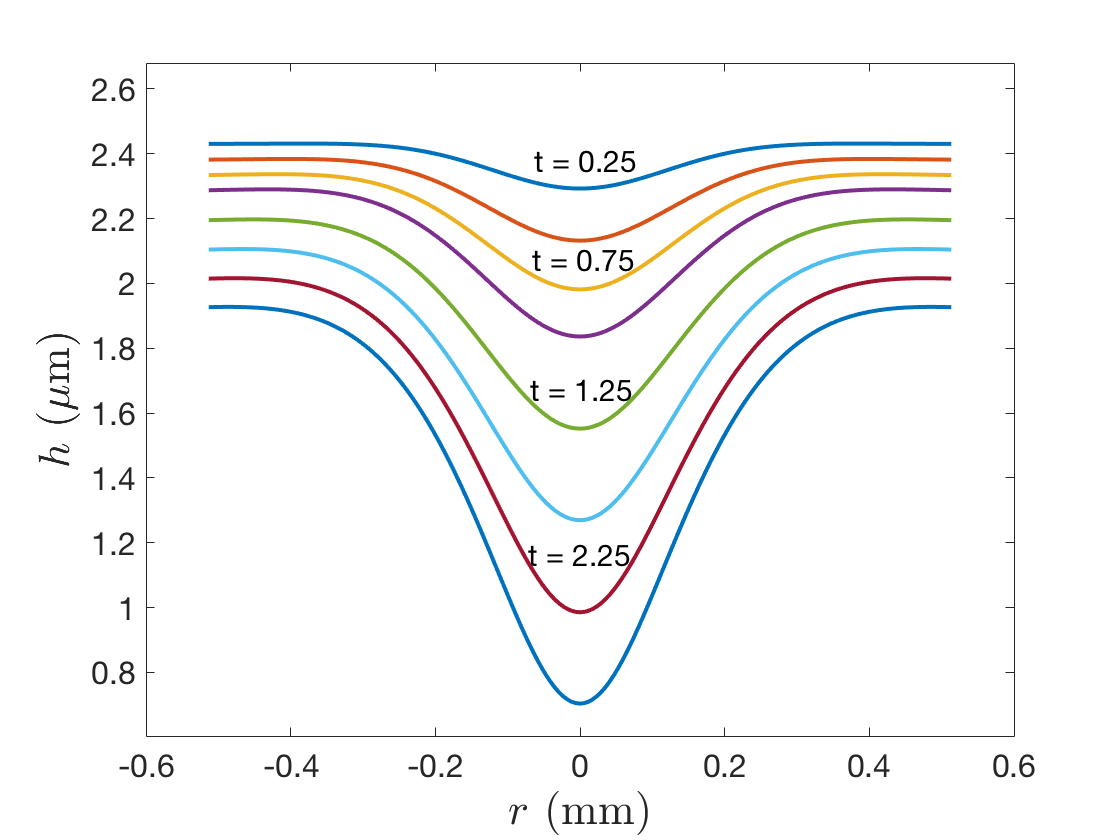}} \\
\subfloat[][Theoretical osmolarity]{\includegraphics[scale=.15]{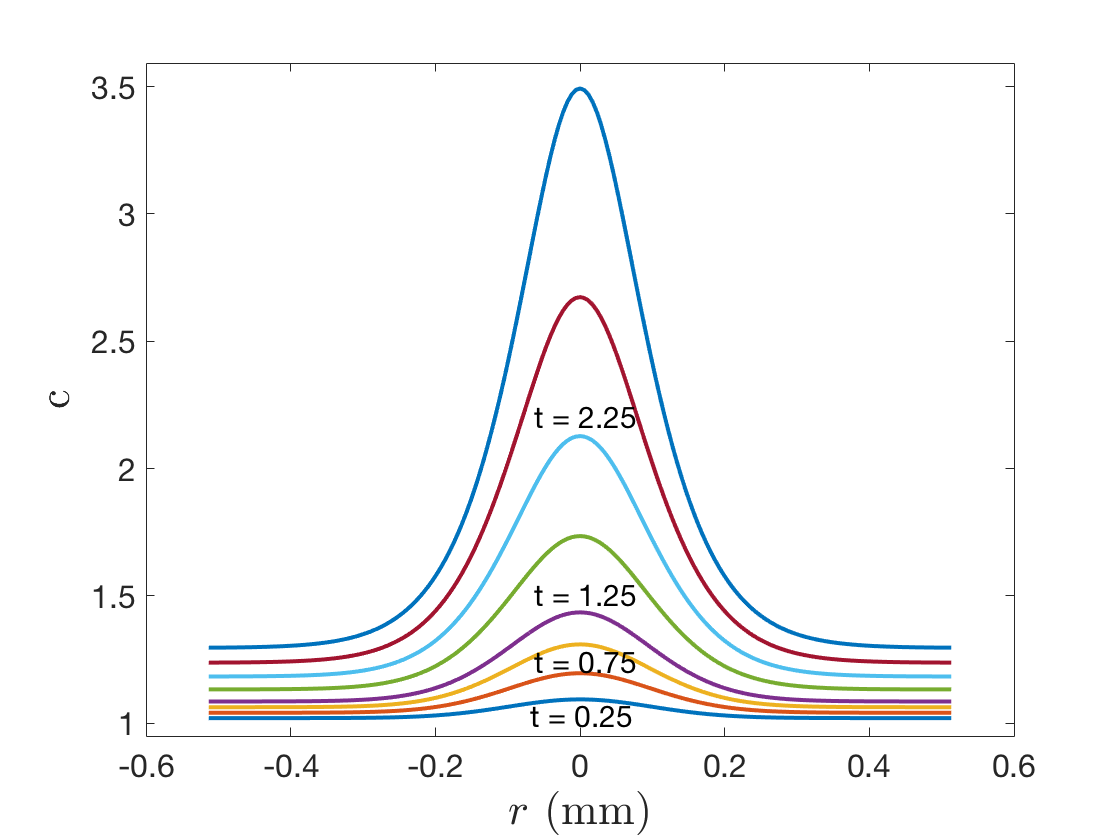}}
\begin{minipage}{18em}
\caption{\footnotesize{S18v2t4 7:30 best spot results (evaporation-only model). FL intensity has been normalized. Theoretical osmolarity is given as a fraction of the isotonic value.}}
\vspace{2cm}
\label{fig:S18v2t4_730_fit_e}
\end{minipage}
\end{figure}

In Figure \ref{fig:S18v2t4_730_fit_e} we show the results of fitting the S18v2t4 7:30 spot instance with an evaporation-only model. This instance is fit well with the mixed-mechanism model; ignoring the Marangoni effect causes the peak evaporation rate to more than double to 53.3 $\mu$m/min, a speed outside experimental ranges. For this reason, we take this as strong evidence that the Marangoni effect plays a crucial role in causing this breakup instance. 

\begin{figure}[H]
\centering
\subfloat[][FL intensity with minima  aligned]{\includegraphics[scale=.12]{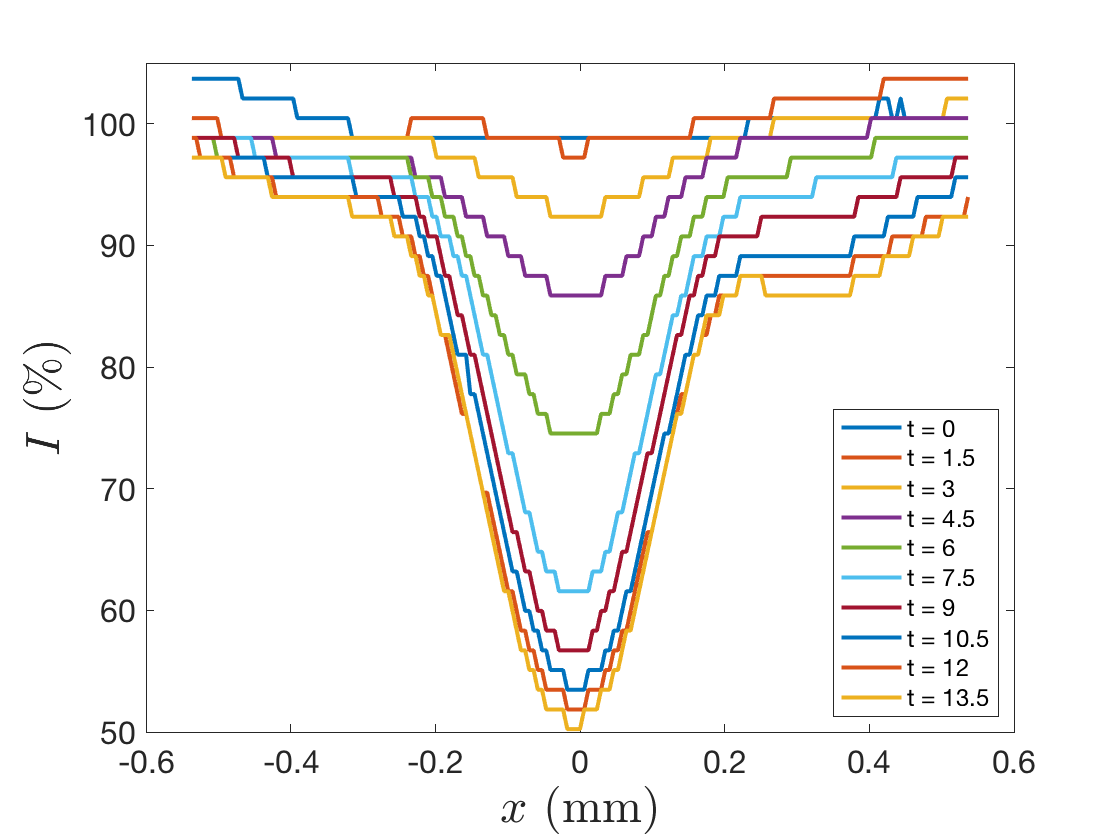}}
\subfloat[][FL intensity decrease]{\includegraphics[scale=.12]{5_11_20_S10v1t2_8_dec.png}}
\subfloat[][FT-TBU data extraction]{\includegraphics[scale=.04]{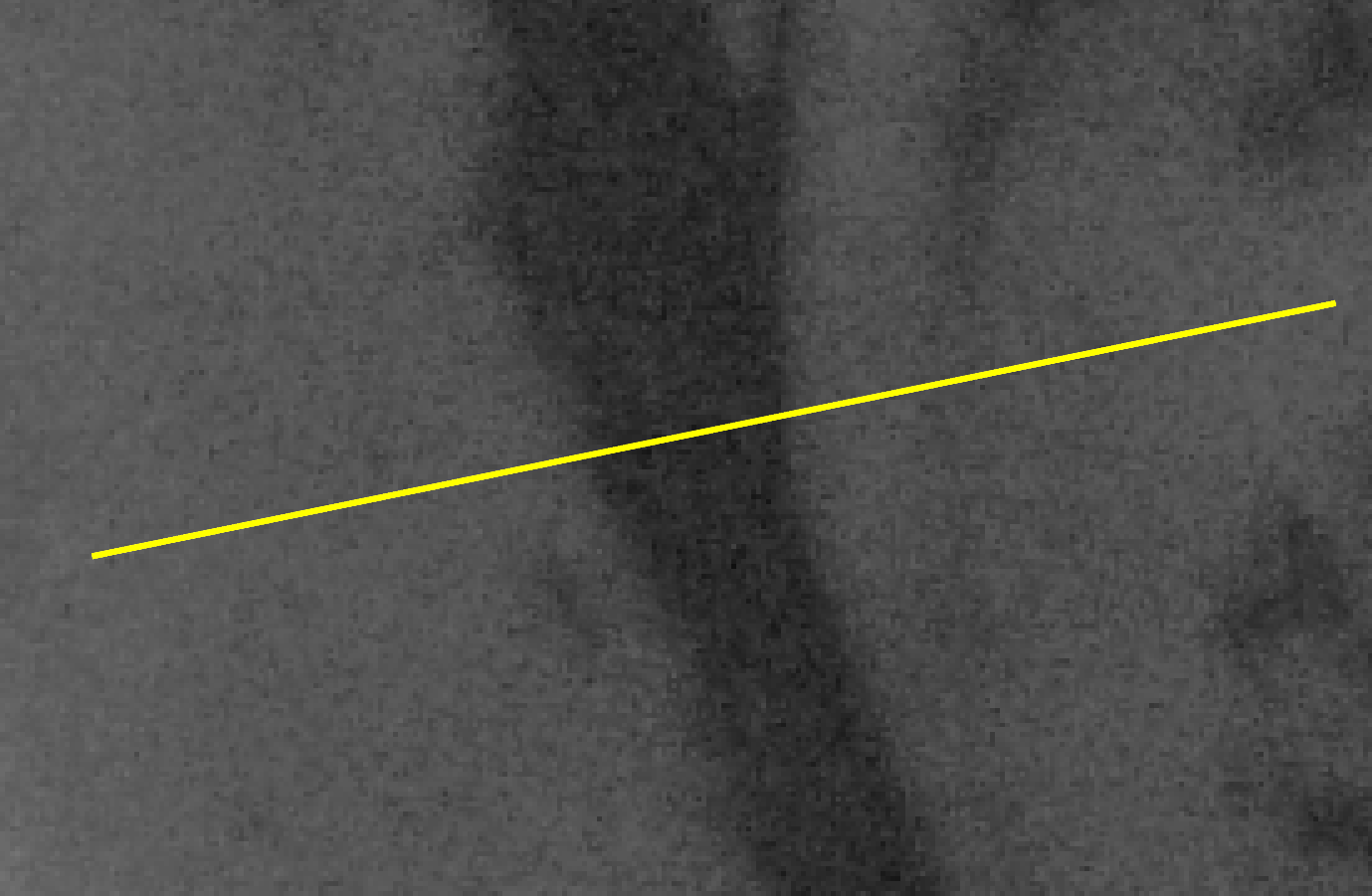}}
\caption{\footnotesize{Extracted data for the S10v1t2 8:00 streak. In (c) the image has been brightened and contrast-enhanced.}}
\label{fig:S10v1t2_8_data}
\end{figure}

\begin{figure}[H]
\centering
\subfloat[][Exp. (---) and best fit th. (- - -) FL intensity]{\includegraphics[scale=.15]{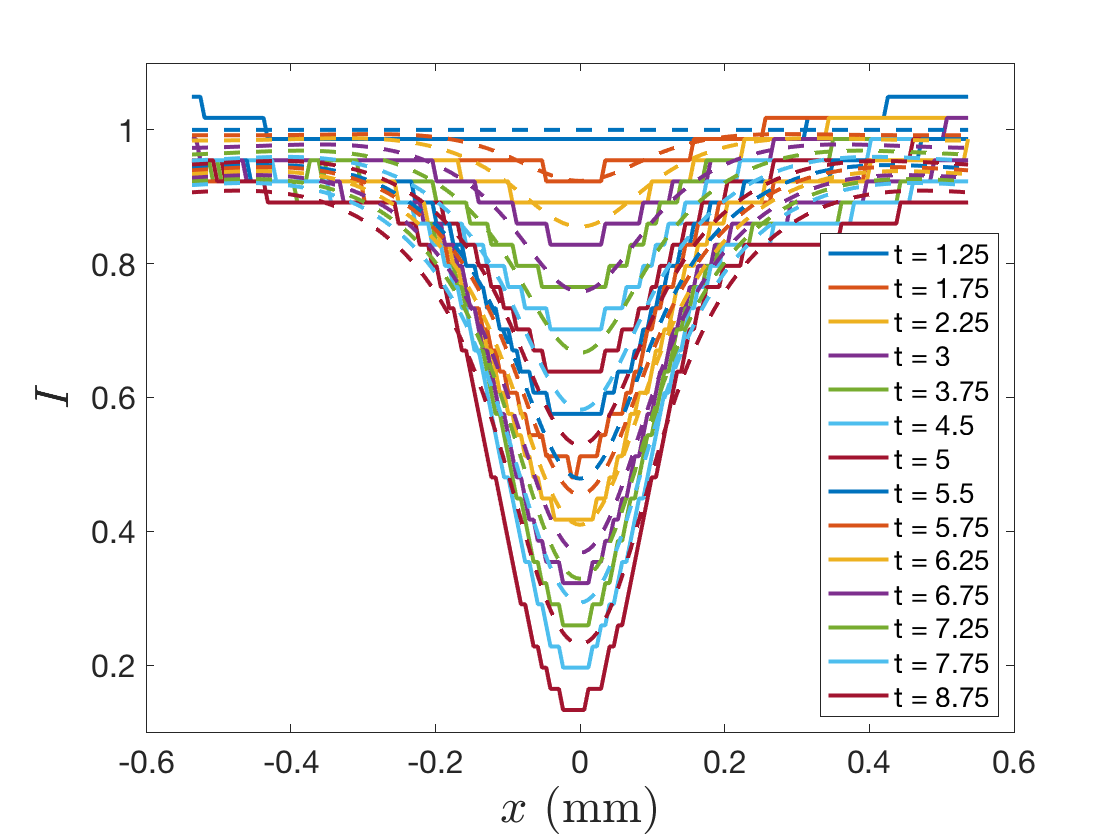}}
\subfloat[][Theoretical TF thickness]{\includegraphics[scale=.15]{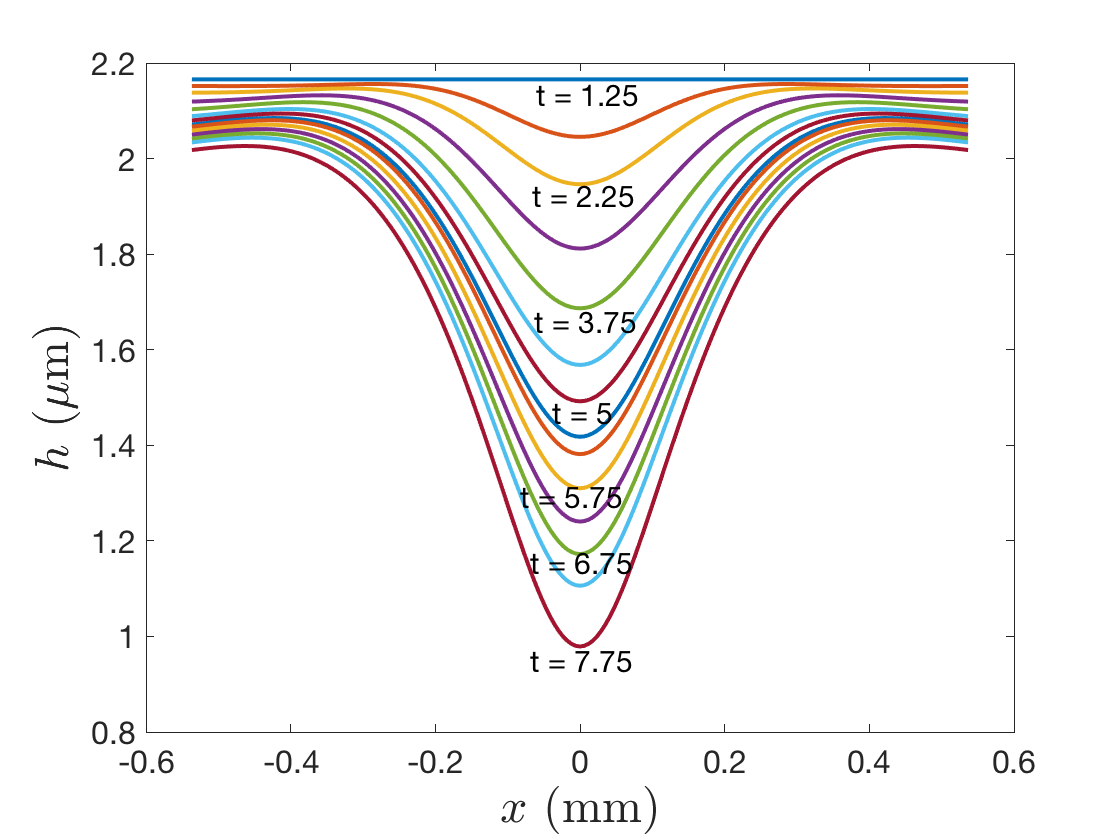}} \\
\subfloat[][Theoretical osmolarity]{\includegraphics[scale=.15]{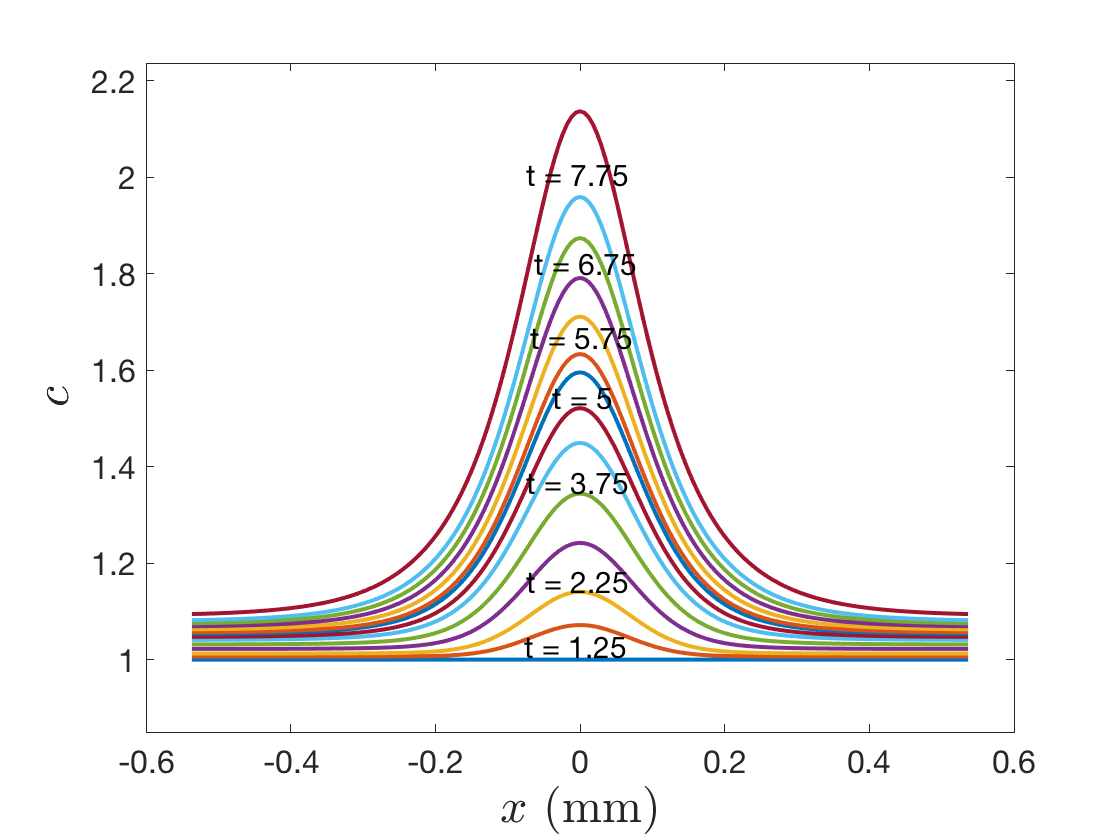}}
\subfloat[][Theoretical depth-averaged fluid velocity]{\includegraphics[scale=.15]{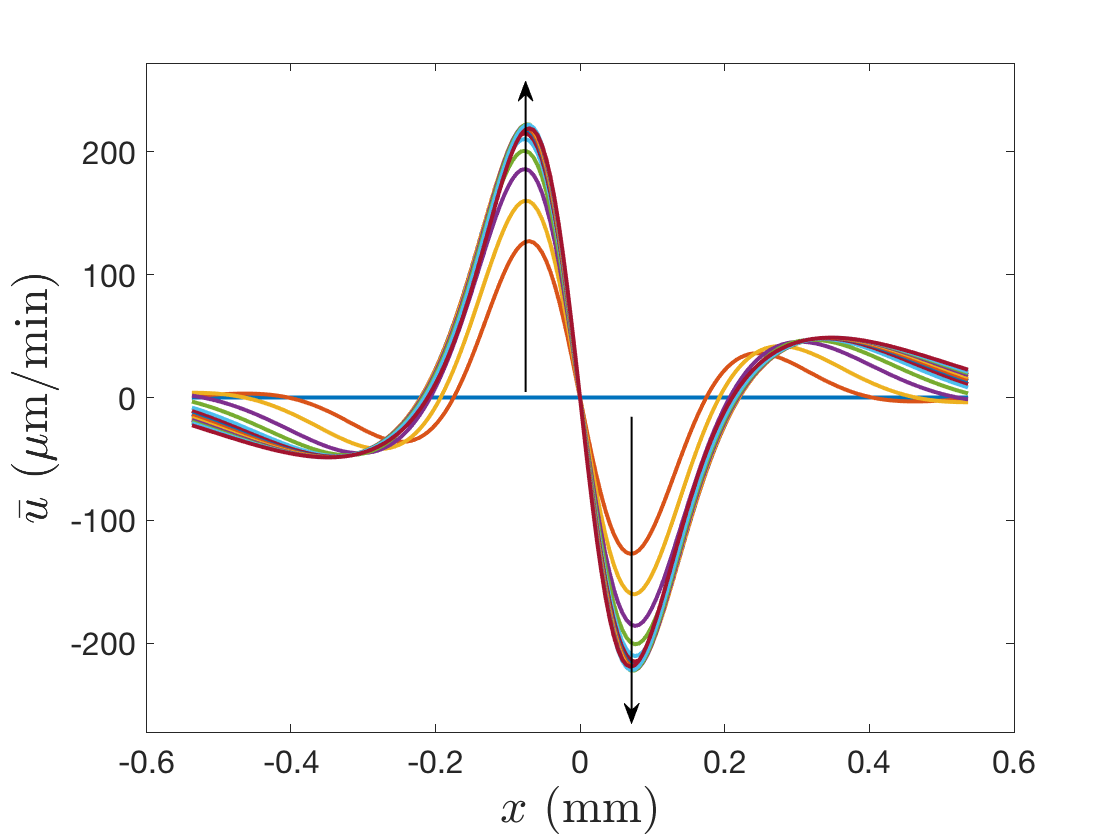}}
\caption{\footnotesize{S10v1t2 8:00 streak best fit results (evaporation-only model). FL intensity has been normalized. Theoretical osmolarity is given as a fraction of the isotonic value. Arrows indicate increasing time.}}
\label{fig:S10v1t2_8_fit}
\end{figure}

In Figure \ref{fig:S10v1t2_8_data} we show the data for the S10v1t2 8:00 streak. This instance is fit with the mixed-mechanism model (not shown) and the evaporation-only model in Figure \ref{fig:S10v1t2_8_fit}. The optimal parameters of the fit with the mixed-mechanism model suggest that the Marangoni effect is essentially nonexistent and therefore evaporation dominates the thinning. This is supported by the realistic peak evaporation rate (19.3 $\mu$m/min) from the evaporation-only model fit. The inward tangential flow seen in Figure \ref{fig:S10v1t2_8_fit}d is characteristic of evaporation-dominated thinning, further support for the lack of importance of the Marangoni effect.

\vspace{-2cm}

\subsection{Marangoni effect-only results}
\label{sec:mar}

\begin{figure}[H]
\centering
\subfloat[][Exp. (---) and best fit th. (- - -) FL intensity]{\includegraphics[scale=.15]{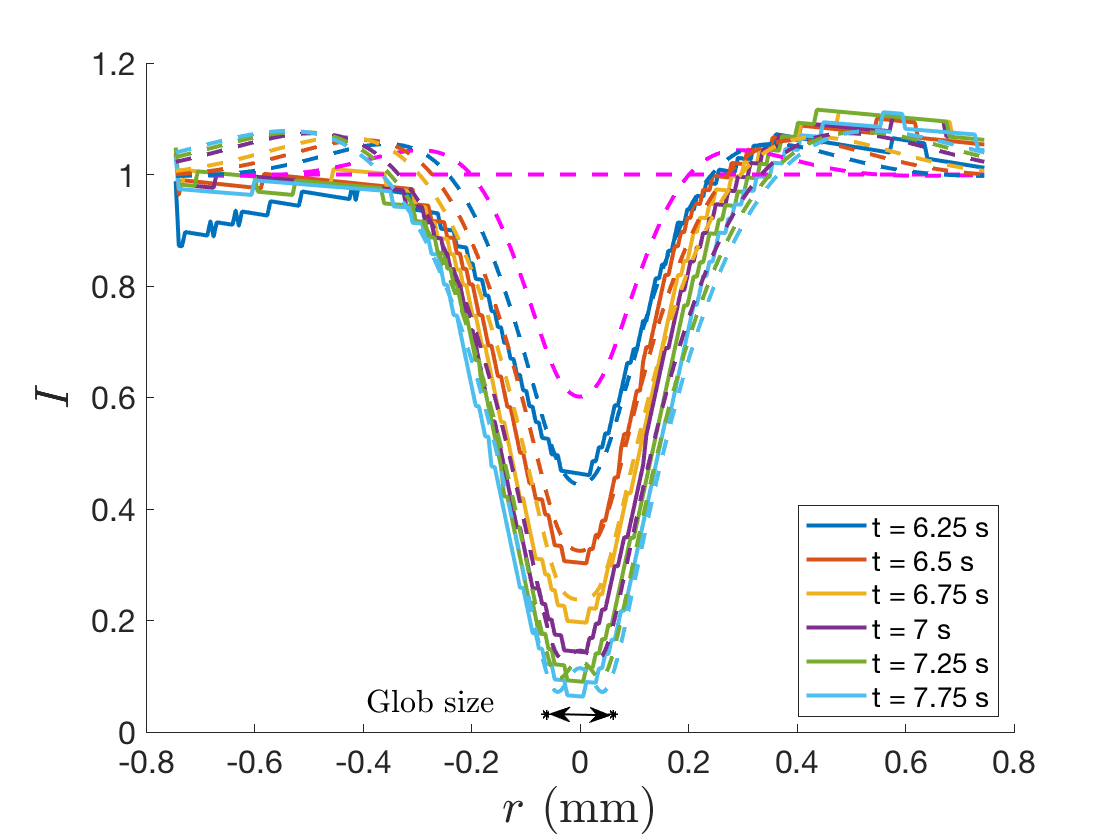}}
\subfloat[][Theoretical TF thickness]{\includegraphics[scale=.15]{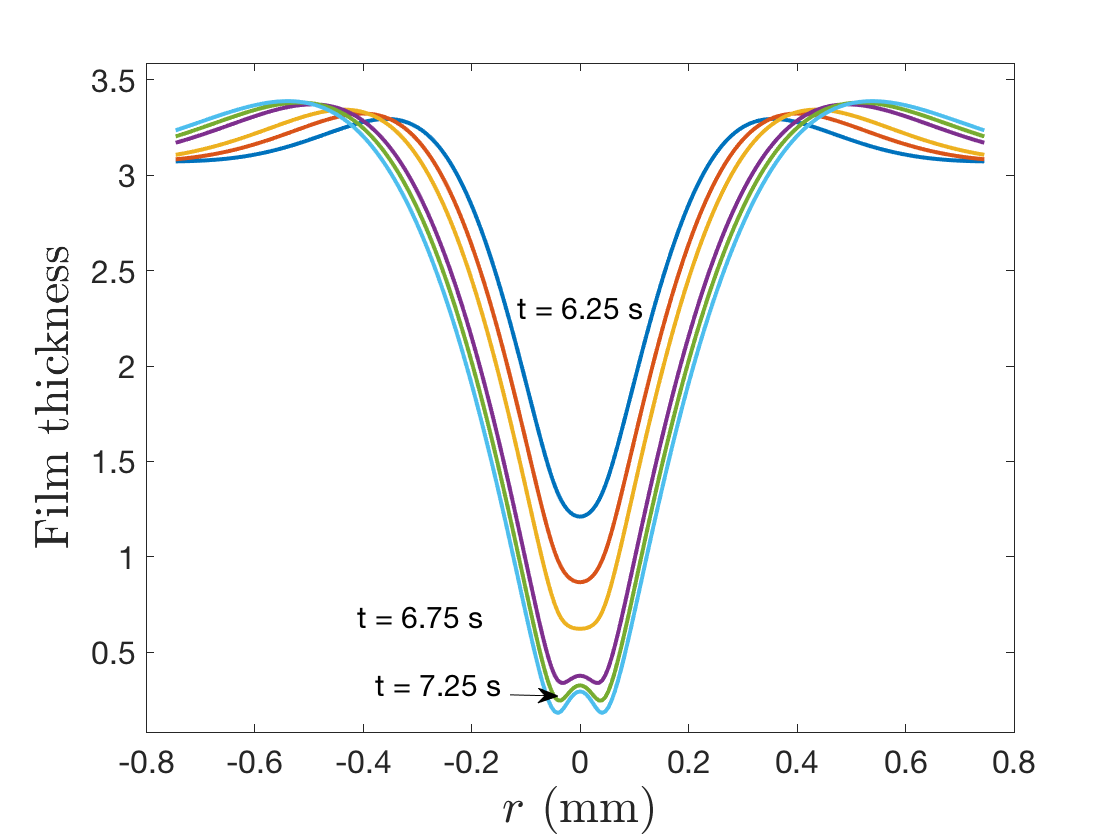}}
\caption{\footnotesize{S10v1t6 12:30 spot best fit results (Marangoni effect-only model). FL intensity has been normalized. 
}}
\label{fig:S10v1t6_1230_fit_m}
\end{figure}

In Figure \ref{fig:S10v1t6_1230_fit_m} we show the results of fitting the S10v1t 12:30 spot instance with a Marangoni effect-only model. Fitting this instance with the mixed-mechanism model gives a 6\% smaller residual and the optimal Marangoni numbers vary by less than 1\%; this is evidence that evaporation plays a relatively weak role in creating the thinning.

\begin{figure}[H]
\centering
\subfloat[][Exp. (---) and best fit th. (- - -) FL intensity]{\includegraphics[scale=.15]{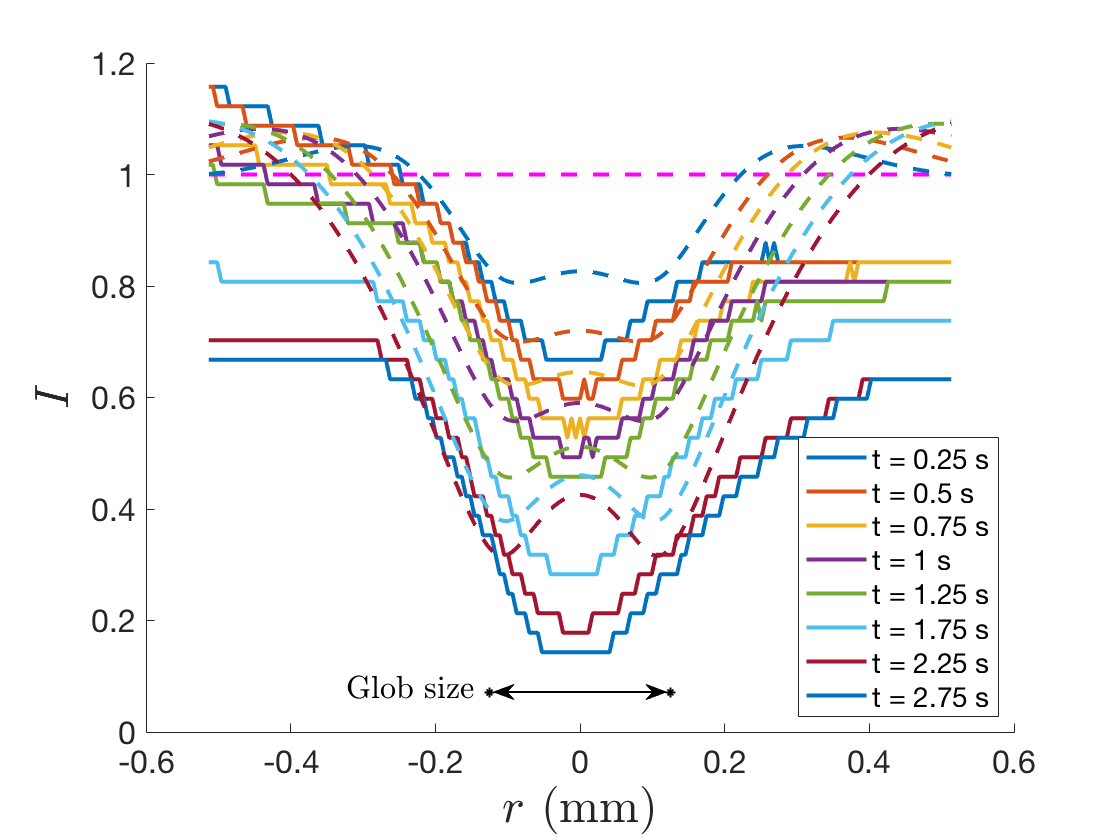}}
\subfloat[][Theoretical TF thickness]{\includegraphics[scale=.15]{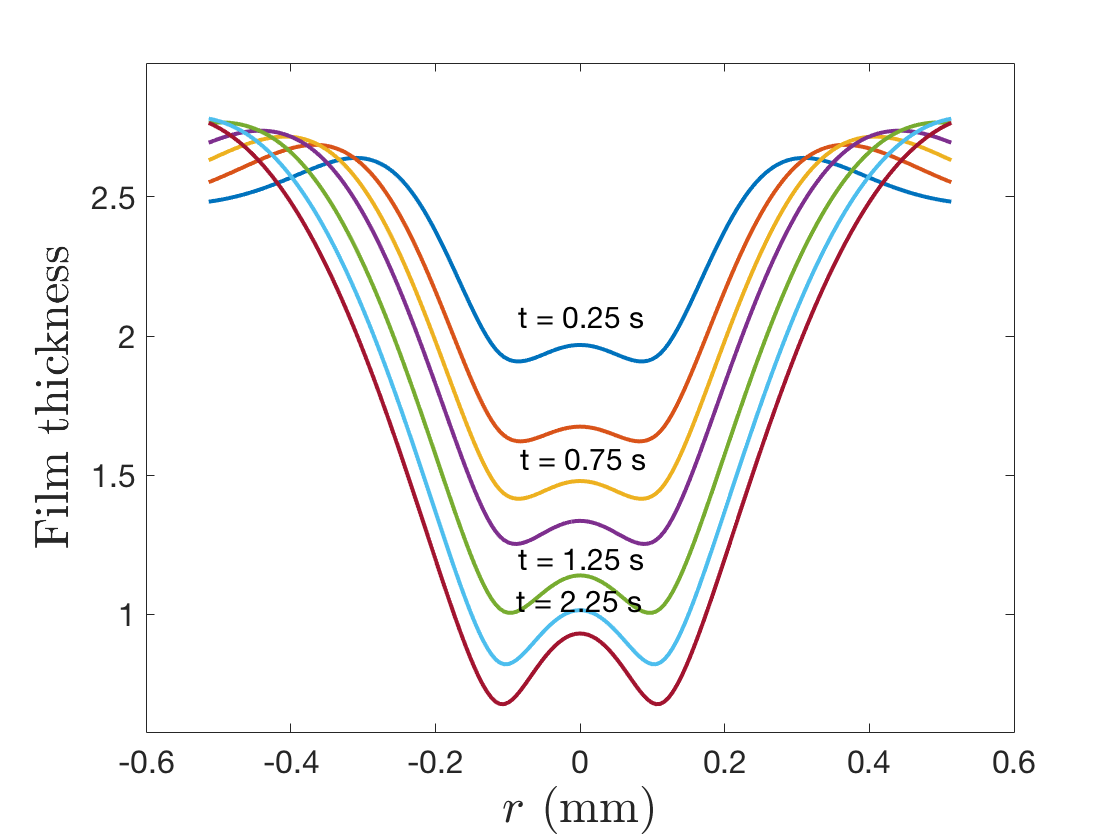}} 
\caption{\footnotesize{S18v2t4 7:30 spot best fit results (Marangoni effect-only model). FL intensity has been normalized.}}
\label{fig:S18v2t4_730_fit_m}
\end{figure}

We show the Marangoni effect-only fit to the S18v2t4 7:30 spot in Figure \ref{fig:S18v2t4_730_fit_m}. The residual is significantly larger in comparison to the fit with the mixed-mechanism model. Both the center and the sides of the experimental FL data in later times are fit very poorly with the theoretical FL intensity from the Marangoni effect-only model. Although we argue in Section \ref{sec:mod} that evaporation plays a small role in thinning in this instance, our results here show that excluding it entirely produces a far worse fit.

\end{document}